\documentclass[11pt,a4paper]{article}
\pdfoutput=1
\usepackage{jheppub}
\usepackage{amsmath,amsfonts,amssymb,bbm, mathtools, mathrsfs}
\usepackage{hyperref, csquotes}
\usepackage{graphicx, color, svg,subcaption}
\usepackage{tikz,pgf,tikz-cd,float}
\usepackage{physics,tensor,upgreek}
\usepackage{rotating}
\usepackage[nottoc,numbib]{tocbibind}
\usetikzlibrary{external}

\definecolor{jred}{rgb}{0.8,0,0}
\definecolor{jgreen}{rgb}{0,0.7,0}
\definecolor{jblue}{rgb}{0,0,0.8}

\usetikzlibrary{arrows,patterns,shapes,positioning,fit}
\tikzstyle{c} =	[coordinate]
\tikzstyle{vc} = [circle, draw=black, line width=1pt, inner sep=0pt, minimum size=2mm]
\tikzstyle{v} =  [circle, draw=black, line width=.2pt, fill=black, inner sep=0pt, minimum size=1.5mm]
\tikzstyle{vb} =[circle, draw=black, line width=.1pt, fill=jred, inner sep=0pt, minimum size=1mm]
\tikzstyle{vh} =[circle, draw=black, line width=.1pt, fill=jblue, inner sep=0pt, minimum size=1.5mm]
\tikzstyle{vs} =[coordinate]
\tikzstyle{e} =	[draw=jred,line width=1pt]
\tikzstyle{eb}= [draw=jgreen,line width=1pt]
\tikzstyle{es}= [dashed, line width=1pt]
\tikzstyle{ev}= [draw=jgreen,line width=1pt]
\tikzstyle{f} = [line width=0.01pt,dotted,fill=blue, fill opacity=.1]
\tikzstyle{cs}= [ellipse, draw=black, line width=.1pt, fill=jred, inner sep=1pt, minimum size=2mm]

\newcommand{\cvfpppp}{
\begin{tikzpicture}[scale=2,x=2ex,y=2ex,baseline={([yshift=-.59ex]current bounding box.center)}] 
\scriptsize{
\node [v]	(1)	at (0,0)	{};
\node [vb, label=above:+]	(b1)	at (-1,1)	{};
\node [vb, label=below:+]	(w1)	at (-1,-1)	{};
\node [vb, label=below:+]	(b2)	at (1,-1)	{};
\node [vb, label=above:+]	(w2)	at (1,1)	{};
\path
\foreach \i/\j in {1/2}{
   	(w\i) edge [ev]   (b\j)
	(b\i) edge [ev]  node [c] {} (w\j)
	}
\foreach \i in {1,2}{
	(w\i)   edge 	[ev]				(b\i)
	(w\i)	edge [ev,bend left=30]	(b\i)
	(w\i)	edge [ev,bend right=30]	(b\i)
	}
\foreach \i in {w1,b1,w2,b2}{
   	(\i) edge [e] (1)
	};
}
\end{tikzpicture}
}

\newcommand{\cvfpppm}{
\begin{tikzpicture}[scale=2,x=2ex,y=2ex,baseline={([yshift=-.59ex]current bounding box.center)}] 
\scriptsize{
\node [v]	(1)	at (0,0)	{};
\node [vb, label=above:\textbf{--}]	(b1)	at (-1,1)	{};
\node [vb, label=below:+]	(w1)	at (-1,-1)	{};
\node [vb, label=below:+]	(b2)	at (1,-1)	{};
\node [vb, label=above:+]	(w2)	at (1,1)	{};
\path
\foreach \i/\j in {1/2}{
   	(w\i) edge [ev]   (b\j)
	(b\i) edge [ev]  node [c] {} (w\j)
	}
\foreach \i in {1,2}{
	(w\i)   edge 	[ev]				(b\i)
	(w\i)	edge [ev,bend left=30]	(b\i)
	(w\i)	edge [ev,bend right=30]	(b\i)
	}
\foreach \i in {w1,b1,w2,b2}{
   	(\i) edge [e] (1)
	};
}
\end{tikzpicture}
}

\newcommand{\cvfppmma}{
\begin{tikzpicture}[scale=2,x=2ex,y=2ex,baseline={([yshift=-.59ex]current bounding box.center)}] 
\scriptsize{
\node [v]	(1)	at (0,0)	{};
\node [vb, label=above:+]	(b1)	at (-1,1)	{};
\node [vb, label=below:+]	(w1)	at (-1,-1)	{};
\node [vb, label=below:\textbf{--}]	(b2)	at (1,-1)	{};
\node [vb, label=above:\textbf{--}]	(w2)	at (1,1)	{};
\path
\foreach \i/\j in {1/2}{
   	(w\i) edge [ev]   (b\j)
	(b\i) edge [ev]  node [c] {} (w\j)
	}
\foreach \i in {1,2}{
	(w\i)   edge 	[ev]				(b\i)
	(w\i)	edge [ev,bend left=30]	(b\i)
	(w\i)	edge [ev,bend right=30]	(b\i)
	}
\foreach \i in {w1,b1,w2,b2}{
   	(\i) edge [e] (1)
	};
}
\end{tikzpicture}
}

\newcommand{\cvfppmmb}{
\begin{tikzpicture}[scale=2,x=2ex,y=2ex,baseline={([yshift=-.59ex]current bounding box.center)}] 
\scriptsize{
\node [v]	(1)	at (0,0)	{};
\node [vb, label=above:\textbf{--}]	(b1)	at (-1,1)	{};
\node [vb, label=below:+]	(w1)	at (-1,-1)	{};
\node [vb, label=below:+]	(b2)	at (1,-1)	{};
\node [vb, label=above:\textbf{--}]	(w2)	at (1,1)	{};
\path
\foreach \i/\j in {1/2}{
   	(w\i) edge [ev]   (b\j)
	(b\i) edge [ev]  node [c] {} (w\j)
	}
\foreach \i in {1,2}{
	(w\i)   edge 	[ev]				(b\i)
	(w\i)	edge [ev,bend left=30]	(b\i)
	(w\i)	edge [ev,bend right=30]	(b\i)
	}
\foreach \i in {w1,b1,w2,b2}{
   	(\i) edge [e] (1)
	};
}
\end{tikzpicture}
}

\newcommand{\cvfppmmc}{
\begin{tikzpicture}[scale=2,x=2ex,y=2ex,baseline={([yshift=-.59ex]current bounding box.center)}] 
\scriptsize{
\node [v]	(1)	at (0,0)	{};
\node [vb, label=above:\textbf{--}]	(b1)	at (-1,1)	{};
\node [vb, label=below:+]	(w1)	at (-1,-1)	{};
\node [vb, label=below:\textbf{--}]	(b2)	at (1,-1)	{};
\node [vb, label=above:+]	(w2)	at (1,1)	{};
\path
\foreach \i/\j in {1/2}{
   	(w\i) edge [ev]   (b\j)
	(b\i) edge [ev]  node [c] {} (w\j)
	}
\foreach \i in {1,2}{
	(w\i)   edge 	[ev]				(b\i)
	(w\i)	edge [ev,bend left=30]	(b\i)
	(w\i)	edge [ev,bend right=30]	(b\i)
	}
\foreach \i in {w1,b1,w2,b2}{
   	(\i) edge [e] (1)
	};
}
\end{tikzpicture}
}

\newcommand{\cvfppmza}{
\begin{tikzpicture}[scale=2,x=2ex,y=2ex,baseline={([yshift=-.59ex]current bounding box.center)}] 
\scriptsize{
\node [v]	(1)	at (0,0)	{};
\node [vb, label=above:+]	(b1)	at (-1,1)	{};
\node [vb, label=below:+]	(w1)	at (-1,-1)	{};
\node [vb, label=below:0]	(b2)	at (1,-1)	{};
\node [vb, label=above:\textbf{--}]	(w2)	at (1,1)	{};
\path
\foreach \i/\j in {1/2}{
   	(w\i) edge [ev]   (b\j)
	(b\i) edge [ev]  node [c] {} (w\j)
	}
\foreach \i in {1,2}{
	(w\i)   edge 	[ev]				(b\i)
	(w\i)	edge [ev,bend left=30]	(b\i)
	(w\i)	edge [ev,bend right=30]	(b\i)
	}
\foreach \i in {w1,b1,w2,b2}{
   	(\i) edge [e] (1)
	};
}
\end{tikzpicture}
}

\newcommand{\cvfppmzb}{
\begin{tikzpicture}[scale=2,x=2ex,y=2ex,baseline={([yshift=-.59ex]current bounding box.center)}] 
\scriptsize{
\node [v]	(1)	at (0,0)	{};
\node [vb, label=above:\textbf{--}]	(b1)	at (-1,1)	{};
\node [vb, label=below:+]	(w1)	at (-1,-1)	{};
\node [vb, label=below:+]	(b2)	at (1,-1)	{};
\node [vb, label=above:0]	(w2)	at (1,1)	{};
\path
\foreach \i/\j in {1/2}{
   	(w\i) edge [ev]   (b\j)
	(b\i) edge [ev]  node [c] {} (w\j)
	}
\foreach \i in {1,2}{
	(w\i)   edge 	[ev]				(b\i)
	(w\i)	edge [ev,bend left=30]	(b\i)
	(w\i)	edge [ev,bend right=30]	(b\i)
	}
\foreach \i in {w1,b1,w2,b2}{
   	(\i) edge [e] (1)
	};
}
\end{tikzpicture}
}

\newcommand{\cvfppmzc}{
\begin{tikzpicture}[scale=2,x=2ex,y=2ex,baseline={([yshift=-.59ex]current bounding box.center)}] 
\scriptsize{
\node [v]	(1)	at (0,0)	{};
\node [vb, label=above:\textbf{--}]	(b1)	at (-1,1)	{};
\node [vb, label=below:+]	(w1)	at (-1,-1)	{};
\node [vb, label=below:0]	(b2)	at (1,-1)	{};
\node [vb, label=above:+]	(w2)	at (1,1)	{};
\path
\foreach \i/\j in {1/2}{
   	(w\i) edge [ev]   (b\j)
	(b\i) edge [ev]  node [c] {} (w\j)
	}
\foreach \i in {1,2}{
	(w\i)   edge 	[ev]				(b\i)
	(w\i)	edge [ev,bend left=30]	(b\i)
	(w\i)	edge [ev,bend right=30]	(b\i)
	}
\foreach \i in {w1,b1,w2,b2}{
   	(\i) edge [e] (1)
	};
}
\end{tikzpicture}
}

\newcommand{\cvftpppp}{
\begin{tikzpicture}[scale=2,x=2ex,y=2ex,baseline={([yshift=-.59ex]current bounding box.center)}] 
\node [v]	(v)	at (0,0) 	{};
\node [vb, label=above:+]	(1)	at (-1,1)	{};
\node [vb, label=below:+]	(2)	at (-1,-1)	{};
\node [vb, label=below:+]	(3)	at (1,-1)	{};
\node [vb, label=above:+]	(4)	at (1,1)	{};
\path
\foreach \i in {1,2,3,4}{
    (\i) edge [e] (v)
    }
\foreach \i/\j in {1/2,2/1,3/4,4/3}{
   (\i)	edge [ev,bend left=15]	(\j)
   (\i)	edge [ev,bend left=40]	(\j)
   };
\end{tikzpicture}
}

\newcommand{\cvftpppm}{
\begin{tikzpicture}[scale=2,x=2ex,y=2ex,baseline={([yshift=-.59ex]current bounding box.center)}] 
\node [v]	(v)	at (0,0) 	{};
\node [vb, label=above:+]	(1)	at (-1,1)	{};
\node [vb, label=below:+]	(2)	at (-1,-1)	{};
\node [vb, label=below:+]	(3)	at (1,-1)	{};
\node [vb, label=above:\textbf{--}]	(4)	at (1,1)	{};
\path
\foreach \i in {1,2,3,4}{
    (\i) edge [e] (v)
    }
\foreach \i/\j in {1/2,2/1,3/4,4/3}{
   (\i)	edge [ev,bend left=15]	(\j)
   (\i)	edge [ev,bend left=40]	(\j)
   };
\end{tikzpicture}
}

\newcommand{\cvftppmma}{
\begin{tikzpicture}[scale=2,x=2ex,y=2ex,baseline={([yshift=-.59ex]current bounding box.center)}] 
\node [v]	(v)	at (0,0) 	{};
\node [vb, label=above:+]	(1)	at (-1,1)	{};
\node [vb, label=below:+]	(2)	at (-1,-1)	{};
\node [vb, label=below:\textbf{--}]	(3)	at (1,-1)	{};
\node [vb, label=above:\textbf{--}]	(4)	at (1,1)	{};
\path
\foreach \i in {1,2,3,4}{
    (\i) edge [e] (v)
    }
\foreach \i/\j in {1/2,2/1,3/4,4/3}{
   (\i)	edge [ev,bend left=15]	(\j)
   (\i)	edge [ev,bend left=40]	(\j)
   };
\end{tikzpicture}
}

\newcommand{\cvftppmmb}{
\begin{tikzpicture}[scale=2,x=2ex,y=2ex,baseline={([yshift=-.59ex]current bounding box.center)}] 
\node [v]	(v)	at (0,0) 	{};
\node [vb, label=above:\textbf{--}]	(1)	at (-1,1)	{};
\node [vb, label=below:+]	(2)	at (-1,-1)	{};
\node [vb, label=below:+]	(3)	at (1,-1)	{};
\node [vb, label=above:\textbf{--}]	(4)	at (1,1)	{};
\path
\foreach \i in {1,2,3,4}{
    (\i) edge [e] (v)
    }
\foreach \i/\j in {1/2,2/1,3/4,4/3}{
   (\i)	edge [ev,bend left=15]	(\j)
   (\i)	edge [ev,bend left=40]	(\j)
   };
\end{tikzpicture}
}

\newcommand{\cvftppmza}{
\begin{tikzpicture}[scale=2,x=2ex,y=2ex,baseline={([yshift=-.59ex]current bounding box.center)}] 
\node [v]	(v)	at (0,0) 	{};
\node [vb, label=above:+]	(1)	at (-1,1)	{};
\node [vb, label=below:+]	(2)	at (-1,-1)	{};
\node [vb, label=below:0]	(3)	at (1,-1)	{};
\node [vb, label=above:\textbf{--}]	(4)	at (1,1)	{};
\path
\foreach \i in {1,2,3,4}{
    (\i) edge [e] (v)
    }
\foreach \i/\j in {1/2,2/1,3/4,4/3}{
   (\i)	edge [ev,bend left=15]	(\j)
   (\i)	edge [ev,bend left=40]	(\j)
   };
\end{tikzpicture}
}
\newcommand{\cvftppmzb}{
\begin{tikzpicture}[scale=2,x=2ex,y=2ex,baseline={([yshift=-.59ex]current bounding box.center)}] 
\node [v]	(v)	at (0,0) 	{};
\node [vb, label=above:0]	(1)	at (-1,1)	{};
\node [vb, label=below:+]	(2)	at (-1,-1)	{};
\node [vb, label=below:+]	(3)	at (1,-1)	{};
\node [vb, label=above:\textbf{--}]	(4)	at (1,1)	{};
\path
\foreach \i in {1,2,3,4}{
    (\i) edge [e] (v)
    }
\foreach \i/\j in {1/2,2/1,3/4,4/3}{
   (\i)	edge [ev,bend left=15]	(\j)
   (\i)	edge [ev,bend left=40]	(\j)
   };
\end{tikzpicture}
}

\newcommand{\cvfnpppp}{
\begin{tikzpicture}[scale=2,x=2ex,y=2ex,baseline={([yshift=-.59ex]current bounding box.center)}] 
\scriptsize{
\node [v]	(1)	at (0,0)	{};
\node [vb, label=above:+]	(b1)	at (-1,1)	{};
\node [vb, label=below:+]	(w1)	at (-1,-1)	{};
\node [vb, label=below:+]	(b2)	at (1,-1)	{};
\node [vb, label=above:+]	(w2)	at (1,1)	{};
\path
\foreach \i/\j in {1/2,2/1}{
   	(w\i) edge [ev,bend left=30]  (b\j)
   	(w\i) edge [ev,bend right=30] (b\j)
	}
\foreach \i in {1,2}{
	(w\i)	edge [ev,bend left=30]  (b\i)
	(w\i)	edge [ev,bend right=30]	(b\i)
	}
\foreach \i in {w1,b1,w2,b2}{
   	(\i) edge [e] (1)
	};
}
\end{tikzpicture}
}

\newcommand{\cvfnpppm}{
\begin{tikzpicture}[scale=2,x=2ex,y=2ex,baseline={([yshift=-.59ex]current bounding box.center)}] 
\scriptsize{
\node [v]	(1)	at (0,0)	{};
\node [vb, label=above:\textbf{--}]	(b1)	at (-1,1)	{};
\node [vb, label=below:+]	(w1)	at (-1,-1)	{};
\node [vb, label=below:+]	(b2)	at (1,-1)	{};
\node [vb, label=above:+]	(w2)	at (1,1)	{};
\path
\foreach \i/\j in {1/2,2/1}{
   	(w\i) edge [ev,bend left=30]  (b\j)
   	(w\i) edge [ev,bend right=30] (b\j)
	}
\foreach \i in {1,2}{
	(w\i)	edge [ev,bend left=30]  (b\i)
	(w\i)	edge [ev,bend right=30]	(b\i)
	}
\foreach \i in {w1,b1,w2,b2}{
   	(\i) edge [e] (1)
	};
}
\end{tikzpicture}
}

\newcommand{\cvfnppmma}{
\begin{tikzpicture}[scale=2,x=2ex,y=2ex,baseline={([yshift=-.59ex]current bounding box.center)}] 
\scriptsize{
\node [v]	(1)	at (0,0)	{};
\node [vb, label=above:+]	(b1)	at (-1,1)	{};
\node [vb, label=below:+]	(w1)	at (-1,-1)	{};
\node [vb, label=below:\textbf{--}]	(b2)	at (1,-1)	{};
\node [vb, label=above:\textbf{--}]	(w2)	at (1,1)	{};
\path
\foreach \i/\j in {1/2,2/1}{
   	(w\i) edge [ev,bend left=30]  (b\j)
   	(w\i) edge [ev,bend right=30] (b\j)
	}
\foreach \i in {1,2}{
	(w\i)	edge [ev,bend left=30]  (b\i)
	(w\i)	edge [ev,bend right=30]	(b\i)
	}
\foreach \i in {w1,b1,w2,b2}{
   	(\i) edge [e] (1)
	};
}
\end{tikzpicture}
}

\newcommand{\cvfnppmmb}{
\begin{tikzpicture}[scale=2,x=2ex,y=2ex,baseline={([yshift=-.59ex]current bounding box.center)}] 
\scriptsize{
\node [v]	(1)	at (0,0)	{};
\node [vb, label=above:\textbf{--}]	(b1)	at (-1,1)	{};
\node [vb, label=below:+]	(w1)	at (-1,-1)	{};
\node [vb, label=below:\textbf{--}]	(b2)	at (1,-1)	{};
\node [vb, label=above:+]	(w2)	at (1,1)	{};
\path
\foreach \i/\j in {1/2,2/1}{
   	(w\i) edge [ev,bend left=30]  (b\j)
   	(w\i) edge [ev,bend right=30] (b\j)
	}
\foreach \i in {1,2}{
	(w\i)	edge [ev,bend left=30]  (b\i)
	(w\i)	edge [ev,bend right=30]	(b\i)
	}
\foreach \i in {w1,b1,w2,b2}{
   	(\i) edge [e] (1)
	};
}
\end{tikzpicture}
}

\newcommand{\cvfnppmza}{
\begin{tikzpicture}[scale=2,x=2ex,y=2ex,baseline={([yshift=-.59ex]current bounding box.center)}] 
\scriptsize{
\node [v]	(1)	at (0,0)	{};
\node [vb, label=above:+]	(b1)	at (-1,1)	{};
\node [vb, label=below:+]	(w1)	at (-1,-1)	{};
\node [vb, label=below:0]	(b2)	at (1,-1)	{};
\node [vb, label=above:\textbf{--}]	(w2)	at (1,1)	{};
\path
\foreach \i/\j in {1/2,2/1}{
   	(w\i) edge [ev,bend left=30]  (b\j)
   	(w\i) edge [ev,bend right=30] (b\j)
	}
\foreach \i in {1,2}{
	(w\i)	edge [ev,bend left=30]  (b\i)
	(w\i)	edge [ev,bend right=30]	(b\i)
	}
\foreach \i in {w1,b1,w2,b2}{
   	(\i) edge [e] (1)
	};
}
\end{tikzpicture}
}

\newcommand{\cvfnppmzb}{
\begin{tikzpicture}[scale=2,x=2ex,y=2ex,baseline={([yshift=-.59ex]current bounding box.center)}] 
\scriptsize{
\node [v]	(1)	at (0,0)	{};
\node [vb, label=above:\textbf{--}]	(b1)	at (-1,1)	{};
\node [vb, label=below:+]	(w1)	at (-1,-1)	{};
\node [vb, label=below:0]	(b2)	at (1,-1)	{};
\node [vb, label=above:+]	(w2)	at (1,1)	{};
\path
\foreach \i/\j in {1/2,2/1}{
   	(w\i) edge [ev,bend left=30]  (b\j)
   	(w\i) edge [ev,bend right=30] (b\j)
	}
\foreach \i in {1,2}{
	(w\i)	edge [ev,bend left=30]  (b\i)
	(w\i)	edge [ev,bend right=30]	(b\i)
	}
\foreach \i in {w1,b1,w2,b2}{
   	(\i) edge [e] (1)
	};
}
\end{tikzpicture}
}

\newcommand{\cvfsppppp}{
\begin{tikzpicture}[scale=2,x=2ex,y=2ex,baseline={([yshift=-.59ex]current bounding box.center)}] 
\scriptsize{
\node [v]	(0)	at (0,0)	{};
\foreach \i in {72,288,360}{
\begin{scope}[rotate=\i]
\node [vb, label=above:+]	(\i)	at (0,1.6)	{};
\end{scope}
}
\foreach \i in {144,216}{
\begin{scope}[rotate=\i]
\node [vb, label=below:+]	(\i)	at (0,1.6)	{};
\end{scope}
}
\path
\foreach \i in {72,144,216,288,360}{
    \foreach \j in {72,144,216,288,360}{
   	    (\i) edge [ev]  (\j)
	    }
    }
\foreach \i in {72,144,216,288,360}{
  	(\i) edge [e] (0)
	};
}
\end{tikzpicture}
}

\newcommand{\cvfsppppm}{
\begin{tikzpicture}[scale=2,x=2ex,y=2ex,baseline={([yshift=-.59ex]current bounding box.center)}] 
\scriptsize{
\node [v]	(0)	at (0,0)	{};
\foreach \i in {72}{
\begin{scope}[rotate=\i]
\node [vb, label=above:\textbf{--}]	(\i)	at (0,1.6)	{};
\end{scope}
}
\foreach \i in {288,360}{
\begin{scope}[rotate=\i]
\node [vb, label=above:+]	(\i)	at (0,1.6)	{};
\end{scope}
}
\foreach \i in {144,216}{
\begin{scope}[rotate=\i]
\node [vb, label=below:+]	(\i)	at (0,1.6)	{};
\end{scope}
}
\path
\foreach \i in {72,144,216,288,360}{
    \foreach \j in {72,144,216,288,360}{
   	    (\i) edge [ev]  (\j)
	    }
    }
\foreach \i in {72,144,216,288,360}{
  	(\i) edge [e] (0)
	};
}
\end{tikzpicture}
}

\newcommand{\cvfspppmm}{
\begin{tikzpicture}[scale=2,x=2ex,y=2ex,baseline={([yshift=-.59ex]current bounding box.center)}] 
\scriptsize{
\node [v]	(0)	at (0,0)	{};
\foreach \i in {288}{
\begin{scope}[rotate=\i]
\node [vb, label=above:+]	(\i)	at (0,1.6)	{};
\end{scope}
}
\foreach \i in {72,360}{
\begin{scope}[rotate=\i]
\node [vb, label=above:\textbf{--}]	(\i)	at (0,1.6)	{};
\end{scope}
}
\foreach \i in {144,216}{
\begin{scope}[rotate=\i]
\node [vb, label=below:+]	(\i)	at (0,1.6)	{};
\end{scope}
}
\path
\foreach \i in {72,144,216,288,360}{
    \foreach \j in {72,144,216,288,360}{
   	    (\i) edge [ev]  (\j)
	    }
    }
\foreach \i in {72,144,216,288,360}{
  	(\i) edge [e] (0)
	};
}
\end{tikzpicture}
}

\newcommand{\cvfspppmz}{
\begin{tikzpicture}[scale=2,x=2ex,y=2ex,baseline={([yshift=-.59ex]current bounding box.center)}] 
\scriptsize{
\node [v]	(0)	at (0,0)	{};
\foreach \i in {288}{
\begin{scope}[rotate=\i]
\node [vb, label=above:+]	(\i)	at (0,1.6)	{};
\end{scope}
}
\foreach \i in {72}{
\begin{scope}[rotate=\i]
\node [vb, label=above:\textbf{--}]	(\i)	at (0,1.6)	{};
\end{scope}
}
\foreach \i in {360}{
\begin{scope}[rotate=\i]
\node [vb, label=above:0]	(\i)	at (0,1.6)	{};
\end{scope}
}
\foreach \i in {144,216}{
\begin{scope}[rotate=\i]
\node [vb, label=below:+]	(\i)	at (0,1.6)	{};
\end{scope}
}
\path
\foreach \i in {72,144,216,288,360}{
    \foreach \j in {72,144,216,288,360}{
   	    (\i) edge [ev]  (\j)
	    }
    }
\foreach \i in {72,144,216,288,360}{
  	(\i) edge [e] (0)
	};
}
\end{tikzpicture}
}

\newcommand{\cvfsppmzz}{
\begin{tikzpicture}[scale=2,x=2ex,y=2ex,baseline={([yshift=-.59ex]current bounding box.center)}] 
\scriptsize{
\node [v]	(0)	at (0,0)	{};
\foreach \i in {288}{
\begin{scope}[rotate=\i]
\node [vb, label=above:0]	(\i)	at (0,1.6)	{};
\end{scope}
}
\foreach \i in {72}{
\begin{scope}[rotate=\i]
\node [vb, label=above:\textbf{--}]	(\i)	at (0,1.6)	{};
\end{scope}
}
\foreach \i in {360}{
\begin{scope}[rotate=\i]
\node [vb, label=above:0]	(\i)	at (0,1.6)	{};
\end{scope}
}
\foreach \i in {144,216}{
\begin{scope}[rotate=\i]
\node [vb, label=below:+]	(\i)	at (0,1.6)	{};
\end{scope}
}
\path
\foreach \i in {72,144,216,288,360}{
    \foreach \j in {72,144,216,288,360}{
   	    (\i) edge [ev]  (\j)
	    }
    }
\foreach \i in {72,144,216,288,360}{
  	(\i) edge [e] (0)
	};
}
\end{tikzpicture}
}

\usepackage{accents}

\definecolor{timelike}{RGB}{227, 11, 91}
\definecolor{spacelike}{RGB}{0, 128, 128}
\definecolor{lightlike}{RGB}{0, 25, 150}

\newcommand{\one}{\mathbbm 1}
\newcommand{\R}{\mathbb R}
\newcommand{\C}{\mathbb C}
\newcommand{\e}{\textrm{e}}
\renewcommand{\i}{\textrm{i}}
\newcommand{\cas}{\textrm{Cas}}     
\newcommand{\SUT}{\mathrm{SU}(2)}
\newcommand{\SUO}{\mathrm{SU}(1,1)}
\newcommand{\ISO}{\mathrm{ISO}(2)}

\newcommand{\SL}{\text{SL$(2,\C)$}}

\newcommand{\Spin}{\mathrm{Spin}(4)}

\newcommand{\defeq}{\vcentcolon=}

\newcommand{\dloc}{d_{\mathrm{loc}}}
\newcommand{\xiloc}{\xi_{\mathrm{loc}}}
\newcommand{\xinloc}{\xi_{\mathrm{nloc}}}

\newcommand{\mfs}{\bar{\Phi}_\plus}
\newcommand{\mfl}{\bar{\Phi}_\zero}
\newcommand{\mft}{\bar{\Phi}_\minus}
\newcommand{\da}{\delta\Phi_\alpha}

\newcommand{\vbr}{\vb*{\rho}}
\newcommand{\vbk}{\vb*{k}}
\newcommand{\vbn}{\vb*{\nu}}
\newcommand{\vbf}{\vb*{\phi}}
\newcommand{\vbi}{\vb*{i}}
\newcommand{\vbg}{\vb*{g}}
\newcommand{\minus}{{\raisebox{-0.5mm}{\scalebox{0.6}[0.6]{$\mathbf{-}$}}}}
\newcommand{\plus}{{\raisebox{-0.5mm}{\scalebox{0.6}[0.6]{$\mathbf{+}$}}}}
\newcommand{\zero}{{\raisebox{-0.5mm}{\scalebox{0.6}[0.6]{$0$}}}}
\newcommand{\np}{n_\plus}
\newcommand{\nz}{n_\zero}
\newcommand{\nm}{n_\minus}
\newcommand{\Vp}{V_\plus}
\newcommand{\Vz}{V_\zero}
\newcommand{\Vm}{V_\minus}
\newcommand{\sumint}{\;\;\mathclap{\displaystyle\int}\mathclap{\textstyle\sum}\;\;\;}

\begin{document}
\title{Phase transitions in TGFT: Landau-Ginzburg analysis of the causally complete Lorentzian Barrett-Crane model}

\author[a]{Roukaya Dekhil,}
\emailAdd{roukaya.dekhil@unifi.it}

\author[b,c,d]{Alexander F. Jercher,}
\emailAdd{alexander.jercher@uni-jena.de}

\author[d]{Andreas G. A. Pithis}
\emailAdd{andreas.pithis@uni-jena.de}

\affiliation[a]{Universit\'a degli Studi di Firenze,\\ Piazza di San Marco, 4, 50121 Firenze FI, Italy, EU}
\affiliation[b]{Munich Center for Quantum Science and Technology (MCQST),\\ Schellingstr. 4, 80799 M\"unchen, Germany, EU}
\affiliation[c]{Arnold Sommerfeld Center for Theoretical Physics,\\ Ludwig-Maximilians-Universit\"at München \\ Theresienstrasse 37, 80333 M\"unchen, Germany, EU}
\affiliation[d]{Theoretisch-Physikalisches Institut, Friedrich-Schiller-Universit\"{a}t Jena\\ Max-Wien-Platz 1, 07743 Jena, Germany, EU}

\date{\today}

\begin{abstract}
{
It is expected that continuum spacetime emerges via phase transition in the tensorial group field theory (TGFT) approach to quantum gravity. Recent work on the application of Landau-Ginzburg mean-field theory to progressively realistic TGFT models has demonstrated how phase transitions can be realized therein. Here, we further develop this setting and consider the causally complete Lorentzian Barrett-Crane (BC) model which includes not only spacelike but also timelike and lightlike tetrahedra as quantum geometric building blocks. In addition, we incorporate discretized scalar fields by $\mathbb{R}$-valued variables of the group fields. In this context, we analyze models with an arbitrary single interaction of simplicial and tensor-invariant type, extend it to the model with the two vertices well-known from causal dynamical triangulations, and also consider a model with colored simplicial interactions. As a main result, we demonstrate for all those cases that a mean-field approximation of a phase transition towards a non-trivial condensate state can always be realized. In particular, we show that the critical behavior is entirely driven by spacelike faces which are characterized by the boost part of the Lorentz group. The latter induces an exponential suppression of fluctuations which then stabilizes the mean-field vacuum. In contrast, timelike faces do not play a role in this as they are characterized by the rotational and thus compact part of the Lorentz group. Since such a state is typically populated by a large number of TGFT quanta, our work lends further considerable support to the existence of a sensible continuum gravitational regime for causally complete TGFT models. Our results also indirectly strengthen the derivation of effective cosmological dynamics and the recently improved study of scalar cosmological perturbations within a mean-field approximation.}
\end{abstract}

\maketitle

\newpage

\section{Introduction}\label{sec:Introduction}

A key challenge of quantum gravity (QG) approaches involving discrete quantum geometric building blocks is to extract effective continuum gravitational physics on macroscopic scales. In analogy to statistical physics~\cite{Goldenfeld:1992qy}, coarse-graining methods are expected to be a crucial tool to describe how smooth spacetime geometries emerge from the underlying fundamental microscopic degrees of freedom. A view shared among various approaches like tensor models~\cite{Gurau:2011xp,Gurau:2016cjo,GurauBook,Eichhorn:2018phj,Gurau:2019qag}, tensorial group field theories (TGFT)~\cite{Freidel:2005qe,Oriti:2006se,Oriti:2007qd,Oriti:2011jm,Krajewski:2011zzu,Carrozza:2013oiy}, loop quantum gravity and spin foam models~\cite{Ashtekar:2004eh,Perez:2012wv,Steinhaus:2020lgb,Asante:2022dnj,Engle:2023qsu,Livine:2024hhc}, causal sets~\cite{Surya:2019ndm}, dynamical triangulations~\cite{Ambjorn:2012jv,Loll:2019rdj,Ambjorn:2022naa,Ambjorn:2024pyv} as well as quantum Regge calculus~\cite{williams2009quantum} is that such a process of emergence is in fact realized at criticality of the underlying quantum system, see also~\cite{Oriti:2013jga,Konopka:2008hp,Koslowski:2011vn}.

The TGFT approach sits at the confluence of these formalisms since it shares various structural connections with most of them. For instance, TGFTs can be understood to provide a completion of the quantum dynamics encoded in spin foam
models~\cite{Perez:2003vx,Perez:2012wv} and a second quantized formulation of loop quantum gravity~\cite{Oriti:2013aqa,Oriti:2014uga} as well as a reformulation of specific simplicial lattice gravity path integrals~\cite{Bonzom:2009hw,Baratin:2010wi,Baratin:2011tx,Baratin:2011hp,Finocchiaro:2018hks}. In particular, TGFTs are characterized by tensor fields, representing quantum geometric building blocks with combinatorially non-local interactions which also underlines their close relation to the tensor model approach~\cite{Gurau:2011xp,Gurau:2016cjo,GurauBook}. More precisely, the rank-$r$ tensor fields live on $r$ copies of a Lie group, turning them into so-called \emph{group} fields. They correspond to $(r-1)$-simplices which are glued together by the interactions to form $r$-dimensional cellular complexes. The specific form of the TGFT action dictates the detailed combinatorial properties of the model at hand and thus its transition amplitudes. In addition, for models designed to generate physical quantum geometries the Lie group corresponds to the local gauge group of gravity. Additional $\mathbb{R}$-valued local arguments of the group fields may be motivated by discretized scalar fields typically employed as a matter reference frame~\cite{Oriti:2016qtz,Li:2017uao,Gielen:2018fqv}. For the analysis of the phase structure of such theories and their critical behavior -- potentially relevant for the emergence of continuum spacetime geometries --  coarse graining methods can be applied. 

As is well-known from local statistical field theory and the literature on critical phenomena~\cite{Zinn-Justin:2002ecy,Zinn-Justin:2007uvz}, the Kadanoff-Wilson formulation of the renormalization group (RG) efficiently implements the required coarse graining operation, the so-called scale transformation. It progressively eliminates short-scale fluctuations towards the infrared (IR) and thus allows us to investigate how a physical theory evolves along scales~\cite{Wilson:1983xri}. After each elimination step, one obtains an effective action that encapsulates the contribution of the so-far eliminated modes. Typically, this procedure allows to search for distinguished loci, mostly points, in the theory space of the system. These are known as fixed points of the renormalization group where such scale transformations become invariant. There, the renormalized couplings and the corresponding effective action can be retrieved which allows to chart in detail the phase diagram of the model at hand and study phase transitions therein. A potent realization of this idea is provided by the functional renormalization group (FRG) methodology~\cite{Delamotte:2007pf,Kopietz:2010zz,Dupuis:2020fhh}. A simpler, yet efficient way to obtain a coarse account of the phase structure is offered by Landau-Ginzburg mean-field theory~\cite{Zinn-Justin:2002ecy,Zinn-Justin:2007uvz,Kopietz:2010zz} which suggests an effective, i.e. coarse-grained, action valid from meso- to macroscales~\cite{Hohenberg:2015jgf} without employing the RG machinery.

Recent works have introduced and advanced the application of these powerful methods to matrix and tensor models~\cite{Sfondrini:2010zm,Eichhorn:2013isa,Eichhorn:2014xaa,BenGeloun:2016tmc,Eichhorn:2017xhy,Eichhorn:2018phj,Eichhorn:2019hsa,Castro:2020dzt,Eichhorn:2020sla} as well as TGFTs~\cite{Benedetti:2015et,BenGeloun:2015ej,BenGeloun:2016kw,Benedetti:2016db,Carrozza:2016tih,Carrozza:2016vsq,Carrozza:2017vkz,BenGeloun:2018ekd,Pithis:2020sxm,Pithis:2020kio,Baloitcha:2020lha,Geloun:2023ray,Juliano:2024rgu,Pithis:2018eaq,Marchetti:2020xvf,Marchetti:2022igl,Marchetti:2022nrf,Dekhil:2024ssa}. Clearly, these works demonstrate from case to case that renormalization and the simpler mean-field approach indeed correspond to a coarse-graining of the lattice dual to the cellular complex generated by such models, see also~\cite{Carrozza:2016vsq,Carrozza:2024gnh}. Notice that this progress is far from trivial due to the combinatorial non-locality of such theories and because quantum gravity needs a manifestly background-independent form of coarse graining~\cite{Pereira:2019dbn,Eichhorn:2021vid}. 

In particular, the application of Landau-Ginzburg mean-field theory to physically compelling quantum geometric TGFT models at rank $r=4$ with Lorentzian signature has considerably advanced recently~\cite{Pithis:2018eaq,Pithis:2019mlv,Marchetti:2020xvf,Marchetti:2022igl,Marchetti:2022nrf,Dekhil:2024ssa} which shows that it is sufficient to investigate their basic phase properties. Indeed, Refs.~\cite{Marchetti:2020xvf,Marchetti:2022igl,Marchetti:2022nrf,Dekhil:2024ssa} have corroborated the conjecture of the existence of condensate phases, i.e. non-perturbative vacua, for such models. They show, just as in local theories~\cite{Strocchi:2013awa,Strocchi:2005yk}, that non-trivial vacua with the non-vanishing expectation value of the field operator can only be obtained if the domain of the group field is non-compact. In particular, non-trivial vacua can be realized in TGFT models for Lorentzian quantum gravity for which the geometric part of the group field domain is necessarily given by (copies of) the non-compact Lorentz group~\cite{Marchetti:2022igl,Marchetti:2022nrf,Dekhil:2024ssa}. Should that part of the domain be compact, as for models for Euclidean quantum gravity, one may extend the TGFT field domain, as mentioned above, by non-compact local directions to that end~\cite{Marchetti:2020xvf,Dekhil:2024ssa}.

Of direct relevance to the present work is how this has explicitly been demonstrated for the Lorentzian Barrett-Crane TGFT model~\cite{Jercher:2022mky} restricted to spacelike building blocks~\cite{Jercher:2021bie} and related models with the same configurations but also featuring tensor-invariant interactions~\cite{Carrozza:2013oiy,Carrozza:2016vsq,Marchetti:2022igl}. This model aims to provide a TGFT quantization of $4d$ Lorentzian Plebanski gravity (reducing to Palatini gravity in first-order formulation upon imposition of constraints)~\cite{Marchetti:2022igl,Marchetti:2022nrf,Dekhil:2024ssa}. Focussing on the geometric aspects, this analysis strikingly showed that not only due to the non-compactness but also due to the hyperbolicity of the geometry of the Lorentz group one finds that phase transitions towards a non-perturbative vacuum (condensate) state exists and that they can be self-consistently described using mean-field theory. More specifically, one finds that near criticality the fluctuations around the non-vanishing mean-field vacuum are exponentially suppressed due to the boost part of the Lorentz group, thus validating the mean-field approximation of the phase transition. This is interesting since micro-causality is enforced by means of the Lorentz group and even more so since such states are typically highly excited by TGFT quanta which indeed together makes a compelling case for an interesting continuum geometric approximation. These results, therefore, provide strong indirect support for the recent extraction of effective
cosmological dynamics from quantum geometric TGFTs in the context of a mean-field approximation which hinges on the existence of a condensate phase in the form of a non-perturbative vacuum~\cite{Gielen:2013kla,Gielen:2013naa,Gielen:2016dss,Oriti:2016acw,deCesare:2016rsf,Pithis:2019tvp,Jercher:2021bie,Oriti:2021rvm,Oriti:2021oux,Marchetti:2021gcv,Jercher:2023nxa,Jercher:2023kfr}. Moreover, one observes that the effective mass of the modes relevant for the critical behavior vanishes not only at criticality but also throughout the entire phase of non-vanishing vacuum expectation value due to the non-locality of the interactions~\cite{Dekhil:2024ssa}. Note that this Landau-Ginzburg analysis required to overcome difficult technical challenges which relate to working with the infinite-dimensional representations of the Lorentz group and its non-compactness necessitating the implementation of a regularization scheme~\cite{Marchetti:2022igl,Marchetti:2022nrf}. In addition, the feature of the effective masslessness required a further minor adaptation of the Landau-Ginzburg method to the context of TGFTs~\cite{Dekhil:2024ssa}.

The goal of our present work is to apply the Landau-Ginzburg method to the more realistic and thus more complex causal completion of the Lorentzian BC model~\cite{Jercher:2022mky}. The perturbative expansion of this model allows us to generate Lorentzian triangulations consisting not only of spacelike tetrahedra but also of timelike and lightlike ones. Clearly, this enlarged set of configurations allows us to better map the causal aspects of discrete and continuum geometry within this model compared to the restricted version. In particular, it renders possible the reconstruction of timelike and lightlike geometric objects without only referring to spacelike configurations. Furthermore, it enables the description not only of quantum geometries with spacelike but also with timelike and lightlike boundaries. For these reasons, it is pressing to investigate the impact of the enlarged set of configurations onto the critical behavior and the phase structure at the mean-field level. 

To achieve this goal, we overcome challenges that arise from the presence of multiple fields corresponding to the different causal configurations, the large number of parameters in the given theory space, and the involved representation theory of the Lorentz group. Concretely, due to the complexity of the theory space, we obtain matrix-valued correlation functions the asymptotic behavior of which we use to extract a correlation length in the local and non-local variables. We explicitly compute the corresponding matrix-valued Ginzburg-$Q$ parameter and show in general that at criticality fluctuations around the mean-field are again exponentially suppressed. Importantly, we show that the presence of timelike faces does not affect the critical behavior of the theory as they are associated with the rotational and thus compact part of the Lorentz group. Hence, the critical behavior is entirely driven by spacelike faces. Our Landau-Ginzburg analysis is conducted in great detail for models with an arbitrary single interaction of simplicial and tensor-invariant types. We subsequently extend the results to a model with the two simplicial interactions well-known from causal dynamical triangulations (CDT). Finally, we consider a colored simplicial model with spacelike tetrahedra and the previously established arguments allow us to conclude that the causally completed and colored setting results in the same critical behavior. Thereby, we significantly extend recent works~\cite{Marchetti:2022igl,Marchetti:2022nrf,Dekhil:2024ssa} to the much more involved case of the causally complete BC model for which we show that also there the Lorentz group leads to an exponential suppression of fluctuations near the critical point thus validating the mean-field approximation of the phase transition towards a causally rich non-perturbative vacuum state. This lends further substantial backing for the existence of a sensible continuum gravitational regime for TGFT models for Lorentzian quantum gravity as
well as in the closely related lattice quantum gravity and spin foam models.

To this end, the article is organized as follows: In Section~\ref{sec:LG of cBC} we introduce the causally complete Barrett-Crane TGFT model and sketch the general idea of the mean-field strategy while highlighting the new aspects arising from the variety of causal configurations. Section~\ref{sec:Fixed but arbitrary interaction} contains a detailed mean-field treatment of the BC model with a single arbitrary interaction of simplicial and tensor-invariant type. We compute the correlation function in local and non-local variables in Sections~\ref{sec:Local correlation function} and~\ref{sec:Non-local correlation function}, respectively. Subsequently, we compute the Ginzburg-$Q$ parameter and study its behavior near criticality. In addition, we discuss further extensions such as the consideration of multiple interactions. We enlarge the theory space in Section~\ref{sec:other interactions}, where we apply the same analysis first to CDT-like interactions and thereafter to colored simplicial interactions. Finally, we summarize and discuss our results in Section~\ref{sec:Conclusion}.

\section{Landau-Ginzburg analysis of the complete Barrett-Crane model}\label{sec:LG of cBC}

\subsection{The complete Barrett-Crane model}\label{sec:complete BC}

The Lorentzian Barrett-Crane model is a TGFT and spin foam model that provides a quantization of first-order Palatini gravity. It was first introduced in~\cite{Barrett:1999qw,Perez:2000ec,Perez:2000ep} and was reformulated in~\cite{Baratin:2011tx,Jercher:2021bie} by introducing an auxiliary timelike normal vector variable that allows to consistently impose the so-called geometricity constraints, discussed momentarily.\footnote{Note that the introduction of an auxiliary normal vector variable~\cite{Jercher:2021bie,Baratin:2011tx} allowed to cure specific deficiencies of the original formulation of the Barrett-Crane model~\cite{Barrett:1999qw,Perez:2000ec,Perez:2000ep}. Other remaining points of criticism of this model are inconclusive, as discussed in detail in Refs.~\cite{Baratin:2011tx,Jercher:2021bie,Jercher:2022mky}.} By introducing this constraint taking into consideration the timelike normal vector, the fundamental building blocks of the extended theory are in turn spacelike tetrahedra, thus posing a restriction of the boundary states and the microscopic causal structure. These limitations have been lifted in a causal completion of the Lorentzian BC model~\cite{Jercher:2022mky} where the normal vector can be timelike, lightlike or spacelike, thus living in the two-sheeted $3$-hyperboloid $\mathrm{H}_\plus$, the light cone $\mathrm{H}_\zero$ or the one-sheeted hyperboloid $\mathrm{H}_{\minus}$, respectively. Thereby, the complete BC model includes all possible causal configurations at the level of a single $4$-dimensional spacetime building block.\footnote{Interestingly, this causal extension was shown to be crucial in improving the dynamics of cosmological perturbations derived from TGFT condensate cosmology~\cite{Jercher:2023kfr,Jercher:2023nxa}. In contrast to former attempts~\cite{Gielen:2022iuu,Gielen:2017eco,Gielen:2018xph,Gerhardt:2018byq,Marchetti:2021gcv}, the inclusion of timelike tetrahedra allowed to causally couple a Lorentzian reference frame and to extract scalar perturbations from entanglement between spacelike and timelike tetrahedra.}$^{,}$\footnote{The glueing of multiple $4$-dimensional building blocks with each other can lead to an irregular light cone structure located at dual vertices, edges or hinges. See~\cite{Benedetti:2008hc} and~\cite{Asante:2021phx,Jercher:2023csk} for a discussion on such causality violations in the context of matrix models and Lorentzian Regge calculus, respectively. Causal regularity in this sense cannot be imposed at the level of a single $4d$ building block but rather in a restriction of the glueing of multiple $4d$ building blocks, see also~\cite{Jercher:2022mky} for a discussion.}

The group field obeys respectively closure and simplicity constraints,
\begin{subequations}\label{eq:geom_constraint}
\begin{align}
\Phi(\vbg, X_\alpha)&=\Phi(\vbg h^{-1} , h \cdot X_\alpha), \quad \forall h \in\SL\, , \\[7pt]
 \Phi(\vbg,X_\alpha)&=\Phi(\vbg\vb*{u} ,X_\alpha), \:\quad\qquad \forall u_1, \ldots, u_4 \in \mathrm{U}_{X_\alpha}\,,
\end{align}
\end{subequations}
which commute in the extended formulation with normal vectors~\cite{Baratin:2011tx,Jercher:2021bie}. Together, the constraints are commonly referred to as geometricity constraints. Here, $\vbg\in\SL^4$, $X_\alpha$ is the normal vector, $\mathrm{U}_{X_\alpha}$ the corresponding stabilizer subgroup of $\SL$ and the index $\alpha\in\{+,0,-\}$ denotes the timelike, null and spacelike signature of the normal vector. We further consider minimally coupling 
 $\dloc$ free massless scalar fields to the group field, following Refs.~\cite{Oriti:2016qtz,Li:2017uao,Gielen:2018fqv}. This is achieved by extending the domain of the field 
\begin{equation}
\label{eq:extended_GFT}
\Phi(\vbg,X_\alpha)\longrightarrow \Phi(\vbg,\vbf,X_\alpha),
\end{equation}
with $\vbf\in\R^{\dloc}$ the scalar field values. Without loss of generality, we consider real-valued group fields hereafter.

\begin{figure}
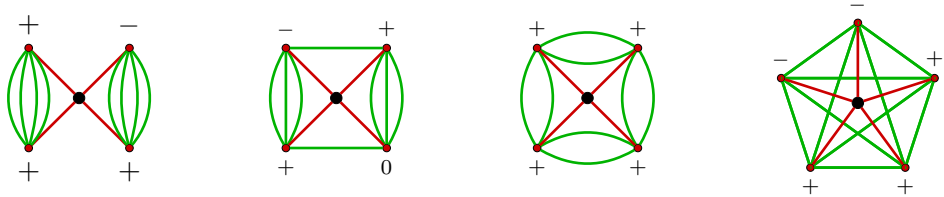

    \centering
\cvftpppm~~~~~~~~~\cvfppmzc~~~~~~~~~\cvfnpppp~~~~~~~~~\cvfspppmm
    \caption{Diagrammatic representation of four exemplary causal vertex graphs $\gamma$. The green half-edges indicate pairwise convolution of the non-local variables $\vbg$ and red vertices represent the fields $\Phi_\alpha(\vbg,\vbf)$, labelled by the causal character $\alpha\in\{+,0,-\}$. The vertex structure of local variables $\vbf$ corresponds to the standard point-like interaction and is indicated by red edges and black vertices. From left to right: double-trace melon with three spacelike and one timelike tetrahedron, quartic melon with two spacelike, one lightlike and one timelike tetrahedron, quartic necklace with four spacelike tetrahedra, $4$-simplex with three spacelike and two timelike tetrahedra.}
    \label{fig:causal vertex graph}
\end{figure}

The dynamics of the causally complete BC model are governed by the action consisting of a kinetic and an interaction term,
\begin{equation}
S[\Phi_\alpha] = K[\Phi_\alpha]+V[\Phi_\alpha],
\end{equation}
where $\Phi_\alpha$ is a short-hand notation for three fields associated with the signatures of the three normal vectors. The kinetic term of the theory $K[\Phi_\alpha]$ is defined as
\begin{equation}
K[\Phi_\alpha] = \frac{1}{2}\sum_\alpha \int\dd{\vbg}\dd{\vbg'}\dd{\vbf}\dd{\vbf'}\dd{X_\alpha}\Phi(\vbg,\vbf,X_\alpha)\mathcal{K}_\alpha(\vbg,\vbf;\vbg',\vbf')\Phi(\vbg',\vbf',X_\alpha),
\end{equation}
with kinetic kernels $\mathcal{K}_\alpha$ being model-dependent. While the BC TGFT model is normally formulated with a quintic interaction term that possesses the combinatorics of a $4$-simplex, in this work we will also consider models subject to the same constraints but also featuring tensor-invariant interactions. For this reason, the setup of the interaction part of the action is given in an even more general form in the following: For a causally complete TGFT there exists in fact a large set of possible interaction terms which in the most general case can be written as
\begin{equation}
V[\Phi_\alpha] = \sum_\gamma\lambda\int\dd{\vb*{X}}\dd{\vbf}\Tr_\gamma\left[\Phi_\plus^{\np}\Phi_\zero^{\nz}\Phi_\minus^{\nm}\right].
\end{equation}
Here, $\gamma$ denotes a \textit{causal vertex graph} which enriches the concept of a vertex graph~\cite{Oriti:2014yla,Marchetti:2020xvf,Marchetti:2022igl} by associating a causal character to its vertices, see Fig.~\ref{fig:causal vertex graph} for four examples. 
The trace $\Tr_\gamma$ encodes the pairwise contraction of group elements according to $\gamma$ which includes in total  $n_\gamma = \np+\nz+\nm$ tetrahedra. Interactions are local in the scalar field variables $\vb{\phi}$, as detailed in~\cite{Oriti:2016qtz,Li:2017uao,Gielen:2018fqv}. Furthermore, the normal vectors are integrated over separately for every field entering the interaction, supporting their interpretation as auxiliary variables~\cite{Jercher:2021bie,Jercher:2022mky}. At the level of the interaction term, it is important to mention that for vertex graphs, tensor-invariant configurations~\cite{Carrozza:2013oiy,GurauBook} such as melons or necklaces are conceivable as well as simplicial ones which are most closely connected to spin foam models~\cite{Perez:2003vx,Perez:2012wv,Engle:2023qsu} and specific simplicial lattice gravity models~\cite{Bonzom:2009hw,Baratin:2010wi,Baratin:2011hp,Baratin:2011tx}. We refer the reader to Refs.~\cite{Carrozza:2016vsq,Marchetti:2022igl} for detailed discussions of the relation between both types of interactions and the general theory space at hand. Suffice to say here that from the spin foam perspective it is by now well-known that the most divergent radiative corrections stem from amplitudes which can be reabsorbed into effective tensor-invariant TGFT interactions~\cite{Carrozza:2016vsq,Engle:2007wy,Freidel:2007py,Perini:2008pd,Krajewski:2010yq,Riello:2013bzw,Dona:2018pxq,Frisoni:2021uwx,Han:2023hbe} thus justifying the investigation of their impact on the phase structure already at mean-field level. As the examples listed in Appendix~\ref{sec:Explicit expressions for chi} show, the set of causal vertex graphs is larger than that of usual vertex graphs. For instance, the quartic melonic graph with two spacelike and two timelike tetrahedra offers three non-equivalent configurations. In Sections~\ref{sec:Fixed but arbitrary interaction} and~\ref{sec:other interactions}, we study feasible examples of such interactions.

Given the total action $S[\Phi_\alpha]$, the dynamics of the three fields $\Phi_\alpha$ are captured by the equations of motion
\begin{equation}\label{eq:GFT_EOM}
0=\fdv{S}{\Phi(\vb*{g},\vb*{\phi},X_\alpha)} = \int\dd{\vbg'}\dd{\vbf'}\mathcal{K}_\alpha(\vbg,\vbf;\vbg',\vbf)\Phi(\vbg',\vbf',X_\alpha)+\fdv{V}{\Phi(\vbg,\vbf,X_\alpha)},
\end{equation}
where the variation of the action is performed with respect to the group field on the extended domain including the normal vector $X_\alpha$. To obtain an equation that is independent of this auxiliary variable, one can integrate out $X_\alpha$.

\paragraph{Spin representation.} To derive a correlation function for the fluctuations of the group field as we aim to do in the present article, it is necessary to work in the corresponding Fourier space. Notice that in this case, we are dealing with local and non-local (geometric) degrees of freedom and hence the decomposition follows separately for each one of them.
For the geometric degrees of freedom, this is achieved via the decomposition of $\Phi(\vbg,\vbf,X_\alpha)$ into unitary irreducible representations of $\SL$, labelled by $(\rho,\nu)\in\R\times\mathbb{Z}/2$, which is commonly referred to as spin representation (see Appendix~\ref{sec:Aspects of SL2C and its Representation Theory} for further details). Taking into account the geometricity constraints in Eqs.~\eqref{eq:geom_constraint}, the spin decomposition of $\Phi_\alpha$ is then given by
\begin{equation}\label{eq:spin rep with normal}
\Phi(\vbg,\vbf,X_\alpha)=\left[\prod_{c=1}^4 \sum_{\nu_c} \int\dd{\rho_c}\left(\rho_c^2+\nu_c^2\right) \sum_{j_c m_c l_c n_c}\right] \Phi_{\vb*{j}\vb*{m}}^{\vbr\vbn, \alpha} (\vbf)\prod_c D_{j_c m_c l_c n_c}^{(\rho_c, \nu_c)}(g_c g_{X_\alpha}) \mathcal{I}_{l_c n_c}^{(\rho_c, \nu_c), \alpha}\,,
\end{equation}
where $\left(\rho_c^2+\nu_c^2\right)$ is the Plancherel measure of functions on $\SL$, $g_{X_\alpha}$ is a representative of the equivalence class $X_\alpha=\left[g_{X_\alpha}\right] \in$ $\SL / \mathrm{U}^{(\alpha)}$ and $\mathcal{I}_{l_c n_c}^{(\rho_c, \nu_c), \alpha}$ are the components of $\mathrm{U}^{(\alpha)}$-invariant states in the canonical basis. For timelike and lightlike normal vectors, the invariant states enforce $\nu_c = 0$, which is consistent with the fact that spacelike and lightlike tetrahedra consist exclusively of spacelike faces. A spacelike normal vector is associated with either $\nu_c = 0$ and thus a spacelike face or with $\rho_c = 0$ and $\nu_c\in \mathbb{N}^+$ and thus a timelike face. For further details, we refer to the explicit discussion outlined in Ref.~\cite{Jercher:2022mky}.

If the normal vector $X_\alpha$ is integrated out, such as in the interaction term, the above-decomposed field takes the form 
\begin{equation}\label{eq:spin rep no normal}
\int\dd{X_\alpha} \Phi(\vbg,\vbf,X_\alpha)=  \left[\prod_{c=1}^4 \sum_{\nu_c} \int \dd{\rho_c}\left(\rho_c^2+\nu_c^2\right) \sum_{j_c m_c l_c n_c}\right] \Phi_{\vb*{j}\vb*{m}}^{\vbr\vbn, \alpha}(\vbf) B_{\vb*{l}\vb*{n}}^{\vbr\vbn, \alpha} \prod_c D_{j_c m_c l_c n_c}^{(\rho_c, \nu_c)}(g_c), 
\end{equation}
where we notice the appearance of the generalized Barrett-Crane intertwiners that are defined as
\begin{equation}\label{eq:BC_intertwiner}
B_{\vb*{j}\vb*{m}}^{\vbr\vbn, \alpha}:=\int\dd{X_\alpha} \prod_{c=1}^4 \sum_{l_c n_c} D_{j_c m_c l_c n_c}^{(\rho_c, \nu_c)}\left(g_{X_\alpha}\right) \mathcal{I}_{l_c n_c}^{(\rho_c, \nu_c), \alpha}\,.
\end{equation}

For the matter degrees of freedom, we consider the standard Fourier transform on $\R^{\dloc}$, such that 
\begin{equation}
    \Phi_{\vb*{j}\vb*{m}}^{\vbr\vbn,\alpha}(\vbf)=\int\frac{\dd{\vbk}}{(2 \pi)^{\dloc}} \Phi_{\vb*{j}\vb*{m}}^{\vbr\vbn,\alpha}(\vbk) \e^{i \vbk\cdot\vbf}.
\end{equation}
This completes the transformation of the fields in group representation $\Phi(\vbg,\vbf,X_\alpha)$ to spin representation $\Phi^{\vbr\vbn,\alpha}_{\vb*{j}\vb*{m}}(\vbk)$.

\subsection{General mean-field strategy}\label{sec:general strategy}

In this section, we sketch the mean-field approach to Lorentzian TGFTs along the lines of~\cite{Marchetti:2020xvf,Marchetti:2022igl,Dekhil:2024ssa}, indicating the key points where the extended causal structure leads to structural differences. The overarching goal is to study phase transitions of the system within a Gaussian approximation. In principle, this follows standard expositions of Landau-Ginzburg mean-field theory, however, the transfer of this method to TGFTs requires extra care due to their non-local interactions. As a first step, the equations of motion, given in their general form in Eq.~\eqref{eq:GFT_EOM}, are evaluated on constant field configurations $\bar{\Phi}_\alpha$.\footnote{Since $\SL$ is non-compact, constant field configurations do not live in an $L^2$-space. In order to derive a consistent treatment of the uniform field configurations, an extension of the space of functions to that of so-called hyperfunctions~\cite{Ruehl1970,hormander2015analysis} is required. In contrast to~\cite{Marchetti:2022igl}, this procedure does therefore not rely on the regularization via a Wick rotation and compactification to the group $\Spin$. This bears the advantage that the causal structure encoded in spacelike, lightlike and timelike building blocks remains clearly visible throughout the ensuing analysis.} Solutions to the resulting set of equations are referred to as the mean-field background\footnote{Notice that for involved interactions that include in particular a sum over different terms, the mean-field equations are given by a system of polynomial equations which is potentially very difficult to solve.}, which is zero above criticality and non-zero below criticality. Fluctuations $\delta\Phi_\alpha$ around the mean-field solutions are studied via the ansatz
\begin{equation}
\Phi_\alpha(\vb*{g},\vb*{\phi},X_\alpha) = \bar{\Phi}_\alpha+\delta\Phi(\vb*{g},\vb*{\phi},X_\alpha)\,.
\end{equation}
Linearizing the equations of motion via this ansatz, one defines an \textit{effective kinetic kernel} $G_{\alpha\beta}$ that reads as
\begin{equation}
G_{\alpha\beta} = \mathcal{K}_\alpha\delta_{\alpha\beta}+F_{\alpha\beta},
\end{equation}
where $F_{\alpha\beta}$ is the Hessian of the interaction term evaluated on the mean-field. As a result, one obtains a field theory of the fluctuations $\delta\Phi_\alpha$, the path-integral of which is a Gaussian. The correlation function associated with the fluctuations, i.e.~the two-point functions of the field theory, defined as   
\begin{equation}
\expval{\delta\Phi_\alpha(\vb*{e},\vb*{0})\delta\Phi_\beta(\vb*{g},\vb*{\phi})} = C_{\alpha\beta}(\vb*{g},\vb*{\phi}),
\end{equation}
are then obtained by inverting the effective kinetic kernel, i.e. $(G_{\alpha\beta})^{-1} = C_{\alpha\beta}$. This is most straightforwardly achieved by going to the spin representation and Fourier space in the geometric and matter degrees of freedom, respectively. Notice that by including tetrahedra with spacelike, lightlike, and timelike signatures, which may interact in general, the Hessian of the interaction term yields in principle a $3\times 3$ matrix. Therefore, the effective kernel is matrix-valued, making the inversion of the correlator more involved. 

From the correlation function $C_{\alpha\beta}(\vbg,\vbf)$, one can extract a local and non-local correlation via the following integration of local and  non-local degrees of freedom
\begin{equation}\label{eq:local and non-local correlator}
C_{\alpha\beta}(\vb*{\phi}) = \int\dd{\vb*{g}}C(\vb*{g},\vb*{\phi}),\qquad C_{\alpha\beta}(\vb*{g}) = \int\dd{\vb*{\phi}}C(\vb*{g},\vb*{\phi}).
\end{equation}
Given an asymptotic exponential decay of $C_{\alpha\beta}(\vbf)$ and $C_{\alpha\beta}(\vbg)$, a local and non-local correlation length, $\xiloc$ and $\xinloc$ can be determined, respectively. What is typical for second-order phase transitions, the correlation length diverges at criticality leading to scale-invariance of the system. 

Validity of the mean-field ansatz in the vicinity of the critical point is given if the fluctuations are small compared to the mean-field. This is quantified by the Ginzburg-$Q$, which is matrix-valued in the present setting,
\begin{equation}\label{eq:definition Q}
Q_{\alpha\beta} = \frac{\int_{\Omega_\xi}\dd{\vb*{g}}\dd{\vb*{\phi}}C_{\alpha\beta}(\vb*{g},\vb*{\phi})}{\int_{\Omega_\xi}\dd{\vb*{g}}\dd{\vb*{\phi}}\bar{\Phi}_\alpha\bar{\Phi}_\beta}.
\end{equation}
where $\Omega_\xi$ is the integration domain defined by the local and non-local correlation length. If the Ginzburg-Levanyuk criterion~\cite{levanyuk1959contribution,ginzburg1961some} $Q_{\alpha\beta}\ll 1$ close to criticality holds, then the mean-field approach constitutes a consistent approximation of the phase transition towards a non-perturbative vacuum state. 

In the following sections, we are interested in explicitly performing the Landau-Ginzburg analysis for several TGFT models with an extended causal structure. First in Section~\ref{sec:Fixed but arbitrary interaction}, we study in great detail a single interaction with double-trace melonic, quartic melonic, necklace (or simplicial) combinatorics and an arbitrary combination of spacelike, lightlike, or timelike tetrahedra. Thereafter in Section~\ref{sec:other interactions}, we consider further examples of TGFT models, being a simplicial CDT-like model and a colored simplicial model with all tetrahedra being spacelike. Since the detailed methods of Section~\ref{sec:Fixed but arbitrary interaction} apply also to these models, we will keep the presentation of Section~\ref{sec:other interactions} rather short, focusing on the main results.

\section{Single interaction with arbitrary causal structure}\label{sec:Fixed but arbitrary interaction}

The kinetic contribution to the TGFT action we are using in this section is defined by the kinetic kernel
\begin{equation}\label{eq:kinetic kernel}
    \mathcal{K}_{\alpha}(\vbg,\vbf;\vbg',\vbf') = \mu_{\alpha}\delta(\vbg^{-1}\vbg') - Z^\phi_\alpha(\vbg^{-1}\vbg')\Delta_{\phi} - Z^g_\alpha(\abs{\vbf-\vbf'})\sum_{c=1}^4 \Delta_c,
\end{equation}
which consists of the following ingredients:
\begin{itemize}
    \item Mass parameters: the $\mu_\alpha$ are parameters playing the role of mass terms which can be motivated via the correspondence of TGFTs perturbative amplitudes with spin foam amplitudes. In the latter context, it would correspond to spin foam edge weights~\cite{Perez:2012wv,Carrozza:2016vsq}. From the perspective of the mean-field framework it simply serves as a control parameter that allows to separate a trivial vacuum state from a nontrivial one. To cover the most general case, we assume different parameters for the three different signatures.
    \item Laplace operators: $\Delta_{\phi}$ is the Laplace operator acting on $\R^{\dloc}$ and $\Delta_c$ is the Laplace operator acting on the group $\SL$. Motivating the local variables via massless, free, and minimally coupled scalar matter fields, $\Delta_\phi$ corresponds to the lowest order of a series expansion of even derivatives. These originate from the discretization of the continuum action of such fields over the cellular geometry, see~\cite{Oriti:2016qtz,Li:2017uao,Gielen:2018fqv} for details. The Laplacian on the Lie group is typically motivated when studying radiative corrections in the renormalization group treatment of the Boulatov and Ooguri TGFT models~\cite{BenGeloun:2011jnm,BenGeloun:2011rc,BenGeloun:2013mgx}. It is generally expected that this argument directly transfers to models subject to simplicity constraints~\cite{Carrozza:2016vsq,Marchetti:2022igl}. In the context of Landau-Ginzburg mean-field theory the Laplacians lead to the notion of a scale which in turn allows us to introduce the notion of correlation length later on.
    \item Weights of Laplacians: we anticipate at this point that one has to enlarge the parameter space of the theory by the functions $Z_\alpha^\phi$ and $Z_\alpha^g$ in order to obtain a well-behaved correlation function, that is, a correlation function that shows an asymptotic exponential fall-off with no oscillations or exponential divergences. Notice that these functions resemble corrections from a wave function renormalization~\cite{WipfRG} with a particular dependence on the geometric and matter variables. From the perspective of discretization of the scalar matter over the cellular geometry the factors $Z_\alpha^\phi$ are also known to encode non-trivial features of the minimal coupling~\cite{Oriti:2016qtz,Li:2017uao,Gielen:2018fqv}. The reciprocity of the matter-gravity coupling suggests in turn factors $Z_\alpha^g$ attached to the Laplacian on the group manifold. Within a full-fledged RG treatment of the models the introduction of such parameters for the wave function renormalization would anyway be required for the consistency of the flow equations. We leave such a deeper analysis to future investigations. The $Z_\alpha^\phi$ only depend on the trace of $\vbg^{-1}\vbg'$ and the functions $Z_\alpha^g$ only depend on the absolute local distance $\abs{\vbf-\vbf'}$ to sustain the symmetries of the kinetic kernel. These symmetries will play an important role in deriving a Fourier transform as commented below. 
\end{itemize}
 
We consider a single causal vertex graph with $\np$ spacelike, $n_0$ lightlike and $\nm$ timelike tetrahedra, yielding the interaction term
\begin{equation}\label{eq:single interaction}
V[\Phi_\alpha] = \lambda\int\dd{\vb*{X}}\dd{\vbf}\Tr_\gamma\left[\Phi_\plus^{\np}\Phi_\zero^{\nz}\Phi_\minus^{\nm}\right].
\end{equation}
Examples of such graphs are given in Fig.~\ref{fig:causal vertex graph} and an exhaustive list of causal vertex graphs with double-trace melonic, quartic melonic, quartic necklace or simplicial combinatorics are given in Appendix~\ref{sec:Explicit expressions for chi}. The parameters of the theory are the masses $\mu_\alpha$, the functions $Z_\alpha^\phi$ and $Z_\alpha^g$, and a single interaction coupling $\lambda$. 

The degree of the interaction $(\np,\nz,\nm)$ determines the symmetries of the model. In particular, the kinetic term is invariant under reflections of the field $\Phi_\alpha\mapsto -\Phi_\alpha$. Thus, if $n_\alpha$ is even, the total action exhibits an invariance under reflections of the field $\Phi_\alpha$. As a consequence, the general interaction in Eq.~\eqref{eq:single interaction} can either yield a $\mathbb{Z}_2^N$-symmetric theory with $N\in\{1,2,3\}$ or no symmetry at all. We also discuss the possibility of defining a $\mathrm{O}(3)$-invariant theory at the very end of this section. Examples of $\mathbb{Z}_2\times\mathbb{Z}_2$-invariant theories in the context of condensed matter physics can be found in~\cite{Kuno:2023xyz,Kennedy1992}. With the Landau-Ginzburg method, one can study the second-order phase transition from $\mu>0$ (unbroken phase) to $\mu<0$ (broken phase). In the absence of any symmetry, studying the degree of the phase transition is less clear and arguments beyond the mean-field method seem to be required. Still, the Landau-Ginzburg method serves also in this case as an approximation method of a phase transition towards a non-vanishing vacuum state which can be checked for self-consistency.\footnote{Irrespective of the symmetries of the action, determining the stability of TGFT vacua is a delicate task due to the non-localities of the interaction. This issue has been considered for instance in~\cite{Freidel:2002tg,Magnen:2009at,BenGeloun:2018ekd} in the context of the $\SUT$-Boulatov GFT model for $3d$ Euclidean quantum gravity. We leave an extension of this analysis to the present context, i.e. mean-field vacua of TGFTs on a non-compact domain and subject to geometricity constraints, as an intriguing task to future research.}  

Evaluating the equations of motion on the uniform field configurations $\bar{\Phi}_\alpha=\mathrm{const.}$ yields the following three equations 
\begin{subequations}\label{eq:mf equations single int}
\begin{align}
    \mfs^{\np-2}\Vp^{\np-1} &= -\frac{\mu_\plus}{\lambda\np}\left(\mfl\Vz\right)^{-\nz}\left(\mft\Vm\right)^{-\nm}\Vp^{4-2n_\gamma},\\[7pt]
    \mfl^{\nz-2}\Vz^{\nz-1} &= -\frac{\mu_\zero}{\lambda\nz}\left(\mft\Vm\right)^{-\nm}\left(\mfs\Vp\right)^{-\np}\Vp^{4-2n_\gamma},\\[7pt]
    \mft^{\nm-2}\Vm^{\nm-1} &= -\frac{\mu_\minus}{\lambda\nm}\left(\mfs\Vp\right)^{-\np}\left(\mfl\Vz\right)^{-\nz}\Vp^{4-2n_\gamma},
\end{align}
\end{subequations}
where the $V_\alpha$ are volume factors that arise from empty $\SL$ or normal vector integrations. Since $\SL$ and all $\mathrm{H}_\alpha$ are non-compact, the volume factors diverge and we consider a regularization by introducing a cutoff in the non-compact direction. The scaling behavior of the $V_\alpha$ in the cutoff and the skirt radius $a$ of the hyperboloids is discussed in Appendix~\ref{sec:Empty integrals}. The mean-field equations in Eqs.~\eqref{eq:mf equations single int} are solved by
\begin{equation}\label{eq:mf solutions}
\bar{\Phi}_\alpha = V_\plus^{-2}\left(-\frac{\mu_\alpha}{n_\alpha\lambda}\right)^{\frac{2+n_\alpha-n_\gamma}{2(n_\gamma-2)}}V_\alpha^{-\frac{n_\alpha}{2(n_\gamma-2)}-\frac{1}{2}}\prod_{\beta\neq \alpha}\left(-\frac{\mu_\beta}{n_\beta\lambda}\right)^{\frac{n_\beta}{2(n_\gamma-2)}}V_\beta^{-\frac{n_\beta}{2(n_\gamma-2)}}.
\end{equation}
Introducing fluctuations $\da$ around the mean-field backgrounds and linearizing the equations of motion, one obtains
\begin{equation}
0 = \sum_\beta\int\dd{\vb*{g}'}\dd{\vb*{\phi}'}\dd{X_\beta'}\left[\delta_{\alpha\beta}\mathcal{K}_\alpha(\vb*{g},\vb*{\phi};\vb*{g}',\vb*{\phi}')\delta(X_\alpha,X_\beta')+F_{\alpha\beta}(\vb*{g},\vb*{\phi};\vb*{g}',\vb*{\phi}')\right]\delta\Phi_\beta(\vb*{g}',\vb*{\phi}',X_\beta').
\end{equation}
The kinetic contribution involves a $\delta$-function imposed on the normal vectors $X_\alpha$ and $X_\beta'$, while the Hessian function $F_{\alpha\beta}$ is independent of those arguments. In spin representation, this leads to an imbalance of Barrett-Crane intertwiners that appear in the Hessian term but are absent in the kinetic term. In order to symmetrize this expression and to eliminate the auxiliary normal vector variable from the equations of motion, we perform an additional $X_\alpha$-integration. As a result, the effective equations of motion for the fluctuating fields take the form
\begin{equation}
0 = \sum_\beta\int\dd{\vb*{g}'}\dd{\vb*{\phi}}G_{\alpha\beta}(\vb*{g},\vb*{\phi};\vb*{g}',\vb*{\phi}')\delta\Phi_\beta(\vb*{g}',\vb*{\phi}'),
\end{equation}
with an effective kinetic kernel $G_{\alpha\beta}$ which can be derived from an effective quadratic action,
\begin{equation}\label{eq:effective action group rep}
S_{\mathrm{eff}}[\delta\Phi_\alpha] = \frac{1}{2}\sum_{\alpha\beta}\int\dd{\vbg}\dd{\vbf}\dd{\vbg'}\dd{\vbf'}\da(\vbg,\vbf)G_{\alpha\beta}(\vbg,\vbf;\vbg',\vbf')\delta\Phi_\beta(\vbg',\vbf').
\end{equation}
The Hessian contribution $F_{\alpha\beta}$ to the effective kinetic kernel is then given by
\begin{equation}\label{eq:Hessian matrix}
V_\alpha F_{\alpha\beta}(\vb*{g},\vb*{\phi};\vb*{g}',\vb*{\phi}')  = \delta(\vb*{\phi}-\vb*{\phi}')
\begin{pmatrix}
-\frac{\mu_\plus}{\np}\chi_{\plus\plus} & -\sqrt{\frac{\mu_\plus\mu_\zero}{\np\nz}}\,\chi_{\plus\zero} & -\sqrt{\frac{\mu_\plus\mu_\minus}{\np\nm}}\,\chi_{\plus\minus}\\[7pt]
-\sqrt{\frac{\mu_\zero\mu_\plus}{\nz\np}}\,\chi_{\zero\plus} & -\frac{\mu_\zero}{\nz}\chi_{\zero\zero} & -\sqrt{\frac{\mu_\zero\mu_\minus}{\nz\nm}}\,\chi_{\zero\minus}\\[7pt]
-\sqrt{\frac{\mu_\minus\mu_\plus}{\nm\np}}\,\chi_{\minus\plus} & -\sqrt{\frac{\mu_\minus\mu_\zero}{\nm\nz}}\,\chi_{\minus\zero} & -\frac{\mu_\minus}{\nm}\chi_{\minus\minus}
\end{pmatrix}\,,
\end{equation}
where we have used the fact that the ratios of volume factors $V_\alpha/V_\beta$ converge to unity after regularization, as shown in Appendix~\ref{sec:Empty integrals}. The matrix $\chi_{\alpha\beta}(\vb*{g},\vb*{g}')$ is the generalization of the function $\mathcal{X}(\vb*{g},\vb*{g}')$ introduced in Refs.~\cite{Marchetti:2020xvf,Marchetti:2022nrf}. In the causally extended setting considered here, it plays an essential role in capturing the interplay of combinatorial non-localities and the different causal characters of the tetrahedra in spite of the projection to uniform field configurations. It is appropriately regularized by volume factors of $\SL$ which, in spin representation, lead to Kronecker-delta-like symbols $\delta_{\rho,i}$, the details of which are explained in Appendix~\ref{sec:Regularization of Dirac delta function}. 

Crucially, the matrix $\chi_{\alpha\beta}$ depends on the combinatorial pattern $\gamma$ governing the interaction, the number of spacelike, lightlike, and timelike tetrahedra as well as their precise gluing. An exhaustive list of this matrix for double-trace melonic, quartic melonic, necklace, and simplicial interactions is given in Appendix~\ref{sec:Explicit expressions for chi}. As an example, the quartic melonic intertaction with two spacelike and two timelike tetrahedra exhibits three different possibilities of gluing, all of which lead to different expressions for $\chi_{\alpha\beta}$. 

Notice that since $F_{\alpha\beta}$ is clearly not diagonal, one obtains a non-trivial correlation between the building blocks of every type of signature. We will make this explicit momentarily. 

\paragraph{Symmetries of the effective kernel.} It is important for further analysis to highlight at this point the symmetries of the effective kernel acting on the geometric and matter variables. The effective kinetic kernel reduces to a function of the absolute value of scalar field differences, $\abs{\vbf-\vbf'}$. Concerning the geometric variables, notice that the matrix $\chi_{\alpha\beta}(\vbg,\vbg')$ is either constant or contains $\delta$-functions on $g$ and $g'$ as shown in Appendix~\ref{sec:Explicit expressions for chi}. Similarly, the Laplace operator $\Delta_{\SL}$ is invariant under left and right translation. Since we assume in addition that the functions $Z^\phi_\alpha(\vbg,\vbg')$ only depend on the trace of $g_c^{-1}g_c$, the effective kinetic kernel $G_{\alpha\beta}$ is invariant under the simultaneous left and right group action of $\SL$ acting on both arguments, i.e. for all $a_1,...,a_4\in\SL$ and $b_1,...,b_4\in\SL$, we have
\begin{equation}
G_{\alpha\beta}(g_1,...,g_4,\vbf;g_1',...,g_4',\vbf') = G_{\alpha\beta}(a_1 g_1 b_1,...,a_4 g_4 b_4,\vbf;a_1 g_1' b_1,...,a_4 g_4'b_4,\vbf').
\end{equation}
Consequently, $G_{\alpha\beta}$ only depends on the trace of $g_c^{-1}g_c'$ and we can conclusively write it in a short form
\begin{equation}\label{eq:2point G}
G_{\alpha\beta}(\vbg,\vbf;\vbg',\vbf') = G_{\alpha\beta}(\vbg^{-1}\vbg',\abs{\vbf-\vbf'}).
\end{equation}
This turns $G_{\alpha\beta}$ into a class function which has important consequences for the spin representation, as discussed in Appendix~\ref{sec:Correlation functions with geometricity constraints}.  Notice, that these symmetries are exactly those of a $2$-point function on a domain with local and non-local variables. This does not correspond to a new feature compared to previous works~\cite{Marchetti:2022igl,Marchetti:2022nrf} but we make these symmetries manifest here for the first time. 

In order to proceed with the program outlined in Section~\ref{sec:general strategy}, a choice of representation basis is required. In the following, we will be working in the spin representation of the extended causal group field presented in Section~\ref{sec:complete BC} and conduct the scheme of the Landau-Ginzburg method in this representation basis.

\subsection{Correlation functions in the spin representation}

As sketched in Section~\ref{sec:general strategy}, the correlation function is obtained by inverting the effective kinetic kernel, which is commonly done in Fourier space. The local degrees of freedom encoded in $\vbf\in\R^{\dloc}$ are transformed employing the standard Fourier transform on $\R^{\dloc}$. The geometric degrees of freedom on the other hand can be expanded in spin representation which was introduced in Section~\ref{sec:complete BC} with further details given in Appendix~\ref{sec:Aspects of SL2C and its Representation Theory}. With the symmetry properties in Eq.~\eqref{eq:2point G}, we observed that $G_{\alpha\beta}$ is a class function. Following Eq.~\eqref{eq:class function}, we, therefore, obtain the following decomposition for the effective kinetic kernel
\begin{equation}\label{eq:G spin rep general form}
G_{\alpha\beta,\vb*{j}\vb*{m}\vb*{l}\vb*{n};\vb*{j'}\vb*{m'}\vb*{l'}\vb*{n'}}^{\vbr\vbn;\vbr'\vbn'}(\vbk;\vbk')\\[7pt]
= G_{\alpha\beta}^{\vbr\vbn}(\vbk)\prod_{c=1}^4\delta(\rho_c-\rho_c')\delta_{\nu_c,\nu_c'}\delta_{j_c,j_c'}\delta_{m_c,m_c'}\delta_{l_c,l_c'}\delta_{n_c,n_c'}\delta(\vb*{k}+\vb*{k}')\,.
\end{equation}
Clearly, $G_{\alpha\beta}$ is diagonal in spin representation, such that its inversion reduces to a matrix inverse. However, to proceed with the computation of $C_{\alpha\beta}$, the interplay of signatures of tetrahedra and faces needs to be made explicit. This is done by computing the spin representation of the effective action in Eq.~\eqref{eq:effective action group rep} where the fluctuation fields respect in particular the simplicity constraint. This yields
\begin{equation}
\begin{aligned}
S_{\mathrm{eff}} &= \frac{1}{2}\sum_{\alpha,\beta}\int\prod_{c=1}^4\dd{\rho_c}\rho_c^2\:\delta\Phi^{\vbr,\alpha}_{\vb*{j}\vb*{m}}(\vbk)B^{\vbr,\alpha}_{\vb*{l}\vb*{n}}G_{\alpha\beta}^{\vbr}(\vbk)\delta\Phi^{\vbr,\beta}_{\vb*{j}\vb*{m}}(\vbk)B^{\vbr,\alpha}_{\vb*{l}\vb*{n}}+\frac{1}{2}\sum_{t=1}^4\sum_{(c_1,...,c_t)}\prod_{u=1}^t\sum_{\nu_{c_u}}\nu_{c_u}^2\\[7pt]
&\times\int\prod_{v=t+1}^4\dd{\rho_{c_v}}\rho_{c_v}^2\delta\Phi_{\vb*{j}\vb*{m}}^{(\vbr\vbn)_t,\minus}(\vbk)B^{(\vbr\vbn)_t,\minus}_{\vb*{l}\vb*{n}}G_{\minus\minus}^{(\vbr\vbn)_t}\left(\vbk\right)\delta\Phi_{\vb*{j}\vb*{m}}^{(\vbr\vbn)_t,\minus}(\vbk)B^{(\vbr\vbn)_t,\minus}_{\vb*{l}\vb*{n}},
\end{aligned}
\end{equation}
where the sum over $(c_1,...,c_t)$ is performed such that the $t$ timelike and $4-t$ spacelike labels are distributed equally across the four possible entries and  $(\vbr\vbn)_t = \nu_{c_1}...\nu_{c_t}\rho_{c_{t+1}}...\rho_{c_4}$. Following the geometric interpretation provided in~\cite{Jercher:2022mky}, the splitting of the action reflects the fact that spacelike faces can be shared between two tetrahedra of any signature, constituting the first term above, whereas timelike faces can only be shared between two timelike tetrahedra, represented by the second term above. Consequently, it is helpful for the remainder of this paper to consider the two terms in the effective action separately. In particular, the inversion of $G_{\alpha\beta}$ is performed for each case individually.

If all faces are spacelike then $G_{\alpha\beta}^{\vbr}$ is matrix-valued, and thus the correlation function is obtained as the matrix inverse, i.e.
\begin{equation}
\sum_\gamma G^{\vbr}_{\alpha\gamma}(\vbk)C_{\gamma\beta}(\vbk) = \delta_{\alpha\beta}.
\end{equation}
As a consequence, the correlation matrix $C_{\alpha\beta}$ contains an inverse factor of the determinant of $G_{\alpha\beta}$, turning it into a rational function in the variables $\vbr$ and $\vbk$. In the presence of at least one timelike face, $G_{\minus\minus}^{(\vbr\vbn)_t}(\vbk)$ is a scalar, and thus, its inverse is simply the multiplicative inverse which can be explicitly given as 
\begin{equation}
C_{\minus\minus}^{(\vbr\vbn)_t}(\vbk) = \frac{1}{Z_\minus^\phi((\vbr\vbn)_t)\vbk^2+\frac{Z_-^g(\vbk)}{a^2}\sum_c\cas_{1,c}+b^\minus}\,.
\end{equation}
Here $a$ is the skirt radius of the two-sheeted hyperboloid (see Appendix~\ref{sec:Empty integrals} for further details) and $b^{\minus} = \mu(1-\chi^{(\vbr\vbn)_t})$ is the effective mass depending on the labels $(\vbr\vbn)_t$. Notice that the function $\chi^{(\vbr\vbn)_t}$ is scalar-valued and of the same form as $\mathcal{X}$ in Refs.~\cite{Marchetti:2022igl,Marchetti:2020xvf}.

The real/direct space correlation function is obtained by performing the inverse of the Fourier transformation for which a detailed derivation is given in Appendix~\ref{sec:A derivation of the correlation function}. Following Eq.~\eqref{eq:spinrepCab}, all but the $(--)$-component contain contributions of spacelike faces only. That is, the Fourier components are given by $C_{\alpha\beta}^{\vbr}(\vbk)$. The $(--)$-component of the real space correlator contains $C_{\minus\minus}^{\vbr}(\vbk)$ as well as the contributions from timelike faces, $C_{\minus\minus}^{(\vbr\vbn)_t}(\vbk)$. An explicit formula is given in Eq.~\eqref{eq:spinrepC--}.

In the following two sections, we study the local and non-local correlation functions, obtained by integrating out the local and non-local variables, respectively. By determining their asymptotic behavior, we extract a correlation length which is a crucial ingredient for the computation of the Ginzburg-$Q$ and thus for testing the self-consistency of the mean-field approximation.   

\subsection{Local correlation function}\label{sec:Local correlation function}

Following Eq.~\eqref{eq:local and non-local correlator}, the local correlation function is obtained by integrating out the geometric variables $\vbg$. As shown in Appendix~\ref{sec:trivial represenation}, this yields a set of projections onto the trivial presentation with the label $\rho = i$, explicitly given in Eq.~\eqref{eq:zero mode projection}. Notice in particular that the timelike labels $\nu$ are constrained to vanish, $\nu = 0$. Since the representations labeled by $(0,\nu)$ entering the correlator $C_{\minus\minus}((\vbr\vbn)_t)$ are restricted by simplicity to values $\nu\in \{2,4,6,...\}$, the condition $\nu = 0$ can never be satisfied. As a result, the contributions to the local correlation of $C_{\minus\minus}$ with $t>0$ timelike faces vanish. Consequently, the local correlator reads as
\begin{equation}\label{eq:loc corr 1}
C_{\alpha\beta}(\vbf) = \int\frac{\dd{\vbk}}{(2\pi)^{\dloc}}\e^{i\vbk\cdot\vbf}C_{\alpha\beta}^{\vbi}(\vbk),
\end{equation}
where $\vbi$ denotes the four labels $\vbr$ being evaluated on the trivial representation $\rho_c = i$. What is essential for the qualitative behavior of the correlation function is that the matrix $\chi_{\alpha\beta}(\vbr)$ evaluated on four trivial representations takes the distinctive values
\begin{equation}
\chi_{\alpha\beta}^{\vbi}=
\begin{cases}
n_\alpha(n_\alpha-1),\qquad &\alpha=\beta,\\[7pt]
n_\alpha n_\beta,\qquad &\alpha\neq \beta.
\end{cases}
\end{equation}
In particular, the combinatorial details of the interaction, captured by the vertex graph $\gamma$, are integrated out and the functions $\chi_{\alpha\beta}^{\vbi}$ depend only on the numbers $\np,\nz$ and $\nm$. We note that this case is in fact equivalent to a local theory with multiple fields.

Since $C^{\vbi}_{\alpha\beta}(\vbk)$ is obtained as the matrix inverse of $G^{\vbi}_{\alpha\beta}(\vbk)$, it contains an an inverse factor of the determinant of $G_{\alpha\beta}$. As a result, $C^{\vbi}_{\alpha\beta}(\vbk)$ is a rational function in $\vbk^2$. To evaluate the integral for the local correlation function in Eq.~\eqref{eq:loc corr 1} explicitly, it is therefore expedient to perform a partial fraction decomposition, that is, we write
\begin{equation}\label{eq:local correlation fraction decomp}
C_{\alpha\beta}^{\vbi}(\vbk) = \sum_{m=1}^3\frac{\varsigma^m_{\alpha\beta}}{\vbk^2+b^m_{\vbi}}.
\end{equation}
The $b_{\vbi}^m$ are interpreted as an \textit{effective mass} evaluated on four trivial representations and are involved functions of $\mu_\alpha$ and $Z_\alpha^\phi(\vbi)$, implicitly depending on the form of $\chi^{\vbi}_{\alpha\beta}$.  In contrast to the case of a single signature, these parameters now contain a mixture of all the $\mu_{\alpha}$ and the explicit evaluation hereafter will only be performed numerically in the following.

The coefficients $\varsigma_{\alpha\beta}^m\in\C$ are constants of the partial fraction decomposition which are independent of the $\mu_\alpha$ and $Z^\phi_\alpha$. Notice that this decomposition depends on the components of the correlator and therefore carries indices $\alpha,\beta$. Since the local interaction is point-wise, all tetrahedra are correlated with one another, and thus, $\varsigma^m_{\alpha\beta}\neq 0$ for all $\alpha,\beta$. As we discuss below, the behavior is different for the non-local variables. 

\begin{figure}
    \centering
    \begin{subfigure}{0.33\textwidth}
    \includegraphics[width=\linewidth]{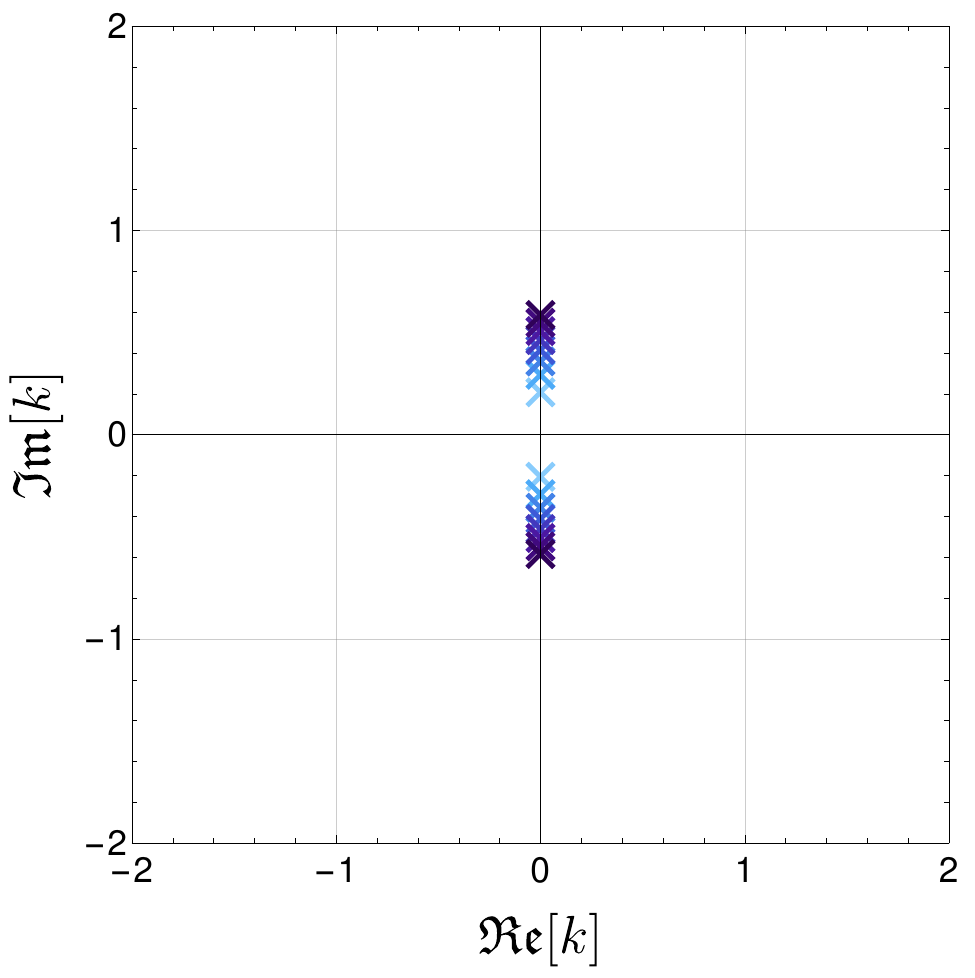}
    \end{subfigure}%
    \begin{subfigure}{0.33\textwidth}
    \includegraphics[width=\linewidth]{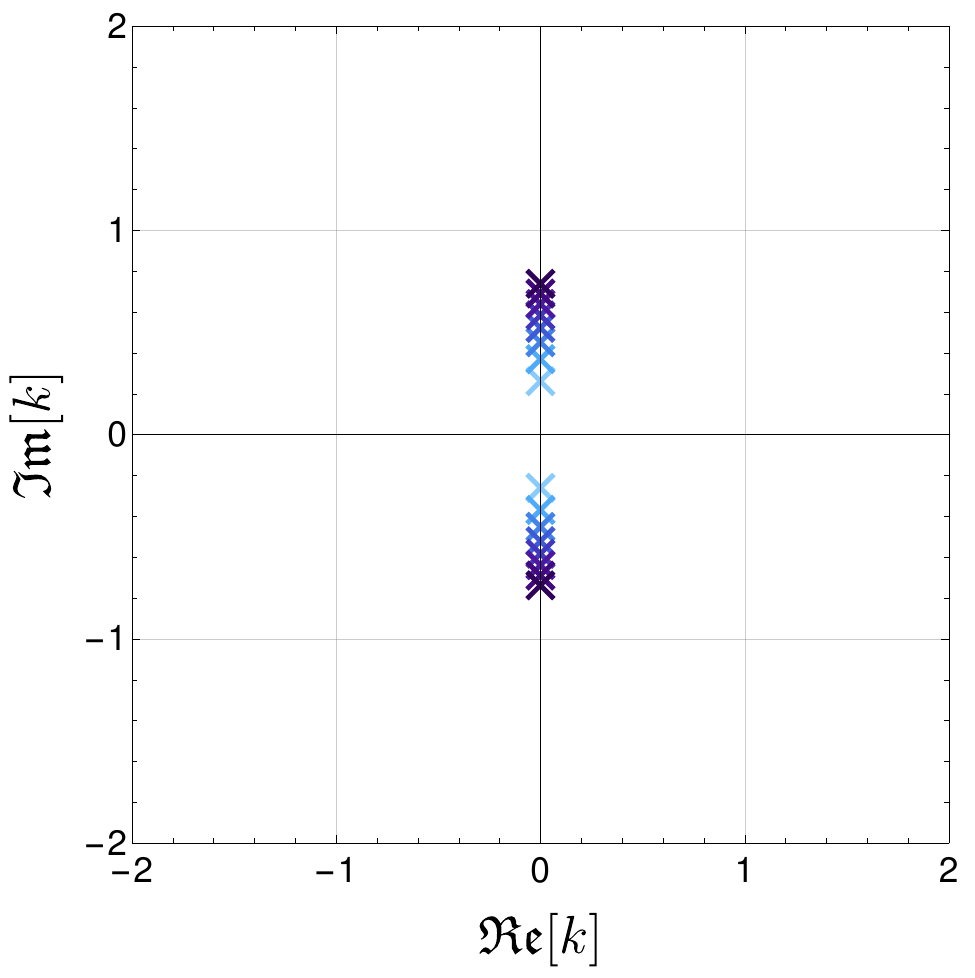}
    \end{subfigure}%
    \begin{subfigure}{0.33\textwidth}
    \includegraphics[width=\linewidth]{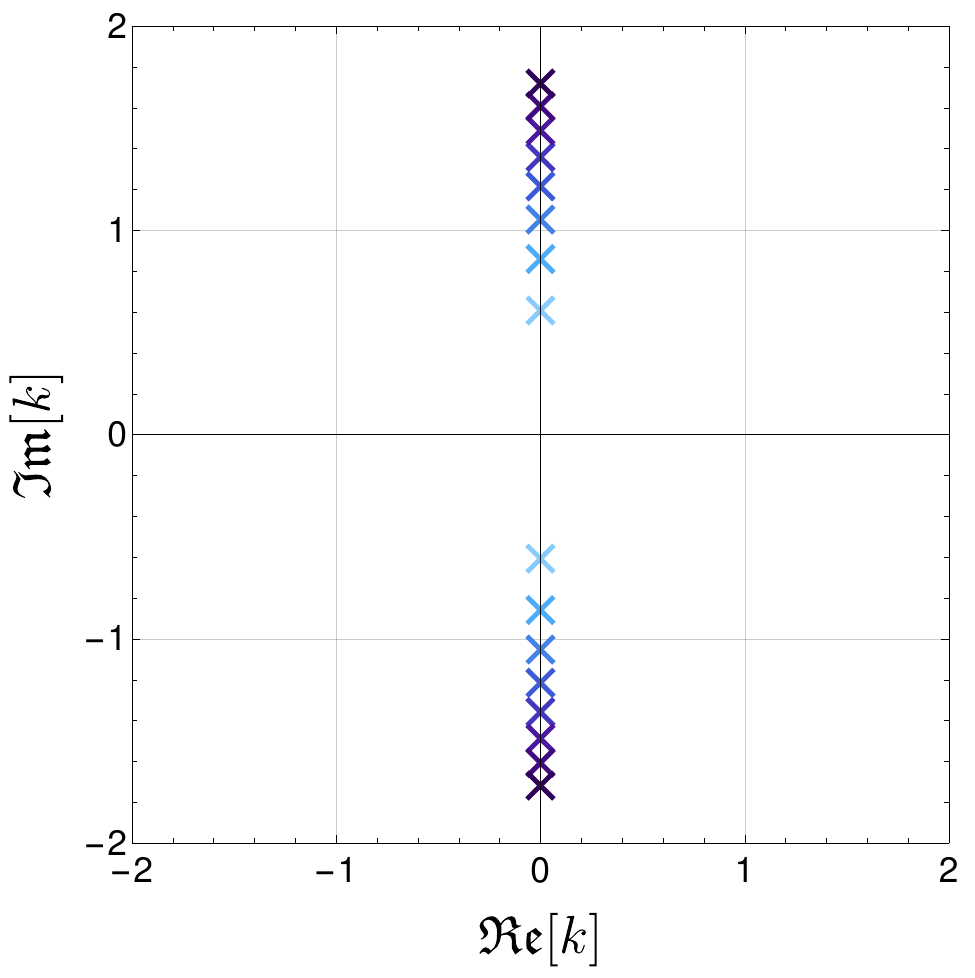}
    \end{subfigure}  
    \caption{Poles in the complex $k$-plane for $b_{\vbi}^m>0$ and a quartic interaction with $\np =2, \nz = 2$ and $\nm = 1$. The parameters are scaled as $(\mu_\plus,\mu_\zero,\mu_\minus) = (\mu,0.8\mu,1.2\mu)$ with $\mu\in\{0.04,\dots,0.32\}$ where darker colors indicate larger values of $\mu$. Crucially, the parameters $Z_\alpha^\phi(\vbi)$ are chosen in the range where all poles lie on the imaginary axis. Here $Z_\alpha^\phi(\vbi) = (-2,-1,0.1)$. The numerical values given here were chosen for demonstrative purposes.}
    \label{fig:kpoles}
\end{figure}

Following~\cite{Marchetti:2022igl}, one introduces spherical coordinates  $\vb*{k}\cdot\vb*{\phi} = kr\cos(\theta)$ that allow to perform explicitly the angular integration. Performing yet another coordinate substitution, $q = kr$, the correlator reduces to
\begin{equation}\label{eq:local correlator q-integral}
C_{\alpha\beta}(r) = \frac{1}{(2\pi)^{\frac{d}{2}}r^{d-2}}\sum_m\varsigma^m_{\alpha\beta}\int\limits_0^\infty\dd{q}\frac{q^{\frac{d}{2}}J_{\frac{d-2}{2}}(q)}{q^2+b_{\vbi}^m r^2},
\end{equation}
where $J_n$ are Bessel functions of the first kind~\cite{GradshteynBook} and $d\equiv \dloc$. Using~\cite[6.566, Formula 2.]{GradshteynBook}, we finally obtain
\begin{equation}\label{eq:local correlator K sum}
C_{\alpha\beta}(r) = \frac{1}{(2\pi)^{\frac{d}{2}}r^{d-2}}\sum_{m}\varsigma_{\alpha\beta}^m\left(\sqrt{b_{\vbi}^m}r\right)^{\frac{d-2}{2}}K_{\frac{d-2}{2}}\left(\sqrt{b_{\vbi}^m}r\right),
\end{equation}
where $K_n$ are the modified Bessel functions of the second type~\cite{GradshteynBook}. The asymptotic behavior of this function in the limit $r\gg 1$ is crucially determined by the $b_{\vbi}^m$, on which we elaborate in the following. A visualization of the pole structure of Eq.~\eqref{eq:local correlator q-integral} is given in Fig.~\ref{fig:kpoles}.

First, if non-zero, the $b_{\vbi}^m$ are homogeneous functions of the mass parameters $\mu_\alpha$, i.e.
\begin{equation}
b_{\vbi}^m(\mu\mu_{\plus},\mu\mu_0,\mu\mu_{\minus}) = \mu\: b_{\vbi}^m(\mu_{\plus},\mu_0,\mu_{\minus}),\qquad \forall \mu\in\R.
\end{equation}
In particular, in the limit $\mu_\alpha\rightarrow 0$, the effective masses $b_{\vbi}^m$ go to zero as well. Second, the  $b_{\vbi}^m$ can in principle vanish, which is entirely determined by the matrix $\chi_{\alpha\beta}^{\vbi}$ and thus by the number $n_{\alpha}$ of spacelike, lightlike and timelike tetrahedra entering the interaction in Eq.~\eqref{eq:single interaction}.\footnote{While a vanishing effective mass can be observed generically in tensorial field theories~\cite{Dekhil:2024ssa}, we emphasize that the vanishing of $b_{\vbi}^m$ has potentially a different reason. Here, it is not the non-localities of the interaction but the fact that one considers a multi-field theory which leads to $b_{\vbi}^m = 0$ for certain cases. In particular, the regularization scheme suggested in~\cite{Dekhil:2024ssa} is not required here to compute the Ginzburg-$Q$ in Section~\ref{sec:arbitrary but fixed Q}.} For $b_{\vbi}^{\bar{m}} = 0$, the corresponding contribution to the correlator decays as a power law, scaling as $C\sim r^{-d+2}$. The remaining $b_{\vbi}^m$ are either positive, negative, or even complex, sensitively depending on the parameter values $Z_\alpha^\phi(\vbi)$. We have checked numerically, that there exists a range of these $Z_\alpha^\phi(\vbi)$ for which the $b^m$ are real and positive and we restrict the theory to this parameter range for the remainder of this work.

Using the asymptotic properties of the Bessel function $K_n$~\cite{GradshteynBook} that read
\begin{equation}
K_n(z)\longrightarrow \sqrt{\frac{\pi}{2z}}\e^{-z},\qquad \abs{\arg(z)}<\frac{3\pi}{2},
\end{equation}
the local correlation function behaves asymptotically as
\begin{equation}
C_{\alpha\beta}(r)\underset{r\gg1}{\longrightarrow}\frac{1}{r^{\frac{d-1}{2}}}\tilde{\varsigma}^{m_*}_{\alpha\beta}\exp\left(-\sqrt{b_{\vbi}^{m_*}}r\right),
\end{equation}
where the index $m_* = \arg\min_m(b_{\vbi}^m)$. Clearly, the positive effective masses yield an exponential suppression. Notice that this is only the case for certain values of the $Z_\alpha^\phi(\vbi)$. Outside this range, the local correlator can exhibit an oscillatory or exponentially decaying behavior or a mixture of those. Notice that this is comparable to Refs.~\cite{Marchetti:2020xvf,Marchetti:2022igl}, wherein the function $\alpha_{\vb*{1,0,0}}$ was chosen to be positive to yield an exponentially decaying correlation function.

A correlation length can be extracted from the asymptotic behavior of the correlation function. In the local case derived here, the correlation length $\xiloc$ is given by
\begin{equation}
\xiloc = \frac{1}{\sqrt{b_{\vbi}^{m_*}}}\:\underset{\mu\rightarrow 0}{\longrightarrow}\: \mu^{-1/2},
\end{equation}
scaling as an inverse square root in a homogeneous limit where all mass parameters are taken to zero, i.e. $(\mu\mu_\plus,\mu\mu_\zero,\mu\mu\minus)$ with $\mu\rightarrow 0$. This is the typical mean-field theory result, where the critical exponent of the correlation length, usually denoted as $\nu_\mathrm{crit}$, is given by $\nu_{\mathrm{crit}} = 1/2$. Furthermore, the scaling of the correlation function at criticality is $C_{\alpha\beta}(r)\sim r^{-d+2}$ which is consistent with standard mean-field theory results. We expect that the scaling of the correlation function in a full RG treatment is modified by an anomalous dimension.

As another consistency check, the case of just one signature, e.g. all tetrahedra spacelike, can be re-obtained by setting $Z^\phi_\zero = Z^\phi_\minus = 0$ and demanding that $Z^\phi_\plus > 0$. Then, the correlation function and correlation length of Refs.~\cite{Marchetti:2022igl,Marchetti:2022nrf} are reproduced.

Concluding the derivation of the local correlation function, we observe that the inclusion of tetrahedra of all signatures leads to a significantly more involved form of the correlation function. In particular, it does not simply consist of either an oscillating or an exponentially decaying behavior as it is known from ferromagnetism or anti-ferromagnetism, respectively, but it actually involves a \textit{superposition} of these terms.\footnote{Correlation functions composed of exponentially decaying and oscillating terms are exotic (see for instance~\cite{Tsirlin2012}) and we leave the interpretation of such cases to future research.} By a suitable choice of $Z_\alpha^\phi(\vbi)$, all of the terms are exponentially decaying from which a correlation length can be extracted. This justifies in the first place the inclusion of the function $Z_\alpha^\phi$ at the beginning of this section as it grants more control over the asymptotics of the local correlation function. The necessity of weighting the Laplace operator $\Delta_\phi$ differently for different causal characters offers a tentative yet intriguing interpretation: it could suggest an intricate interplay of the scalar fields $\vbf$ and the causal character of tetrahedra. Notice that such an interplay has been exploited in Refs.~\cite{Jercher:2023kfr,Jercher:2023nxa}, where scalar clock and rod fields  have been coupled to the underlying causal structure to form a physical Lorentzian reference frame. 

We close by noting that a local theory of multiple fields shows the same behavior as the local correlation function we derived here. In particular, the extension of the parameter space by the functions $Z_\alpha^\phi$ is also in this case necessary to obtain an exponentially decaying correlation function.

\subsection{Non-local correlation function}\label{sec:Non-local correlation function}

To obtain a correlation length for the non-local variables, the scalar field dependence of $C_{\alpha\beta}(\vbg,\vbf)$ is integrated out as in Eq.~\eqref{eq:local and non-local correlator}. In Fourier space, this leads to evaluating the components on vanishing scalar field momenta $\vbk = 0$, while the dependence on the representation labels is non-trivial. The full form of the $(--)$-correlator contains a contribution with all faces spacelike and a contribution with a mixture of spacelike and timelike faces, as the expressions in Appendix~\ref{sec:A derivation of the correlation function} show. 

As argued in~\cite{Marchetti:2020xvf,Marchetti:2022igl}, for further analysis of the correlation function and the computation of the Ginzburg-$Q$ below, it is useful to expand $C_{\alpha\beta}$ in zero modes. Since zero modes are a priori not part of the spin decomposition of $L^2$-functions on $\SL$, we need to extend for this purpose the correlation function to 
\begin{equation}
C^{\mathrm{ext}}_{\alpha\beta}(\vbg) = \sum_{s=s_0}^4V_\plus^{-s}\sum_{(c_1,...,c_s)}\int\prod_{c=c_1}^{c_s}\dd{g_{c}}C_{\alpha\beta}(\vbg)\equiv \sum_{s=s_0}^4V_\plus^{-s}\sum_{(c_1,...,c_s)}C_{\alpha\beta}^s(\vbg_{4-s}),
\end{equation}
where $s$ labels the number of zero modes.\footnote{Volume factors have been included for regularization and can be derived from a de-compactification from $\Spin$ to $\SL$, as shown in~\cite{Marchetti:2022igl}.} Hereafter, we restrict to $s\geq s_0$ zero modes for which the matrix $\chi_{\alpha\beta}$ is non-vanishing. That is because for $s < s_0$, the mass corrections vanish, generically leading to an oscillating correlation function. As discussed in~\cite{Marchetti:2022igl,Marchetti:2022nrf,Marchetti:2020xvf}, these terms correspond to long-range corrections that are present irrespective of the phase transition, justifying their exclusion in the analysis of the critical behavior. The number $s_0$ depends on the combinatorics and is given by $s_0 = 0$ for double-trace melons, $s_0 = 1$ for quartic melons, $s_0 =2$ for necklaces and $s_0 = 3$ for simplicial interactions~\cite{Marchetti:2020xvf}.  Due to the projection onto $s$ trivial representations, the \textit{residual correlation function} $C_{\alpha\beta}^s$ only depends on the $4-s$ remaining group variables, $\vbg_{4-s} = (g_{c_{s+1}},\dots,g_{c_4})$.

In the following, we split the analysis of this function into the contributions to $C^s_{\alpha\beta}$ that contain only spacelike labels $\rho$ and those contributions that contain at least one timelike label $\nu$. 

\subsubsection{Contributions with only spacelike faces}\label{sec:Contributions with only spacelike faces}

In the case where all faces are spacelike, the contribution to the correlation function $C_{\alpha\beta}^s(\vbg_{4-s})$ is given in terms of an integral
\begin{equation}\label{eq:nonlocal integrand all sl}
\eval{C^s_{\alpha\beta}(\vbg_{4-s})}_{\mathrm{sl}} = \int\prod_{c=c_{s+1}}^{c_4}\dd{\rho_{c}}\rho_{c}^2D^{(\rho_{c},0)}_{j_{c}m_{c}j_{c}m_{c}}(g_{c})\bar{C}_{\alpha\beta}^{s,\vbr_{4-s}}(\vb*{0}).
\end{equation}
Notably, the matrix $\chi_{\alpha\beta}^{c_1\dots c_s}$ entering ${C}^s$ is evaluated on $s$ zero modes and therefore takes a constant value, depending on the combination of the combinatorics of $\gamma$ and the distribution of signatures $\{+,0,-\}$, see Appendix~\ref{sec:Explicit expressions for chi} for explicit expressions. While components of $C_{\alpha\beta}^s$ can in principle vanish, we focus in the following on the asymptotic behavior of the non-vanishing components, keeping the expression for ${C}_{\alpha\beta}$ general. 

To extract the asymptotic behavior for large distances on the group manifold, we perform a Cartan decomposition as in Eqs.~\eqref{eq:Cartan decomp g}--\eqref{eq:Cartan decomp D}. Since the resulting $\SUT$-Wigner matrices are in fact independent of $\rho_{c_1},...,\rho_{c_{4-s}}$, they can be factorized from the integral, yielding
\begin{equation}
\eval{C^s_{\alpha\beta}(\vbg_{4-s})}_{\mathrm{sl}} = C^s_{\alpha\beta}\left(\vb*{\eta}_{4-s}\right)_{\vb*{j}_{4-s}\vb*{m}_{4-s}}\prod_{u=1}^{4-s}D^{j_{c_u}}_{m_{c_u}m_{c_u}}(u_{c_u}^{-1}v_{c_u}),
\end{equation}
where $u,v\in\SUT$ and $\eta\in\R_{+}$ is interpreted as the rapidity parameter of a Lorentz boost along the $z$ axis, see also Appendix~\ref{sec:Aspects of SL2C and its Representation Theory}. The function $C^s_{\alpha\beta}\left(\vb*{\eta}_{4-s}\right)_{\vb*{j}_{4-s}\vb*{m}_{4-s}}$ is given by
\begin{equation}\label{eq:C(eta) integral}
 C^s_{\alpha\beta}\left(\vb*{\eta}_{4-s}\right)_{\vb*{j}_{4-s}\vb*{m}_{4-s}} = \int\prod_{c=c_{s+1}}^{c_4}\dd{\rho_{c}}\rho_{c}^2\: d_{j_cj_cm_c}^{(\rho_c,0)}\left(\frac{\eta_c}{a}\right)C_{\alpha\beta}^{s,\vbr_{4-s}}.
\end{equation}
In the following, we suppress the dependence of the factorized  $C^s_{\alpha\beta}(\vb*{\eta}_{4-s})$ on the magnetic indices because they are not relevant for the extraction of the scaling behavior of the correlator.

\begin{figure}
    \centering
    \begin{subfigure}{0.33\textwidth}
    \includegraphics[width=\linewidth]{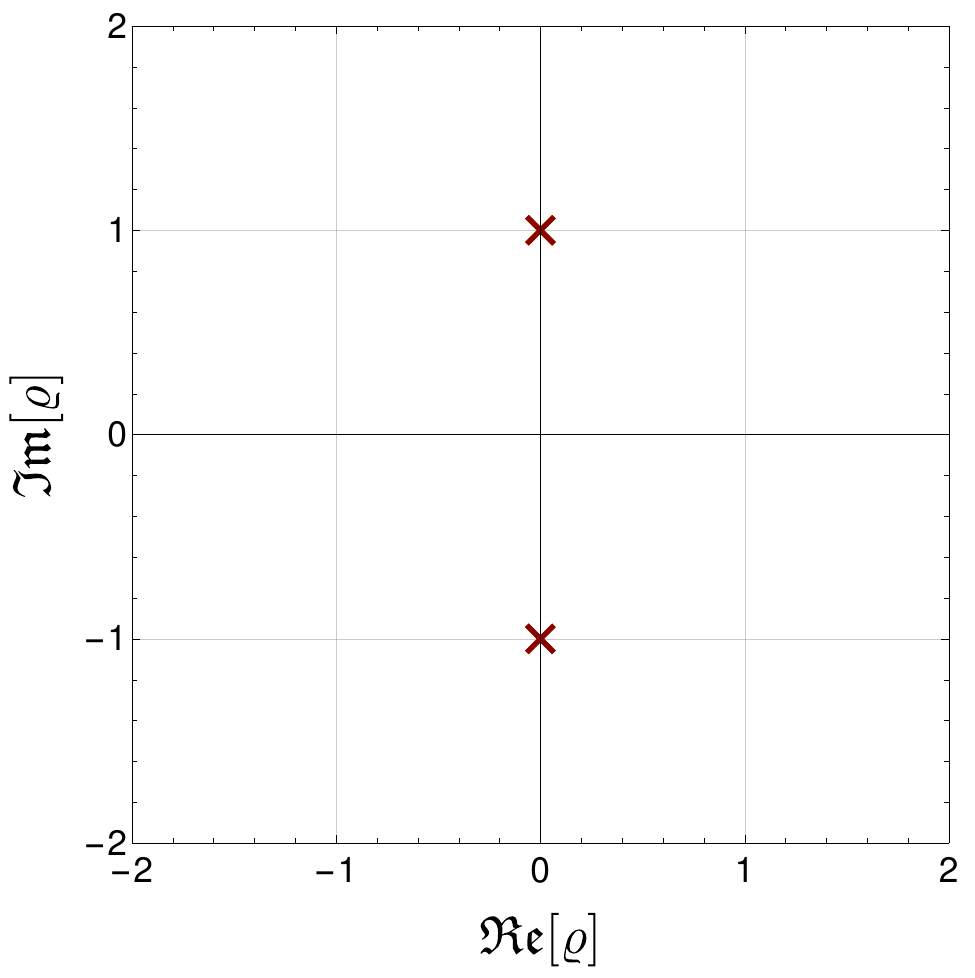}
    \end{subfigure}%
   \begin{subfigure}{0.33\textwidth}
    \includegraphics[width=\linewidth]{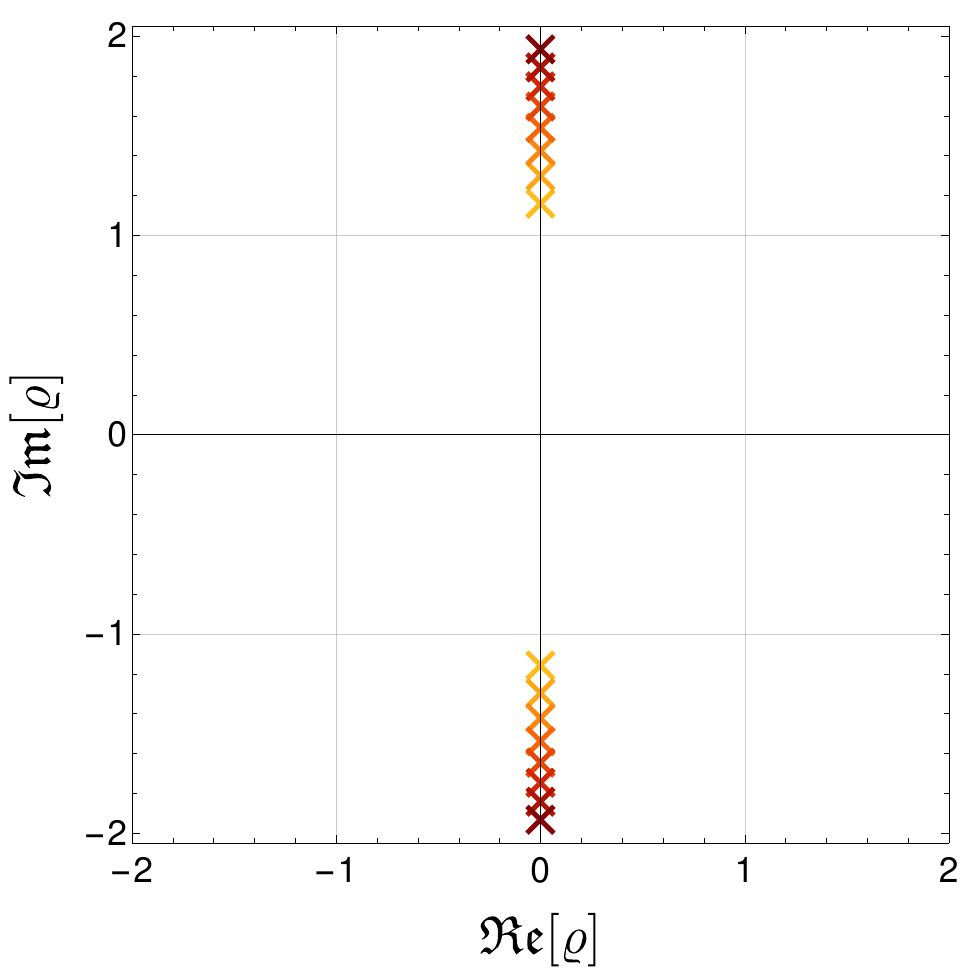}
    \end{subfigure}%
    \begin{subfigure}{0.33\textwidth}
    \includegraphics[width=\linewidth]{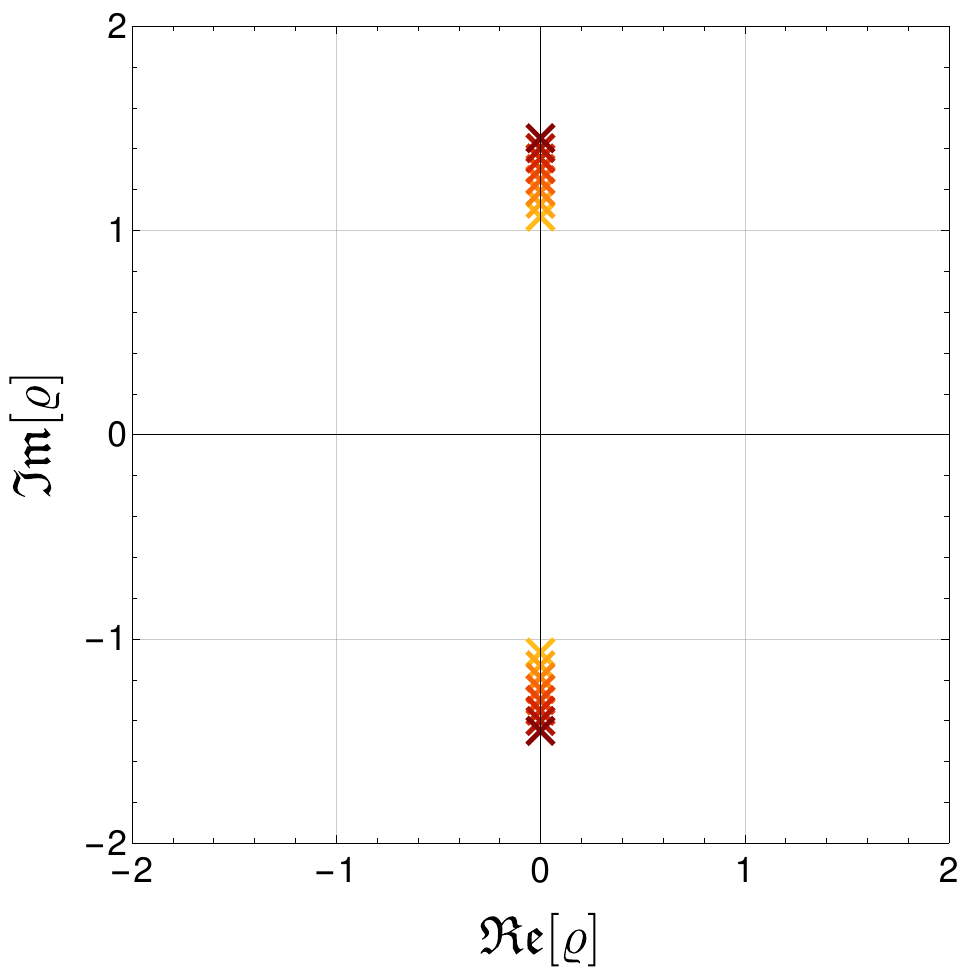}
    \end{subfigure}
    \caption{Poles of the non-local correlation function for $s=3$ zero modes in the complex $\rho$-plane with $\chi_{\alpha\beta}$ given as e.g. the second entry in Tab.~\eqref{tab:211}, which is common to several quartic interactions. The parameters $\mu_{\alpha}$ are scaled as $(\mu_\plus,\mu_\zero,\mu_\minus) = (\mu,0.8\mu,1.2\mu)$ with $\mu\in\{0.2,\dots,1.6\}$ where darker colors corresponds to larger values of $\mu$. Here, $a=1$ and the functions $Z_\alpha^g(\vb*{0}) = (1,2,0.5)$ lie in a range that yields poles above the value $\rho = i$. Notice that in the left picture, the poles are given by $\rho = \pm i$, corresponding to a vanishing effective mass $b_{c_1\dots c_s}^m = 0$. The numerical values given here were chosen for demonstrative purposes.}
    \label{fig:rhopoles3s}
\end{figure}

Following~\cite{Marchetti:2022igl}, the integrals in Eq.~\eqref{eq:C(eta) integral} can be evaluated by performing a contour integration for one of the variables, say $\rho_{c_1}$ and performing a stationary phase analysis for the remaining $3-s$ variables $\rho_{c_2},...,\rho_{4-s}$. To actually conduct these two steps, it is advantageous to carry out a partial fraction decomposition of $C_{\alpha\beta}^{s,\vbr_{4-s}}(\vb*{0})$. That is, the residual correlation function in spin representation is re-written as
\begin{equation}
 C^s_{\alpha\beta}\left(\vb*{\eta}_{4-s}\right) = \sum_{m=1}^3 \int\prod_{c=c_{s+1}}^{c_4}\dd{\rho_{c}}\rho_{c}^2\:d_{j_cj_cm_c}^{(\rho_c,0)}\left(\frac{\eta_c}{a}\right)\frac{\varrho_{\alpha\beta}^m}{\frac{1}{a^2}\sum_u (\rho_{c_u}^2+1)+b^m_{c_1\dots c_s}},
\end{equation}
where $a$ is the skirt radius of the two-sheeted hyperboloid. The $b_{c_1\dots c_s}^m$ are interpreted as the \textit{effective mass} evaluated on $s$ zero modes which are intricate functions of the $\mu_{\alpha}$ and $Z_\alpha^g(\vb*{0})$ that implicitly depend on the matrix $\chi_{\alpha\beta}^{c_1\dots c_s}$. These masses determine the pole structure of the integrand above. In Fig.~\ref{fig:rhopoles3s}, the poles are depicted for $s=3$ zero modes and an exemplary choice of combinatorics for illustrative purposes.

The coefficients $\varrho_{\alpha\beta}^m\in\C$ arise from the partial fraction decomposition and are independent of the parameters $\mu_\alpha$ and $Z^g_\alpha$. Furthermore, the $\varrho_{\alpha\beta}^m$ explicitly depend on the components $\alpha,\beta$. In particular, if tetrahedra of causal character $\bar{\alpha}$ and $\bar{\beta}$ are uncorrelated, then $\varrho_{\bar{\alpha}\bar{\beta}}= 0$. As an example, the spacelike and timelike tetrahedra of the first vertex depicted in Tab.~\ref{tab:202} are entirely uncorrelated such that $\varrho^m_{\plus\minus} = \varrho^m_{\minus\plus} = 0$, while the diagonal components $\varrho^m_{\plus\plus}$ and $\varrho^m_{\minus\minus}$ are non-zero.

Two properties of the $b_{c_1\dots c_s}^m$ are determining for the asymptotic behavior of the non-local correlation function. First, if non-zero, the $b_{c_1\dots c_s}^m$ are homogeneous functions of the parameters $\mu_\alpha$, i.e.
\begin{equation}
b_{c_1\dots c_s}^m(\mu\mu_{\plus},\mu\mu_0,\mu\mu_{\minus}) = \mu\: b_{c_1\dots c_s}^m(\mu_{\plus},\mu_0,\mu_{\minus}),\qquad \forall \mu\in\R.
\end{equation}
In particular, in the limit $\mu_\alpha\rightarrow 0$, the effective masses scale to zero, $b_{c_1\dots c_s}^m\rightarrow 0$. 

Second, the effective mass $b_{c_1\dots c_s}^{\bar{m}}$ for one of the $m$, say $\bar{m}$,  will generically evaluate to zero, depending on the combinatorics and the distribution of signatures of the interaction. Such a vanishing of the effective mass in general tensorial field theories with just one type of field has been highlighted in Ref.~\cite{Dekhil:2024ssa}. In particular, it was shown therein that the Landau-Ginzburg method can be applied also to this case by introducing a regularization of the vanishing effective mass via a small parameter $\epsilon>0$, i.e. 
\begin{equation}\label{eq:b regularization}
b^{\bar{m}}_{c_1\dots c_s} = \epsilon f(\{\mu_\alpha\}),
\end{equation}
where $f$ is any positive homogeneous function of the $\mu_\alpha$. The computation of the correlation function and the Ginzburg-$Q$ parameter can then be carried through with the limit $\epsilon\rightarrow 0$ taken at the end of the calculation. For all non-vanishing $b_{c_1\dots c_s}^m$, there exists a range of parameter values $Z_\alpha^g(\vb*{0})$ such that $b_{c_1\dots c_s}^m>0$. This will lead to an exponentially decaying correlation function and we therefore restrict to this range for the remainder. Overall, we continue the evaluation of $C_{\alpha\beta}$ with positive $b_{c_1\dots c_s}^m$.

Utilizing the asymptotic behavior of the residual $\SL$-Wigner matrices $d^{(\rho,0)}_{jjm}(\frac{\eta}{a})$, derived in~\cite{Marchetti:2022igl},  
\begin{equation}\label{eq:asymptotic Wigner rho}
d^{(\rho,0)}_{jlm}\left(\frac{\eta}{a}\right)\underset{\frac{\eta}{a}\gg 1}{\longrightarrow} c_{\rho}(j,l,m)\e^{-\frac{\eta}{a}}\e^{i\rho\frac{\eta}{a}},
\end{equation}
we apply the residue theorem for the $\rho_{c_{s+1}}$-contour integral, yielding a sum of exponential terms
\begin{equation}
\e^{i\phi_m\frac{\eta}{a}} = \exp\left(i\frac{\eta}{a}\left(\bar{\rho}_{m,+}+\sum_{u=s+2}^{4}\rho_{c_u}\right)\right),
\end{equation}
with 
\begin{equation}
\bar{\rho}_{m,+} = i\sqrt{\sum_{u=s+2}^4\rho_{c_u}+4-s+a^2b_{c_1\dots c_s}^m}.
\end{equation}
The extrema of the phase $\phi_m$ satisfy the equation
\begin{equation}
\pdv{\phi_m}{\rho_{c_u}} \overset{!}{=} 0
\end{equation}
such that at leading order, the stationary phase approximation yields the labels $\rho_{c_u}$ evaluated on
\begin{equation}
\bar{\rho}_{m} = i\sqrt{1+\frac{a^2b_{c_1\dots c_s}^m}{4-s}}.
\end{equation}

The asymptotics of the correlation function is considered in the \enquote{isotropic} limit, where $\eta_{c_u} = \eta\gg 1$ for all $u\in\{1,...,4-s\}$. The expectation that this still captures the asymptotic behavior of the non-local correlation function is supported by the results of~\cite{Marchetti:2022igl,Marchetti:2022nrf}. Altogether, we obtain 
\begin{equation}
\eval{C^s_{\alpha\beta}(\vb*{\eta}_{4-s})}_{\eta_c\equiv\eta}\:\underset{\eta\gg 1}{\longrightarrow}\: \sum_{m=1}^3 \tilde{\varrho}^m_{\alpha\beta}\exp\left(-(4-s)\frac{\eta}{a}\left(1+\sqrt{1+\frac{a^2b_{c_1\dots c_s}^m}{4-s}}\right)\right)\,,
\end{equation}
where we remind the reader that this correlation function contains only terms with spacelike labels $\rho$. Clearly, the correlation function exhibits an exponential decay independent of the value of the parameters $\mu_{\alpha}$, in particular close to the critical point where $\mu_{\alpha}\rightarrow 0$. As proposed in~\cite{Marchetti:2022igl}, we re-scale the correlation function by the Jacobian determinant on the hyperboloid as present in the Haar measure,
\begin{equation}
\tilde{C}^s_{\alpha\beta}(\vb*{\eta}_{4-s}) = \sinh^{2(4-s)}\left(\frac{\eta}{a}\right)C^s_{\alpha\beta}(\vb*{\eta}_{4-s}).
\end{equation}
Expanding the square root for small values of $a^2b^m_{c_1\dots c_s}$, one finally extracts
\begin{equation}\label{eq:asymptotics of Cs}
\tilde{C}^s_{\alpha\beta}(\vb*{\eta}_{4-s})\:\underset{\eta/a\gg 1}{\longrightarrow}\:\exp\left(-\frac{1}{2}a b^{m_*}_{c_1\dots c_s}\eta\right),
\end{equation}
where $m_* = \arg \min_m(b^m_{c_1\dots c_s})$. The non-local correlation length is thus defined as
\begin{equation}
\xinloc = \frac{2}{a b^{m_*}_{c_1\dots c_s}}\underset{\mu\rightarrow 0}{\longrightarrow} (a\mu)^{-1},
\end{equation}
exhibiting the same scaling behavior as in~\cite{Marchetti:2022igl,Marchetti:2022nrf} extracted with the same method if the mass parameters $\mu_\alpha$ are homogeneously taken to zero. Finally, we note that the derivation of the correlation length could alternatively be pursued using the second-moment method~\cite{Marchetti:2020xvf} as also done in Ref.~\cite{Marchetti:2022igl} for the Lorentzian BC model restricted to spacelike tetrahedra. It can be straightforwardly expected that the application of that method to the present case will lead to the same result as produced here. 

In retrospect, including the functions $Z_\alpha^g$ turned out to be essential to obtain a regime where the correlation function exhibits an exponential suppression. For $Z_\alpha^g$ lying outside the chosen regime above, there exist combinations of signature and combinatorics, such that the correlation function is a mixture of exponentially decaying and exponentially diverging terms. Exponentially diverging correlation functions indicate long-range correlations which we exclude for studying phase transitions. This exponential behavior, diverging or converging, is characteristic of the underlying hyperbolic geometry of $\SL$. We note that these results are not surprising as they would be obtained also in a local multi-field theory on the two-sheeted hyperboloid $\mathrm{H}_\plus$.

This concludes the investigation of the asymptotic behavior of purely spacelike contributions to the non-local correlation function evaluated on $s$ zero modes. In the following section, we consider the $(--)$-component of the non-local correlator in the presence of one or more timelike faces, after which we give a combined summary of the non-local results.

\subsubsection{Contributions with timelike faces}\label{sec:Contributions with timelike faces}

To complete the analysis of the non-local correlation function, we study in this section the component $C^s_{\minus\minus}$ if at least one of the labels is associated with a timelike face. For $t>0$ labels $\nu_{c_1},...,\nu_{c_t}$ this results in evaluating the following expression
\begin{equation}\label{eq:nonlocal integrand tl}
\eval{C^s_{\minus\minus}(\vb*{\eta}_{4-s})}_{\mathrm{tl}} = \prod_{c=c_{s+1}}^{c_{s+t}}\sum_{\nu_{c}}\nu_{c}^2\prod_{c'=s+t+1}^{4}\int\dd{\rho_{c'}}\rho_{c'}^2\frac{d^{(0,\nu_{c})}_{j_{c}j_{c}m_{c}}\left(\frac{\eta_c}{a}\right)d^{(\rho_{c'},0)}_{j_{c'}j_{c'}m_{c'}}\left(\frac{\eta_{c'}}{a}\right)}{\frac{1}{a^2}Z_\minus^g(\vb*{0})\left(\sum_{c'}\rho_{c'}^2-\sum_c\nu_{c}^2+(4-s)\right)+b^{\minus}_{c_1\dots c_s}},
\end{equation}
where $b^{\minus}_{c_1\dots c_s} $ is the effective mass evaluated on $s$ zero modes with $t$ timelike labels.

For further evaluation, the asymptotic form of the reduced Wigner matrices restricted to timelike labels is essential. It is shown in Appendix~\ref{sec:asymptotics of d} that the scaling of $d^{(0,\nu)}(\eta)$ is in fact independent of $\nu$, i.e.
\begin{equation}\label{eq:asymptotic Wigner nu}
d^{(0,\nu)}_{jlm}\left(\frac{\eta}{a}\right)\:\underset{\eta\gg 1}{\longrightarrow}\:\e^{-\frac{\eta}{a}}.
\end{equation}
This property of the Wigner matrix has crucial consequences for the asymptotic behavior of the correlation function, which is discussed separately for the cases $t=4-s$ and $t<4-s$ in what follows.

\paragraph{$\vb*{t = 4-s}$ timelike labels.} In this case, the asymptotic behavior for large $\frac{\eta}{a}$ fully decouples from the representation labels $\nu_{c_1},...,\nu_{c_{4-s}}$, i.e. 
\begin{equation}
\eval{C_{\minus\minus}^s(\vb*{\eta}_{4-s})}_{t=4-s}\sim \e^{-(4-s)\frac{\eta}{a}},
\end{equation}
which is readily independent of the effective mass $b^{\minus}_{c_1\dots c_s}$. As a result, also the correlation function re-scaled by the Jacobian determinant on the hyperboloid is independent of the mass parameters $\mu_\alpha$. We conclude that due to the independence of the mass parameters the contributions with $4-s$ timelike labels do not affect the critical behavior of the correlation function. 

\paragraph{$\vb*{t < 4-s}$ timelike labels.} If one or more of the $4-s$ remaining labels are spacelike, one can apply the strategies presented in detail in Section~\ref{sec:Contributions with only spacelike faces}. That is, one performs a contour integration for one of the spacelike labels by applying the residue theorem. The remaining $\rho$-integrals are then performed via a stationary phase approximation. 

To be explicit, we consider representation labels $\nu_{c_{s+1}},...,\nu_{c_{s+t}},\rho_{c_{s+t+1}},...\rho_{c_{4}}$, or $\nu_{c},\rho_{c'}$ for short. Performing the contour integration for the variable $\rho_{c_{t+1}}$, the corresponding poles of the integrand in Eq.~\eqref{eq:nonlocal integrand tl} are given by
\begin{equation}
\bar{\rho}_{c_{t+1}} = \pm\sqrt{\sum_{c}\nu_{c}^2-\sum_{c'}\rho_{c'}^2-(4-s)-\frac{a^2b^{\minus}_{c_1\dots c_s}}{Z^g_\minus(\vb*{0})}}.
\end{equation}
Evaluating the exponential function $\e^{i\rho_{c_{t+1}}}$ on the pole leads to an overall oscillating exponential $\e^{i\phi\eta/a}$ with maxima satisfying
\begin{equation}
\bar{\rho} = \pm\sqrt{\frac{\sum_c\nu_{c}^2-(4-s)-\frac{a^2b^{\minus}_{c_1\dots c_s}}{Z^g_\minus(\vb*{0})}}{4-s-t}}.
\end{equation}
Since $\nu_{c}\in 2 \mathbb{N}^+$ and $b^\minus_{c_1\dots c_s}\rightarrow 0$ near criticality, the solutions $\bar{\rho}$ are real near the phase transition. 

Finally, contributions to $C^s_{\minus\minus}$ with $t<4-s$ timelike faces show the following asymptotic behavior
\begin{equation}
\eval{C_{\minus\minus}^s(\vb*{\eta}_{4-s})}_{t<4-s} \:\underset{\eta\gg 1}{\longrightarrow}\:\sum_{\nu_{c_{s+1}}...\nu_{c_{s+t}}}\exp\left(\frac{\eta}{a}\left(-(4-s)+i(4-s-t)\bar{\rho}\right)\right).
\end{equation}
As $\mu_\alpha\rightarrow 0$, the function $\bar{\rho}$ does not change qualitatively and it remains finite and real for any $\nu_c$. We conclude that the asymptotics of the correlation function $C^s_{\minus\minus}$ or its re-scaled form $\tilde{C}^s_{\minus\minus}$ remain unaffected in the critical region $\mu_\alpha\rightarrow 0$. This suggests the overall interpretation that contributions which contain at least one timelike face are insensitive to the critical behavior which is instead driven by contributions from spacelike labels $\rho$.

Summarizing the asymptotic analysis of the non-local correlation function, there exists an exponential decay behavior close to criticality which is induced by contributions from spacelike faces. A correlation length can be extracted which in a homogeneous limit scales as $\xinloc \sim (a\mu)^{-1}$. Timelike faces on the other hand do not contribute to this exponential decay behavior and therefore do not drive the critical behavior of the system. These results offer an elegant geometric explanation: The group $\SL$ is topologically given as $\SL\cong_\mathrm{top}\mathrm{H}_\plus\times S^3$, i.e. a product of the two-sheeted hyperboloid $\mathrm{H}_\plus$ and the $3$-sphere $S^3$, which in turn is diffeomorphic to $\text{SU}(2)$. Continuous labels $\rho$ of spacelike faces are associated with the non-compact hyperbolic part, while discrete timelike faces are associated with the compact spherical/rotational part. Following~\cite{Benedetti:2014gja} as well as the results of~\cite{Pithis:2020sxm,Marchetti:2020xvf,Marchetti:2022igl,Marchetti:2022nrf}, phase transitions require a non-compact domain of the fields, which is precisely what we have found in this section. 

\subsection{Ginzburg-$Q$}\label{sec:arbitrary but fixed Q}

The validity of the mean-field approach for a single arbitrary interaction studied in this section is tested through the derivation of the Ginzburg-$Q$, defined in Eq.~\eqref{eq:definition Q}. The range of integration $\Omega_{\xi}$ is determined by the local- and non-local correlation lengths. Explicitly, $\Omega_{\xi}$ is defined as
\begin{equation}
\Omega_{\xi} = \SL^4_{\xi}\times [-\xiloc,\xiloc]^{\dloc},
\end{equation}
where $\SL_{\xi}$ denotes the non-compact part of $\SL$, parametrized by the variable $\eta$, which is restricted to $\eta\in [0,\xinloc]$. 

In the following, we identify volume factors of different signatures $V_\alpha\equiv V$ justified by the arguments in Appendix~\ref{sec:Empty integrals}. Furthermore, the $\SL$ volume factors are regulated by cutoffs $L$ and $\xinloc$, and are differentiated as $V_L$ and $V_\xi$, respectively. For large values of $L$ and $\xi$, the volume factors scale as
\begin{equation}
V_L\sim\e^{2L/a}~~~\text{and}~~~V_\xi\sim\e^{2\xinloc/a},
\end{equation}
with $a$ the skirt radius of the two-sheeted hyperboloid.

First, we compute the denominator of $Q_{\alpha\beta}$ given in Eq.~\eqref{eq:definition Q}. Since the mean-field background solutions $\bar{\Phi}_\alpha$ are constant, the integration yields four volume factors of $V_\xi$ restricted by the non-local correlation length $\xinloc$ and $\dloc$ volume factors of $\R$, similarly cut off by the local correlation length $\xiloc$. Furthermore, from the mean-field background solutions in Eq.~\eqref{eq:mf solutions}, we extract
\begin{equation}
\bar{\Phi}_\alpha\bar{\Phi}_\beta =  V_L^{-4}V_L^{-2\frac{n_\gamma-1}{n_\gamma-2}}\lambda^{-\frac{2}{n_\gamma-2}}m_{\alpha\beta}(\mu_\plus,\mu_\zero,\mu_\minus).
\end{equation}
The function $m_{\alpha\beta}$ is explicitly given by
\begin{equation}
\begin{aligned}
& m_{\alpha\beta}^\gamma(\mu_\plus,\mu_\zero,\mu_\minus)\\[7pt]
=& \left(-\frac{\mu_\alpha}{n_\alpha}\right)^{\frac{2+n_\alpha-n_\gamma}{2(n_\gamma-2)}}\prod_{\alpha'\neq \alpha}\left(-\frac{\mu_{\alpha'}}{n_{\alpha'}}\right)^{\frac{n_{\alpha'}}{2(n_\gamma-2)}}\left(-\frac{\mu_\beta}{n_\beta}\right)^{\frac{2+n_\beta-n_\gamma}{2(n_\gamma-2)}}\prod_{\beta'\neq \beta}\left(-\frac{\mu_{\beta'}}{n_{\beta'}}\right)^{\frac{n_{\beta'}}{2(n_\gamma-2)}}.
\end{aligned}
\end{equation}
While $m_{\alpha\beta}$ has an intricate form, we consider in the remainder only its scaling behavior in a homogeneous limit in the parameters $\mu_\alpha$. To that end, we notice that $m_{\alpha\beta}$ is a homogeneous and non-zero function, scaling as
\begin{equation}
m_{\alpha\beta}(\mu\mu_\plus,\mu\mu_\zero,\mu\mu_\minus) = \mu^{\frac{2}{n_\gamma-2}}m_{\alpha\beta}(\mu_\plus,\mu_\zero,\mu_\minus).
\end{equation}
Combined with the empty integrations, we obtain the following expression for the denominator
\begin{equation}
\int_{\Omega_\xi}\dd{\vbg}\dd{\vbf}\bar{\Phi}_{\alpha}\bar{\Phi}_\beta\: \sim\: \xiloc^{\dloc}\left(\frac{V_\xi}{V_L}\right)^4 V_L^{-2\frac{n_\gamma-1}{n_\gamma-2}}\lambda^{-\frac{2}{n_\gamma-2}}m_{\alpha\beta}(\mu_\plus,\mu_\zero,\mu_\minus).
\end{equation}
Using the scaling behavior of the function $m_{\alpha\beta}$ and $\xiloc\sim(a\xinloc)^{1/2}$, the homogeneous limit, $(\mu\mu_\plus,\mu\mu_\zero,\mu\mu_\minus)$ with $\mu\rightarrow 0$, yields
\begin{equation}\label{eq:denominator scaling}
\int_{\Omega_\xi}\dd{\vbg}\dd{\vbf}\bar{\Phi}_{\alpha}\bar{\Phi}_\beta \:\sim\:(a\xinloc)^{\frac{\dloc}{2}}\left(\frac{V_\xi}{V_L}\right)^4 V_L^{-2\frac{n_\gamma-1}{n_\gamma-2}}\lambda^{-\frac{2}{n_\gamma-2}}\mu^{\frac{2}{n_\gamma-2}}.
\end{equation}
Since $m_{\alpha\beta}$ is a non-zero function, we neglect its matrix structure which would simply yield different constant proportionality factors for the denominator. 

The numerator of $Q_{\alpha\beta}$, given in Eq.~\eqref{eq:definition Q}, includes the extended correlation function that contains the zero mode expansion. Since the parameters $Z_\alpha^\phi(\vbi)$ are chosen such that the local correlation function exhibits an exponential suppression (see Section~\ref{sec:Local correlation function}), the local integration domain can be extended to all of $\R^{\dloc}$. This yields
\begin{equation}
\int_{\SL_\xi^4}\dd{\vbg}C_{\alpha\beta}^{\mathrm{ext}}(\vbg) = \sum_{s=s_0}^4 \left(\frac{V_\xi}{V_L}\right)^s\sum_{(c_1\dots c_s)} \int\dd{\vb*{g}_{4-s}}C_{\alpha\beta}^s(\vbg_{4-s}),
\end{equation}
where we notice that the contributions from the timelike faces with labels $(0,\nu)$ vanish due to the projection onto the trivial representation $(i,0)$. This is another indication that timelike faces do not affect correlations near criticality. From the remaining contributions, we obtain
\begin{equation}\label{eq:numerator scaling}
\int\dd{\vb*{g}_{4-s}}C_{\alpha\beta}^s(\vbg_{4-s})\: \sim \:\sum_m \frac{\varrho^m_{\alpha\beta}}{b^m_{c_1\dots c_s}}\:\longrightarrow\: \frac{\varrho^{m_*}_{\alpha\beta}}{b_{c_1\dots c_s}^{m_*}},    
\end{equation}
where the contribution with the smallest effective mass $b_{c_1\dots c_s}^{m_*}$ dominates. Depending on the combinatorics governing the interaction of the model, we notice that this term is possibly vanishing and thus regulated as prescribed in Eq.~\eqref{eq:b regularization}. We keep the constants $\varrho^m_{\alpha\beta}$ as they encode the matrix structure of the correlation function, capturing also vanishing entries as explained above. 

Combining numerator and denominator, given in Eqs.~\eqref{eq:denominator scaling} and~\eqref{eq:numerator scaling}, respectively, the Ginzburg-$Q$ scales as
\begin{equation}
Q_{\alpha\beta}\:\sim\: \varrho^{m_*}_{\alpha\beta}\lambda^{\frac{2}{n_\gamma-2}}V_L^{2\frac{n_\gamma-1}{n_\gamma-2}}(a\xinloc)^{-\frac{\dloc}{2}+\frac{n_\gamma}{n_\gamma-2}}\sum_{s=s_0}^4\left(\frac{V_\xi}{V_L}\right)^{-(4-s)}.
\end{equation}
Taking the limit of large group volume first, $L\rightarrow \infty$, we notice that the $s_0$-term of the zero mode sum dominates. After absorbing volume factors and the skirt radius into the coupling,
\begin{equation}\label{eq:coupling redefinition}
\bar{\lambda} = V_L^{2(n_\gamma-1)+(4-s_0)(n_\gamma-2)} a^{\frac{n_\gamma}{2}-\frac{\dloc(n_\gamma-2)}{4}}\lambda,
\end{equation}
which is consistent with the results of Ref.~\cite{Marchetti:2022igl}, the Ginzburg-$Q$ is finally given by
\begin{equation}\label{eq:Q arbitrary interaction}
Q_{\alpha\beta}\sim\varrho^{m_*}_{\alpha\beta}\bar{\lambda}^{\frac{2}{n_\gamma-2}}(\xinloc)^{-\frac{\dloc}{2}+\frac{n_\gamma}{n_\gamma-2}}\e^{-2(4-s_0)\frac{\xinloc}{a}}.
\end{equation}

In the homogeneous limit $(\mu\mu_\plus,\mu\mu_\zero,\mu\mu_\minus)$ with $\mu\rightarrow 0$, for which the non-local correlation length diverges, $\xinloc\rightarrow \infty$, the non-zero entries of $Q_{\alpha\beta}$ constantly approach zero. This proves the validity of the mean-field approach. Remarkably, the general result of the exponential suppression holds irrespective of the details of the underlying interactions which only affect its overall strength via the number $s_0$. We highlight that the matricial factor $\varrho_{\alpha\beta}^{m^*}$ encodes the interplay of combinatorial non-locality with the causal character of the tetrahedra, i.e. it encodes whether and how tetrahedra of different signatures are correlated.

The determining factor for the asymptotic behavior of $Q_{\alpha\beta}$ is the suppressing exponential function that arises from the hyperbolic structure of the Lorentz group $\SL$ which is responsible for the Lorentz boosts. The same result has been found previously~\cite{Marchetti:2022igl,Marchetti:2022nrf}, where all tetrahedra were assumed to be spacelike. Our results go far beyond this restricted setting as we include tetrahedra of arbitrary signature (spacelike as well as timelike faces). As a secondary result, we find that timelike faces do not contribute to the critical behavior of the theory as they are characterized by the rotational -- and thus compact -- part of the Lorentz group. This is apparent from the non-local correlation function discussed in Section~\ref{sec:Non-local correlation function}. In the same vein, contributions to the Ginzburg-$Q$ emanating from timelike faces vanish as a result of the projection onto zero modes. 

Furthermore, we would like to add that the possibility of a vanishing mass has been discussed at length in Ref.~\cite{Dekhil:2024ssa}. Transferring these results to the present setting, we notice above all, that the asymptotic behavior of $Q_{\alpha\beta}$ does not change if the $b_{c_1\dots c_s}^m$ vanish. This can be seen by explicitly using the $\epsilon$-regularization suggested in Eq.~\eqref{eq:b regularization}. Furthermore, due to the presence of three signatures, there is in fact a set of three effective mass parameters $\{b_{c_1\dots c_s}^m\}$, not all of which are vanishing for the combinatorics we considered here. 

Finally, we note that in the limit of infinite skirt radius $a$, i.e. the flat limit,  the $3$-hyperboloid is turned into $\mathbb{R}^3$. Applying this limit right before computing the integrals in $Q_{\alpha\beta}$ would yield results perfectly consistent with those obtained in~\cite{Marchetti:2022igl} and~\cite{Marchetti:2020xvf}.

\paragraph{Multiple interactions.} The analysis of the present section relied on the restriction to a single interaction term. Including a sum of interactions leads, in the most general case, to analytically unsolvable mean-field equations therefore obscuring any further investigation. In Section~\ref{sec:CDT-like model}, we consider two interactions in the analysis of a CDT-like model which is still analytically feasible. However, the most straightforward generalization consists of interactions of the same number of spacelike, lightlike, and timelike tetrahedra, $(\np,\nz,\nm)$, but with different combinatorics $\gamma$. These will lead to the same mean-field equations. We briefly sketch in the following how the results of this section are generalized to this case.

Multiple interactions of the same degree $(\np,\nz,\nm)$ are included by modifying the interaction term in Eq.~\eqref{eq:single interaction} to
\begin{equation}
V[\Phi_\alpha] = \sum_\gamma \lambda_\gamma\int \dd{\vb*{X}}\dd{\vbf}\Tr_\gamma\left[\Phi_\plus^{\np}\Phi_\zero^{\nz}\Phi_\minus^{\nm}\right].
\end{equation}
Solutions of the mean-field equations are obtained by replacing $\lambda$ in Eq.~\eqref{eq:mf solutions} with $\sum_\gamma\lambda_\gamma$ and the Hessian matrix $F_{\alpha\beta}$ in Eq.~\eqref{eq:Hessian matrix} is obtained by replacing every entry $\chi_{\alpha\beta}$ with $\sum_\gamma\tilde{\lambda}_\gamma\chi_{\alpha\beta}^\gamma$, where $\tilde{\lambda}_\gamma$ is given by
\begin{equation}
\tilde{\lambda}_\gamma = \frac{\lambda_\gamma}{\sum_{\gamma'}\lambda_{\gamma'}}.
\end{equation}
These modifications affect the pole structure of the correlation functions, that is whether the effective masses vanish and which parameter ranges of $Z_\alpha^\phi(\vbi)$ and $Z_\alpha^g(\vb*{0})$ lead to positive effective masses. Irrespective of these differences, the non-vanishing components of $Q_{\alpha\beta}$ are still subject to exponential suppression close to criticality, and thus, the mean-field approximation is valid also for multiple interactions of the same degree.

\paragraph{O(3)-invariance in the space of causal characters.} We close this section by commenting on the idea of defining a theory that is $\mathrm{O}(3)$-invariant with respect to the causal character index $\alpha$.  A textbook example of multi-field theories is given by the so-called vector models where a collection of real-valued scalar fields $\{\phi_i\}_{i\in\{1,\dots,N\}}$ on the domain $\mathcal{M}$ are considered. Such a collection can be summarized as a multiplet of fields, forming a vector $\vec{\phi}\in\R^N$. An action that only depends on the Euclidean vector product $\vec{\phi}\cdot\vec{\phi}$ then exhibits a continuous $\mathrm{O}(N)$-symmetry with respect to the internal indices $i\in\{1,\dots, N\}$. For such a system, one can study the second-order phase transition arising from the spontaneous breaking of the $\mathrm{O}(N)$-symmetry. What is characteristic of the breaking of continuous symmetries is that one finds a massive radial mode $|\vec{\phi}|$ and $N-1$ massless Goldstone modes that are associated with the angular variables.  Furthermore, the vacuum of the theory is $\mathrm{O}(N)$-degenerate~\cite{WipfRG}. 

It is tempting to proceed in a similar fashion with the causal Barrett-Crane model and interpret the fields $\Phi_\alpha$ as a triplet $\vec{\Phi}$ that transforms under $\mathrm{O}(3)$ acting on the causal character indices $\alpha$. This would allow us to study the continuous breaking of symmetries in the context of a causal TGFT. However, there exist multiple obstructions that render this direction unfruitful. 

First, the simplicity constraints in Eq.~\eqref{eq:geom_constraint} imply that the fields $\Phi_\alpha$ live on distinct domains, being the different homogeneous spaces of $\SL$. This is already in contrast to local field theories where the fields $\phi_i$ are all defined on the same domain. Second the spin representation of the field $\Phi_\minus$ comes with spacelike components, $(\rho,0)$ and timelike components, $(0,\nu)$, spoiling a straightforward definition of $\mathrm{O}(3)$ transformations. One could still define such transformations to only act on field coefficients with spacelike labels. However, these first two obstructions together imply that one cannot write down an $\mathrm{O}(3)$-invariant action. This is because the vertex amplitudes crucially depend on the distributions of causal characters to the tetrahedra. Consider for instance two signatures $\alpha\in\{+,-\}$ and the interaction in spin representation
\begin{equation}
V =\Tr_\gamma\left[a\Phi_\plus^4+2b\Phi_\plus^2\Phi_\minus^2+c\Phi_\minus^4\right],
\end{equation}
for a given vertex graph $\gamma$. This interaction is $\mathrm{O}(2)$-invariant only if $a=b=c$. However, in spin representation, these three different vertices are weighted by different vertex amplitudes which depend on the choice of signature~\cite{Jercher:2022mky}. We therefore conclude that one cannot define a $\mathrm{O}(3)$-invariant action. This is to be expected from a physical point of view since causally distinct building blocks should have different quantum geometric weights associated with them.

\section{Extending the theory space}\label{sec:other interactions}

Apart from the brief discussion just given, the interactions studied in the previous section focused on the analysis of a variety of models with only one interaction term. However, inspiration from other physically intriguing models of quantum gravity tells us that one could consider more than one type of spacetime building block. One such example is CDT, which can be translated to a CDT-like TGFT model~\cite{Jercher:2022mky} defined by two interaction terms of different degrees. We study the basic phase structure of this model with the Landau-Ginzburg method in Section~\ref{sec:CDT-like model}. Another application of the formalism developed in the previous section is to consider a multi-field theory where instead of a causal character, the index of the field is associated with a so-called color degree of freedom~\cite{Gurau:2010nd}. This allows us to study a colored simplicial model which is closely related to colored tensor models~\cite{Gurau:2011xp,Gurau:2016cjo} and tensor-invariant interactions~\cite{Bonzom:2012hw,Gurau:2011tj}. We study the critical behavior of this model with the mean-field machinery in Section~\ref{sec:colored model}.

\subsection{Simplicial CDT-like interactions}\label{sec:CDT-like model}

\begin{figure}
    \centering
    \includegraphics[width=0.8\textwidth]{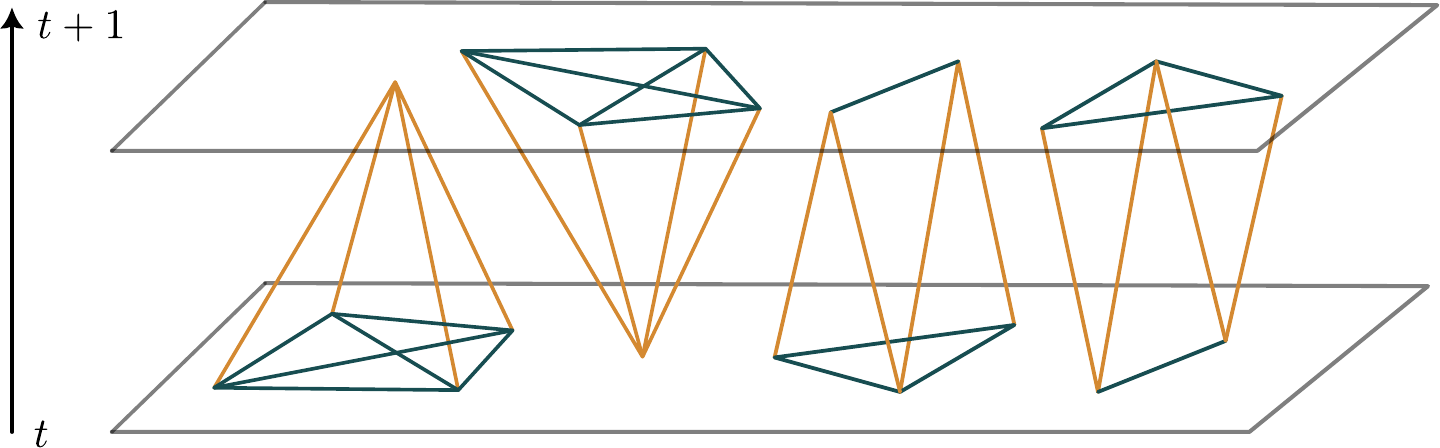}
    \caption{In CDT a triangulation is built from two fundamental building blocks and their time-reflection, i.e. $(4,1)$ or $(1,4)$-simplices (left) and $(3,2)$- or $(2,3)$-simplices (right).  These interpolate between two consecutive spacelike hypersurfaces at integer times $t$ and $t+1$. The numbers in brackets indicate the number of vertices in the triangulations of the respective constant-time slices. In this sense, the tetrahedron which forms the base of the $(4,1)$-simplex lies entirely in the $t$-hypersurface and thus is spacelike, indicated by teal blue edges. Timelike edges are colored in orange. Consequently, all the other tetrahedra are timelike, which can be mimicked by the TGFT interaction in Eq.~\eqref{eq:CDT interaction}. See also~\cite{Ambjorn:2012jv,Loll:2019rdj} for further details.}
    \label{fig:CDT}
\end{figure}

Causal dynamical triangulations~\cite{Ambjorn:2001cv,Ambjorn:2012jv,Loll:2019rdj} is a quantum gravity path integral approach based on the Regge action with fixed spacelike and timelike edge length and a varying triangulation. The configurations that are being summed over consist of four $4$-simplices denoted as the $(4,1)$, $(1,4)$, $(3,2)$ and $(2,3)$-simplices, depicted in Fig.~\ref{fig:CDT}. The restriction to this small subset of all possible causal simplicial configurations implements the so-called \textit{foliation constraint} which ensures that the gluing of these building blocks results in a discrete spacetime foliated into spacelike hypersurfaces.\footnote{An alleviation of this constraint has been studied in two and three dimensions in~\cite{Jordan:2013awa,jordan2013globally}, called locally causal dynamical triangulations, showing results which are similar to the more restrictive case. An extension of this work to four dimensions has not been conducted yet.} Despite the conceptual and mathematical differences between CDT and TGFTs, which have been discussed in~\cite{Jercher:2022mky}, we perform in this section a mean-field analysis of a CDT-like TGFT model.

To that end, we consider the complete Barrett-Crane TGFT model with spacelike and timelike tetrahedra. The overall interaction consists of two terms, one for the $(4,1)$ and one for the $(3,2)$-type simplices,
\begin{equation}\label{eq:CDT interaction}
V[\Phi_\plus,\Phi_\minus]
=
\int\dd{\vb*{X}}\dd{\vbf}\left(\lambda_{(4,1)}\Tr_\gamma\left[\Phi_\plus\Phi_\minus^4\right]+\lambda_{(3,2)}\Tr_\gamma\left[\Phi_{\minus}^5\right]\right),
\end{equation}
where $\gamma$ is the vertex graph corresponding to simplicial interactions. Notice that this captures also the remaining two simplices $(1,4)$ and $(2,3)$ related by time reversal since the TGFT model includes a sum over spacetime orientation~\cite{Livine:2002rh,Jercher:2022mky}. The kinetic term considered in this section is the one given in Eq.~\eqref{eq:kinetic kernel}, where $\alpha\in\{+,-\}$ in this section. Consequently, kinetic kernels and correlation functions are $2\times 2$-matrices. Clearly, the theory does not exhibit a $\mathbb{Z}_2$ or $\mathbb{Z}_2\times\mathbb{Z}_2$-symmetry. Still, we can apply the Landau-Ginzburg method and study if it provides a valid description of the phase transition towards a non-trivial condensate state in the mean-field background picture. 

The equations of motion at the mean-field level are given by the following set of coupled polynomial equations
\begin{align}
0&=\mu_\plus\bar{\Phi}_{\plus}+\lambda_{(4,1)}\bar{\Phi}_\minus^4V_L^{10},\\[7pt]
0&=\mu_\minus\bar{\Phi}_\minus+4\lambda_{(4,1)}\bar{\Phi}_{\plus}\bar{\Phi}_{\minus}^3V_L^{10}+5\lambda_{(3,2)}\bar{\Phi}_{\minus}^4V_L^{10}.
\end{align}
For $\mu_\plus,\mu_\minus>0$, the solutions are given by $\bar{\Phi}_\plus = \bar{\Phi}_\minus = 0$.  If both mass parameters are turned negative, the point $(\Phi_\plus,\Phi_\minus) = (0,0)$ becomes a local maximum of the mean-field action and the non-zero solutions are given by
\begin{align}
\bar{\Phi}_\plus &= -q\frac{\left(5\mu_\plus\lambda_{(3,2)}\pm\sqrt{\mu_\plus(16\mu_\minus\lambda_{(4,1)}^2+25\mu_\plus\lambda_{(3,2)}^2)}\right)^{\frac{4}{3}}}{16\mu_\plus\lambda_{(4,1)}^{5/3}V_L^{10/3}},\label{eq:CDTmfsol1}\\[7pt]
\bar{\Phi}_\minus &=
\frac{q}{2}\frac{\left(5\mu_\plus\lambda_{(3,2)}\pm\sqrt{\mu_\plus(16\mu_\minus\lambda_{(4,1)}^2+25\mu_\plus\lambda_{(3,2)}^2)}\right)^{\frac{1}{3}}}{\lambda_{(4,1)}^{1/3}V_L^{10/3}},\label{eq:CDTmfsol2}
\end{align}
where $q$ is a third root of unity. This involved form of the mean-field background solution is a direct consequence of the sum of two interaction terms with different powers of fields. Still, one can extract that $\bar{\Phi}_\alpha\sim V_L^{-10/3}\mu^{1/3}\lambda^{-1/3}$ if the masses and the couplings are homogeneously re-scaled as $(\mu\mu_\plus,\mu\mu_\minus)$ and $(\lambda\lambda_{(4,1)},\lambda\lambda_{(3,2)})$, respectively. We will exploit this scaling behavior to give a compact form of the Ginzburg-$Q$, computed below.

Performing the linearization $\Phi_\alpha(\vbg,\vbf,X_\alpha) = \bar{\Phi}_\alpha+\delta\Phi(\vbg,\vbf,X_\alpha)$, the effective kinetic kernel $G_{\alpha\beta}$ takes the form
\begin{equation}
G_{\alpha\beta}(\vbg,\vbf;\vbg'\vbf') = \delta_{\alpha\beta}\mathcal{K}_{\alpha}(\vbg,\vbf;\vbg'\vbf')+V_LF_{\alpha\beta}(\vbg,\vbf;\vbg',\vbf'),
\end{equation}
where the Hessian yields
\begin{equation}
F_{\alpha\beta} = \delta(\vbf-\vbf')
\begin{pmatrix}
0 \;& \bar{\Phi}_{\minus}^3\chi^{(4,1)}_{\plus\minus}\\[7pt]
\bar{\Phi}_{\minus}^3\chi^{(4,1)}_{\plus\minus}\; & \bar{\Phi}_\plus\bar{\Phi}_{\minus}^2\chi^{(4,1)}_{\minus\minus}+\bar{\Phi}_{\minus}^4\chi^{(3,2)}
\end{pmatrix}.
\end{equation}
The bi-local functions $\chi^{(4,1)}_{\alpha\beta}$ and $\chi^{(3,2)}$ are defined as
\begin{equation}
\chi^{(3,2)} = \chi_{\mathrm{simpl}},\qquad 
\chi^{(4,1)}_{\alpha\beta} =  
\begin{pmatrix}
0\; & \frac{1}{5}\chi_{\mathrm{simpl}}\\[7pt]
\frac{1}{5}\chi_{\mathrm{simpl}} & \frac{3}{5}\chi_{\mathrm{simpl}}
\end{pmatrix},
\end{equation}
with $\chi_{\mathrm{simpl}}(\vbg,\vbg')$ for the simplicial interactions is defined as 
\begin{equation}
    \chi_{\mathrm{simpl}} = 5 V_L^{-3}\sum_{c=1}^4\delta(g_c,g_c').
\end{equation} 
In contrast to a single interaction, inserting the mean-field solutions into the Hessian matrix $F_{\alpha\beta}$ does not simply yield a factor of $-\mu$. Instead, $F_{\alpha\beta}$ is an involved function of the two mass parameters $\mu_\alpha$ and the two couplings $\lambda_{(4,1)}$ and $\lambda_{(3,2)}$. 

In the next two paragraphs, we compute the correlation function $C_{\alpha\beta}$ as the inverse of the effective kinetic kernel. Thereafter, we compute the Ginzburg-$Q$ and study its behavior in the limit $\mu\rightarrow 0$.

\paragraph{Local correlation function.} We remind the reader that the local correlation function does not contain contributions from timelike faces with labels $\nu$ as a result of projecting onto four zero modes. Following the steps of Section~\ref{sec:Local correlation function}, the local correlation function in momentum space can be written as 
\begin{equation}
C_{\alpha\beta}^{\vbi}(\vbk) = \sum_{m=1}^2\frac{\varsigma^m_{\alpha\beta}}{\vbk^2+b_{\vbi}^m}.
\end{equation}
The effective masses  $b_{\vbi}^m$ explicitly depend on the functions $Z^\phi_\alpha(\vbi)$ and the two interaction couplings $\lambda_{(4,1)},\lambda_{(3,2)}$. Furthermore, the effective masses depend quantitatively (though not qualitatively) on the ``$\pm$'' choice of the vacuum in Eqs.~\eqref{eq:CDTmfsol1} and~\eqref{eq:CDTmfsol2}. This is an immediate consequence of the more involved interactions of different degrees in the fields. The $\varsigma^m_{\alpha\beta}$ are non-zero constant coefficients arising from the partial fraction decomposition. 

As discussed in Section~\ref{sec:Local correlation function}, the sign of $b^m_{\vbi}$ determines whether the correlations exhibit an exponential decay or long-ranging oscillations. If $Z^\phi_\plus$ and $Z^\phi_{\minus}$ are both positive or both negative, then one of the effective masses is negative while the other is positive. This holds for any choice of $\lambda_{(4,1)},\lambda_{(3,2)}$. Moreover, there exists a range of $Z^\phi_\plus,Z^\phi_{\minus}$ with opposite signs such that both effective masses are positive. This case can be obtained only if the couplings have opposite signs. For studying the critical behavior of the model in the limit $\mu_\alpha\rightarrow 0$, we restrict to the latter range of parameters. Then, the local correlation function exhibits an exponential decay
\begin{equation}
C_{\alpha\beta}(r)\:\underset{r\gg 1}{\longrightarrow}\: \e^{-r/\xiloc},
\end{equation}
where we identified $\xiloc^{-2} \sim b^{m_*}_{\vbi}$ with $m_*=\arg\min_m b^m_{\vbi}$. In a homogeneous limit $(\mu\mu_\plus,\mu\mu_{\minus})$ with $\mu\rightarrow 0$, the correlation length diverges as a square root.

\paragraph{Non-local correlation function.} As a direct consequence of the interactions being of simplicial form, we identify $s_0=3$ as the lower bound of zero modes, below which the Hessian vanishes. As explained in Section~\ref{sec:Non-local correlation function}, we do not consider zero modes $s<s_0$ as they generically lead to exponentially diverging correlation functions on $\SL$ independent of the parameter $\mu$. Consequently, we focus on the case of $s=3$ zero modes in this paragraph. Furthermore, we use the result of Section~\ref{sec:Non-local correlation function} where we have shown that the timelike faces do not contribute to the critical behavior. Thus, in the following, we only consider components with all faces being spacelike. 

The correlation function for all faces being spacelike is computed as detailed in Section~\ref{sec:Non-local correlation function}. In particular, the residual correlation function $C^{s=3,\rho_c}_{\alpha\beta}$ evaluated on $3$ zero modes can be decomposed via a partial fraction decomposition,
\begin{equation}
C_{\alpha\beta}^{s=3,\rho_c} = \sum_{m=1}^2\frac{\varrho^m_{\alpha\beta}}{\frac{1}{a^2}(\rho_c^2+1)+b^{m}_{c_1\dots c_s}}.
\end{equation}
The coefficients $\varrho^m_{\alpha\beta}$ are constants all of which are non-zero. This follows directly from the CDT-like interaction in Eq.~\eqref{eq:CDT interaction} which correlates spacelike and timelike tetrahedra. One of the effective masses vanishes which, as argued in~\cite{Dekhil:2024ssa}, arises generically for combinatorially non-local interactions. We adopt an $\epsilon$-regularization for this case, just as in Section~\ref{sec:Contributions with only spacelike faces}. The other effective mass $b_{c_1\dots c_s}$ is a homogeneous function of the $\mu_\alpha$ and depends on the parameters $Z^g_{\alpha}(\vb*{0})$ as well as the couplings $\lambda_{(4,1)},\lambda_{(3,2)}$. In the previous paragraph, we restricted the couplings to be of opposite sign. In this case, we identify numerically a range of values for the parameter $Z^g_\alpha$ for which the effective mass is positive. In this range, the $Z^g_\alpha$ are either both negative or of opposite sign. 

In the regime of positive effective mass,  the non-local correlation function (re-scaled by the Jacobian determinant on the residual domain) exhibits an exponential decay in the isotropic limit of $\eta_c\equiv \eta\rightarrow \infty$. We extract a non-local correlation length that scales in a homogeneous limit of $(\mu_\plus,\mu_\minus)\rightarrow(0,0)$ as $\xinloc\sim(a b)^{-1}$. Here $b$ is the smallest effective mass. Clearly, the non-local correlation length diverges as $\mu\rightarrow 0$. 

\paragraph{Ginzburg-$\mathbf{Q}$.} Following the procedure detailed in Section~\ref{sec:arbitrary but fixed Q}, we compute the Ginzburg-$Q$ as in Eq.~\eqref{eq:definition Q}. To that end, the integration domain is defined by the local and non-local correlation lengths determined in the previous paragraphs. The denominator of $Q_{\alpha\beta}$ is determined by the empty group- and $\R$-integrations as well as the mean-field background solutions. To extract the scaling of $Q$ in the homogeneous mass limit, we notice that the mean-field background solutions scale as
\begin{equation}
\bar{\Phi}_\alpha \rightarrow \mu^{\frac{1}{3}}V_L^{-\frac{10}{3}}f_\alpha(\mu_\plus,\mu_\minus,\lambda_{(4,1)},\lambda_{(3,2)}),
\end{equation}
in the limit $(\mu\mu_\plus,\mu\mu_\minus)$ with $\mu \rightarrow 0$. The functions $f_\alpha$ are residual functions that carry a non-trivial but homogeneous dependence on the couplings $\lambda_{(4,1)},\lambda_{(3,2)}$, meaning that in a homogeneous re-scaling of the couplings, $(\lambda\lambda_{(4,1)},\lambda\lambda_{(3,2)})$, we have $f_\alpha\sim \lambda^{-\frac{1}{3}}$. Together with the empty group- and $\R$-integrations on the domain $\Omega_\xi$, we find that the denominator of $Q$ scales as
\begin{equation}
\int_{\Omega_\xi}\dd{\vbg}\dd{\vbf}\bar{\Phi}_\alpha\bar{\Phi}_\beta\sim (a\xinloc)^{\dloc}\left(\frac{V_\xi}{V_L}\right)^4V_L^{-2\frac{4}{3}}\lambda^{-\frac{2}{3}}\mu^{\frac{2}{3}}.
\end{equation}
Notice the structural similarity to the denominator of $Q$ for a single arbitrary interaction in Eq.~\eqref{eq:denominator scaling} if $n_\gamma = 5$ is inserted. 

Evaluation of the numerator of $Q$ yields
\begin{equation}
\int_{\SL^4_\xi}\dd{\vbg}C_{\alpha\beta}^\mathrm{ext}(\vbg) = \sum_{s=s_0}^4\left(\frac{V_\xi}{V_L}\right)^s\frac{\varrho^{m_*}_{\alpha\beta}}{b^{m_*}_{c_1\dots c_s}},
\end{equation}
where the contributions of timelike faces to the non-local correlation function are integrated out by the projection onto zero modes. $b_{c_1\dots c_s}^{m_*}$ being the dominant effective mass contribution and $\varrho^{m_*}_{\alpha\beta}$ coefficients obtained in the partial fraction decomposition of the non-local correlation function. As a result, the Ginzburg-$Q$ is given by
\begin{equation}
Q_{\alpha\beta}\sim \varrho_{\alpha\beta}^{m_*}\lambda^{\frac{2}{3}}V_L^{2\frac{4}{3}}(a\xinloc)^{-\frac{\dloc}{2}+\frac{5}{3}}\sum_{s=s_0}^4\left(\frac{V_\xi}{V_L}\right)^{-4-s}. 
\end{equation} 
Taking the limit $L\rightarrow \infty$, the contribution with $s = s_0 = 3$ of the above sum dominates. If one redefines the scaling parameter of the couplings $\lambda$ as previously done in Eq.~\eqref{eq:coupling redefinition} for $n_\gamma =5$, the final form of $Q$ is given by
\begin{equation}
Q_{\alpha\beta}\sim\varrho^{m_*}_{\alpha\beta}\bar{\lambda}^{\frac{2}{3}}(\xinloc)^{-\frac{\dloc}{2}+\frac{5}{3}}\e^{-2\frac{\xinloc}{a}}.
\end{equation}

We observe an exponential suppression of $Q$ in the limit of $\xinloc\rightarrow \infty$ which is induced by the hyperbolic structure of $\SL$. Furthermore, we find that $Q_{\alpha\beta}$ scales as in the case of a fixed but arbitrary interaction with $n_\gamma = 5$, see Eq.~\eqref{eq:Q arbitrary interaction} for direct comparison. The main difference to the model discussed in Section~\ref{sec:Fixed but arbitrary interaction} is the intricate dependence of the mean-field on the mass parameters and the two couplings $\lambda_{(4,1)}$ and $\lambda_{(3,2)}$. However, in the homogeneous limit of mass and coupling, $Q$ takes the same form as for a single simplicial interaction. 

Also for a CDT-like model with two types of simplicial interactions, we find self-consistency of the mean-field approach, thus proving it to be a valid approximation of the theory close to $\mu_\plus = \mu_\minus = 0$. While the model does not exhibit a $\mathbb{Z}_2$-symmetry and the approximation method here does not describe a phase transition due to spontaneous symmetry breaking, the model still undergoes a phase transition to a non-perturbative vacuum state. We leave for future work, potentially involving arguments beyond mean-field theory, which type of phase transition this represents.  

\subsection{Colored simplicial model}\label{sec:colored model}

It has been shown in~\cite{DePietri:2000ii,Gurau:2010nd,Gurau:2010mhz} that the Feynman graphs of simplicial group field theories are in general not dual to simplicial pseudo-manifolds but exhibit topological singularities. To remedy this behavior, a coloring of the group fields has been suggested, ensuring indeed that the generated graphs are dual to pseudo-manifolds~\cite{Gurau:2009tw,Gurau:2011xp,GurauBook}. In this section, we extend the Landau-Ginzburg analysis of the spacelike Barrett-Crane model, first conducted in~\cite{Marchetti:2022igl}, to the case of colored simplicial interactions. The methods developed in the previous sections prove helpful in this endeavor since the additional color index $i$ of the field will introduce similar structures as the index $\alpha$ marking the causal character of the normal vector. In particular, the correlation function as well as the Ginzburg-$Q$ will carry a pair of color indices. 

A group field theory for four spacetime dimensions is colored by considering not only one but five group fields, labelled by an index $i$, i.e.
\begin{equation}
\Phi(\vbg,\vbf,X_\plus)\longrightarrow\Phi_i(\vbg,\vbf,X_\plus),
\end{equation}
with $i\in\{1,\dots ,5\}$. The kinetic term encodes the gluing of tetrahedra carrying the same color, expressed at the level of the fields as
\begin{equation}
K[\Phi_i] = \frac{1}{2}\sum_{i=1}^5\int\dd{\vbg}\dd{\vbg'}\dd{\vbf}\dd{\vbf'}\dd{X_\plus}\Phi_i(\vbg,\vbf,X_\plus)\mathcal{K}(\vbg,\vbf;\vbg'\vbf')\Phi_i(\vbg,\vbf,X_\plus),
\end{equation}
with kinetic kernel
\begin{equation}
\mathcal{K}(\vbg,\vbf;\vbg'\vbf') = \mu\delta(\vbg^{-1}\vbg')-Z^\phi(\vbg^{-1}\vbg')\Delta_\phi-Z^g(\abs{\vbf-\vbf'})\sum_c\Delta_c.
\end{equation}
Notice in particular that we associate with every color the same mass and the same functions $Z^\phi$ and $Z^g$, i.e. $\mathcal{K}$ does not depend on the color $i$. The interaction term of the colored simplicial model is given by
\begin{equation}
\begin{aligned}
&V[\Phi] = \lambda\int\dd[10]{g}\dd{\vbf}\dd[5]{X}\Phi_1(g_{12},g_{13},g_{14},g_{15},\vbf,X_1)\Phi_5(g_{15},g_{25},g_{35},g_{45},\vbf,X_5)\\[7pt]
\times &\Phi_4(g_{45},g_{14},g_{24},g_{34},\vbf,X_4)\Phi_3(g_{34},g_{35},g_{13},g_{23},\vbf,X_3)\Phi_2(g_{23},g_{24},g_{25},g_{12},\vbf,X_2),
\end{aligned}
\end{equation}
such that the five tetrahedra that make up a $4$-simplex are all of a different color. 

The action exhibits a set of different $\mathbb{Z}_2$-based symmetries. More precisely, it is invariant under such a symmetry if either two fields or four fields are simultaneously reflected. Choosing two, respectively four fields out of five yields ten, respectively five possible transformations. Together with the identity map, there are in total $16$ transformations. The presence of symmetries in this model implies in particular, that the transition from $\mu> 0$ to $\mu< 0$ will result in its breaking and this can be explored via the Landau-Ginzburg method. We note here that while the action clearly does not possess a simultaneous $\mathbb{Z}_2$-symmetry of all the five fields, the mean-field approach may still be used, as above, to study the transition towards a non-trivial vacuum.

Varying the action with respect to the field $\Phi_i(\vbg,\vbf,X_\plus)$ and evaluating the resulting equations on constant field configurations, $\bar{\Phi}_i = \mathrm{const}$, one obtains the mean-field background equations given by
\begin{equation}
\mu\bar{\Phi}_i+\lambda V_L^{10}\prod_{j\neq i}\bar{\Phi}_j = 0,\quad\forall i\in\{1,\dots,5\}.
\end{equation}
The non-zero solutions are given by tuples of five mean-fields $(\bar{\Phi}_i)_{i\in\{1,\dots,5\}}$ which are of the form 
\begin{equation}\label{eq:colored mf sol}
\bar{\Phi}_i = -q_i \mu^{\frac{1}{3}}\lambda^{-\frac{1}{3}}V_L^{-\frac{10}{3}},
\end{equation}
where the $q_i$'s are a third root of unity. In total, there are $48$ solutions. Given the $16$ $\mathbb{Z}_2$-transformations, there are therefore three distinct solutions from which all the other ones can be generated once the symmetry group acts upon them.

After performing a linearization, $\Phi_i(\vbg,\vbf,X_\plus) = \bar{\Phi}_i+\delta\Phi_i(\vbg,\vbf,X_\plus)$, the equations of motion are of the form
\begin{equation}
\sum_{j=1}^5\int\dd{\vbg'}\dd{\vbf'}G_{ij}(\vbg,\vbf;\vbg',\vbf)\delta\Phi_j(\vbg',\vbf'),
\end{equation}
with effective kinetic kernel $G_{ij} = \delta_{ij}\mathcal{K}+V_L F_{ij}$, defined via the Hessian $F_{ij}$. Here, the $G_{ij}$ and thus also the correlator $C_{ij}$, are $5\times 5$ matrices. The Hessian matrix $F_{ij}$ is of the form
\begin{equation}
V_L F_{ij} = -\mu\delta(\vbf-\vbf')\chi_{ij},
\end{equation}
with
\begin{equation}
\chi_{ij} = \frac{1}{4} 
\begin{pmatrix}
0 & \epsilon_{12}\prod\limits_{c\neq 1}\delta_{\rho_c,i} & \epsilon_{13}\prod\limits_{c\neq 2}\delta_{\rho_c,i} & \epsilon_{14}\prod\limits_{c\neq 3}\delta_{\rho_c,i} & \epsilon_{15}\prod\limits_{c\neq 4}\delta_{\rho_c,i}\\
\epsilon_{12}\prod\limits_{c\neq 4}\delta_{\rho_c,i} & 0 & \epsilon_{23}\prod\limits_{c\neq 1}\delta_{\rho_c,i} & \epsilon_{24}\prod\limits_{c\neq 2}\delta_{\rho_c,i} & \epsilon_{25}\prod\limits_{c\neq 3}\delta_{\rho_c,i}\\
\epsilon_{13}\prod\limits_{c\neq 3}\delta_{\rho_c,i} & \epsilon_{23}\prod\limits_{c\neq 4}\delta_{\rho_c,i} & 0 & \epsilon_{34}\prod\limits_{c\neq 1}\delta_{\rho_c,i} & \epsilon_{35}\prod\limits_{c\neq 2}\delta_{\rho_c,i}\\
\epsilon_{14}\prod\limits_{c\neq 2}\delta_{\rho_c,i} & \epsilon_{24}\prod\limits_{c\neq 3}\delta_{\rho_c,i} & \epsilon_{34}\prod\limits_{c\neq 4}\delta_{\rho_c,i} & 0 & \epsilon_{45}\prod\limits_{c\neq 1}\delta_{\rho_c,i}\\
\epsilon_{15}\prod\limits_{c\neq 1}\delta_{\rho_c,i} & \epsilon_{25}\prod\limits_{c\neq 2}\delta_{\rho_c,i} &  \epsilon_{35}\prod\limits_{c\neq 3}\delta_{\rho_c,i} &  \epsilon_{45}\prod\limits_{c\neq 4}\delta_{\rho_c,i} & 0
\end{pmatrix}
.
\end{equation}
The symbols $\epsilon_{ij}$ depend on the choice of vacuum $\bar{\Phi}_i$ with $\epsilon_{ij} = \mathrm{sgn}(\bar{\Phi}_k\bar{\Phi}_l\bar{\Phi}_m)$ for $k,l,m\neq i,j$. Since $\chi_{ij}$ explicitly depends on the chosen vacuum $\bar{\Phi}_i$ via the $\epsilon_{ij}$, also the effective kinetic kernel $G_{ij}$ will carry this dependence. We emphasize, however, that the correlation functions computed in the following paragraphs do not depend on this choice. 

\paragraph{Local correlation function.} The local correlation function $C_{ij}$ is obtained by inverting the effective kinetic kernel $G_{ij}$ and integrating out the non-local variables.  In Fourier representation, $C_{ij}(\vbk)$ can be decomposed via a partial fraction decomposition, yielding
\begin{equation}
C_{ij}^{\vbi}(\vbk) = \frac{\varsigma_{ij}^1}{\vbk^2+3\frac{\abs{\mu}}{Z^\phi(\vbi)}}+\frac{\varsigma_{ij}^2}{\vbk^2-2\frac{\abs{\mu}}{Z^\phi(\vbi)}},
\end{equation}
where the $\varsigma$ are non-zero constant coefficients. From this form of the correlation function, the two effective masses can be immediately extracted,
\begin{equation}
b^1_{\vbi} = 3\frac{\abs{\mu}}{Z^\phi(\vbi)},\qquad b^2_{\vbi}=-2\frac{\abs{\mu}}{Z^\phi(\vbi)}.
\end{equation}
Notice that this result is entirely independent of the choice of vacuum. Clearly, for any sign of $Z^\phi(\vbi)$ one effective mass is positive while the other is negative. As discussed at the end of Section~\ref{sec:Local correlation function}, the resulting correlation function is then a sum of an oscillating and an exponentially decaying term. 

This behavior is a direct consequence of considering the same kinetic kernel for every color $i$, in particular considering $Z^\phi(\vbi)$ to be independent of $i$. A similar result would have been observed in Section~\ref{sec:Local correlation function} if we had considered the restriction $Z^\phi_\alpha\equiv Z^\phi$ for all $\alpha\in\{+,0,-\}$. One possible interpretation of this behavior is that the mean-field approximation cannot be applied to the colored simplicial model in the presence of local variables $\vbf\in\R^{\dloc}$. If one however assumes the applicability of the mean-field approximation, this suggests to enlarge the parameter space of the theory, allowing for color-dependent functions $Z^\phi_i(\vbi)$. Up to this point, the interplay of colors and matter degrees of freedom has not been studied and was assumed to be decoupled. Therefore, colorizing weights of the local Laplace operator $\Delta_\phi$ constitutes a novel proposal. A related extension of the theory space has been considered in~\cite{Juliano:2024rgu}, where the couplings of tensor-invariant interactions have been assumed to depend on the respective color. We leave further exploration of the intriguing connection between colors and matter to future research. For the remainder, we consider an extension $Z^\phi\rightarrow Z^\phi_i$, where a range of parameters can be identified that leads to two positive effective masses and thus to an exponentially decaying correlation function. The resulting correlation function scales as $\xiloc^{-2}\sim \mu$. 

\paragraph{Non-local correlation function.} We compute in this section the non-local correlation function for $s=3$ zero modes. The case of $s=4$ zero modes yields the local result while $s<s_0=3$ zero modes yield exponentially diverging correlation functions. Notice that the labels of the spin representation are only given by $(\rho,0)$ since we assume that the tetrahedra are spacelike, thus carrying only spacelike faces.

For three zero modes, the residual non-local correlation function is of the form
\begin{equation}
C_{ij}^{s=3,\rho_c} = \sum_{m=1}^5\frac{\varrho^m_{ij}}{\frac{1}{a^2}(\rho_c^2+1)+b_{c_1\dots c_3}^m},
\end{equation}
where the zero modes are injected in the arguments $c'\neq c$. Here, $\varrho^m_{ij}$ are coefficients of the partial fraction decomposition which are all non-zero since all the colored tetrahedra interact with each other and are thus correlated. The $b_{c_1\dots c_s}^m$ are the effective masses that depend on $Z^g(\vb*{0})$. Notice that the effective masses are independent of the choice of the background/vacuum $\bar{\Phi}_i$. Furthermore, one of the effective masses vanishes which we observed similarly in the previous sections, and is to be expected from the arguments made in~\cite{Dekhil:2024ssa}. We proceed with the regularization proposed in~\cite{Dekhil:2024ssa} for this case.

\begin{figure}
    \centering
    \includegraphics[width=0.33\textwidth]{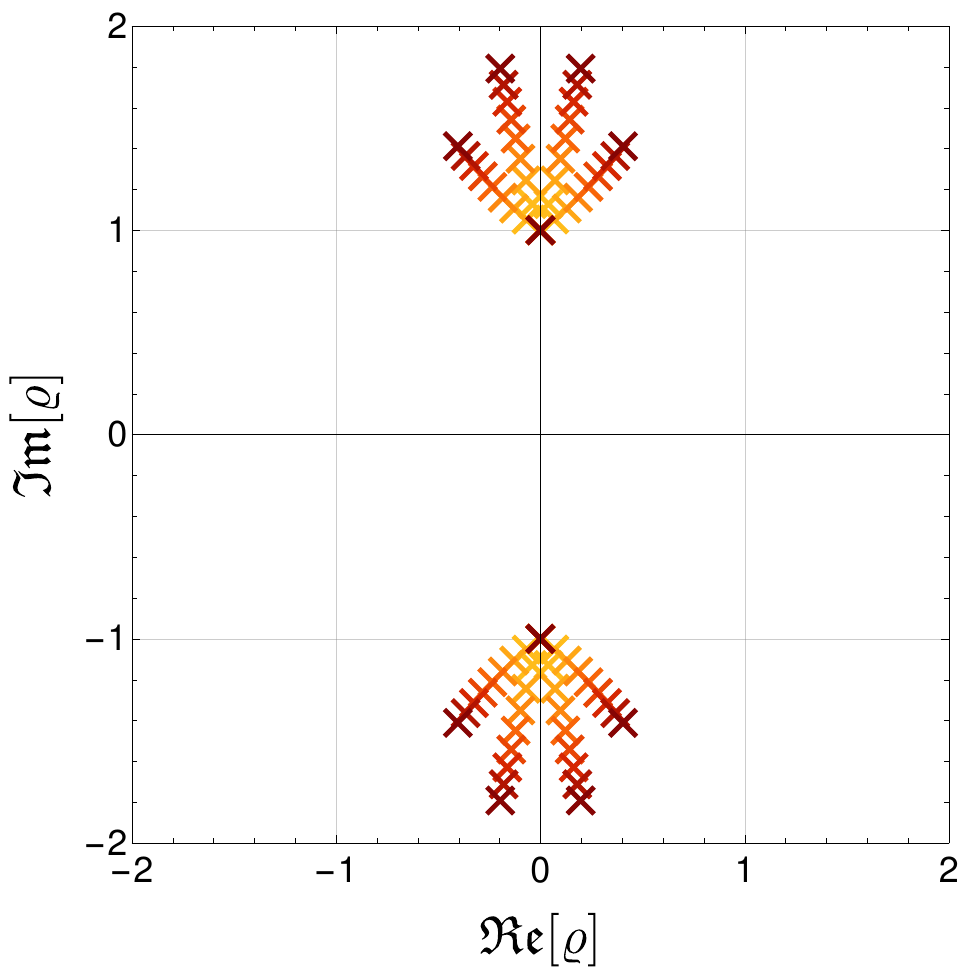}
    \caption{Poles of the non-local correlator in the complex $\rho$-plane. The parameter $\mu$ lies in the range $\abs{\mu}\in\{0.15,\dots,1.1\}$ where darker colors corresponds to larger values of $\abs{\mu}$. Here, $a=1$ and the parameter $Z^g(\vb*{0}) = -1$ were chosen for demonstrative purposes. Notice that for these choices, the imaginary part of the poles satisfies $\abs{\mathfrak{Im}{\rho\}}} > 1$ which translates to a positive real part of the effective mass, $b^m_R > 0$. As in the uncolored case~\cite{Dekhil:2024ssa}, one of the effective masses vanishes.}
    \label{fig:coloredrhopoles}
\end{figure}

A particularity of the colored model is that the poles of the non-local correlation function, $\bar{\rho} = i\sqrt{1+a^2b}$, are genuine complex numbers that do not lie on the real or imaginary axes. An exemplary plot of poles in the limit $\mu\rightarrow 0$ is shown in Fig.~\ref{fig:coloredrhopoles}. Notice that the sign of $Z^g(\vb*{0})$ determines whether the poles lie in the plane $\mathfrak{Im}\{z\}>i$ or $\mathfrak{Im}\{z\}<i$, or equivalently, if the effective mass $b$ has positive or negative real part. This has important consequences for the asymptotic behavior of the non-local correlation function as we show momentarily. 

Following the steps detailed in Section~\ref{sec:Non-local correlation function}, the re-scaled non-local correlation function on the hyperboloid exhibits an asymptotic behavior given by
\begin{equation}
\tilde{C}^{s=3}_{ij}(\eta_c)\:\underset{\eta/a\gg 1 }{\longrightarrow} \:\sum_{m=1}^5 c_m \exp\left(-\frac{1}{2}ab^{m}_{c_1\dots c_s}\eta\right).
\end{equation}
Splitting the effective mass into real and imaginary part, $b^m_{c_1\dots c_s} = b^m_R+ib^m_I$, we obtain
\begin{equation}
\tilde{C}^{s=3}_{ij}(\eta_c)\:\underset{\eta/a\gg 1 }{\longrightarrow}\: \sum_{m=1}^5 c_m \exp\left(-\frac{1}{2}ab^{m}_R\eta\right)\exp\left(-\frac{i}{2}ab^{m}_I\eta\right).
\end{equation}
The presence of an oscillatory multiplicative term is a peculiar feature of the colored simplicial model that is not present in the models of the previous sections. We furthermore emphasize that it is different from the potential oscillatory terms in the local correlation function of the previous paragraph. There, the oscillations enter as additive terms that do not fall off exponentially, thus constituting long-range correlations. In contrast, here we observe oscillating correlations that are multiplied with an exponential term, therefore still allowing for the extraction of a correlation length. More precisely, for $Z^g(\vb*{0}) < 0$ we find $b^m_R > 0$ for all $m$ and thus, the exponential term decays. Consequently, we define the correlation length as $\xinloc \sim (a b^{m_*}_R)^{-1}$, where $m_* = \arg\min_{m}(b^m_R)$. The one vanishing effective mass, which can also be observed in Fig.~\ref{fig:coloredrhopoles}, is regularized as $\epsilon\abs{\mu}$~\cite{Dekhil:2024ssa}.

\paragraph{Ginzburg-$\mathbf{Q}$.} The Ginzburg-$Q_{ij}$, with color indices $i,j\in\{1,\dots,5\}$, is obtained by adapting the computation of Section~\ref{sec:arbitrary but fixed Q} to the colored case. Inserting the mean-field background solution of Eq.~\eqref{eq:colored mf sol} into the denominator of $Q_{ij}$ yields
\begin{equation}
\int_{\Omega_\xi}\dd{\vbg}\dd{\vbf}\bar{\Phi}_i\bar{\Phi}_j \sim (a\xinloc)^{\dloc}\left(\frac{V_\xi}{V_L}\right)^4V_L^{-2\frac{4}{3}}\lambda^{-\frac{2}{3}}\mu^{\frac{2}{3}},
\end{equation}
which only depends on $i,j$ via third roots of unity $q_i,q_j$ which are negligible for the scaling of $Q_{ij}$. The numerator of $Q_{ij}$ is given by
\begin{equation}
\int_{\SL_\xi^4}\dd{\vbg}C_{ij}^{\mathrm{ext}}(\vbg) = \sum_{s=3}^4\left(\frac{V_\xi}{V_L}\right)^s\frac{\varrho^{m_*}_{ij}}{b^{m_*}_{c_1\dots c_s}}.
\end{equation}
As a result, the Ginzburg-$Q$ evaluates to
\begin{equation}
Q_{ij} \sim \varrho_{ij}^{m_*}\lambda^{\frac{2}{3}}V_L^{2\frac{4}{3}}(a\xinloc)^{-\frac{\dloc}{2}+\frac{5}{3}}\sum_{s=3}^4\left(\frac{V_\xi}{V_L}\right)^{-4-s},
\end{equation}
which in the limit $L\rightarrow \infty$ scales as
\begin{equation}
Q_{ij}\sim\varrho_{ij}^{m_*}\bar{\lambda}^{\frac{2}{3}}(\xinloc)^{-\frac{\dloc}{2}+\frac{5}{3}}\e^{-2\frac{\xinloc}{a}}.
\end{equation}
The coupling $\lambda$ has been re-scaled according to Eq.~\eqref{eq:coupling redefinition}. In the limit $\mu\rightarrow 0$, the components of the Ginzburg-$Q$ exhibit an exponential suppression induced by the boost part of the Lorentz group. Therefore, the mean-field approximation of the phase transition towards the non-perturbative vacuum proves valid in the regime of parameters that we have discussed above. The scaling of the components $Q_{ij}$ in the coupling and non-local correlation length are the same as in the previous cases of CDT-like interactions and a fixed but arbitrary simplicial interaction where the hyperbolic part of the Lorentz group leads to an exponential suppression. 

\paragraph{Connection to tensor-invariant interactions.} Colored models can in principle be uncolored by integrating out all but one field. In the context of tensor models, it has been shown in~\cite{Bonzom:2012hw,Gurau:2011tj} that uncoloring a simplicial colored model yields tensor-invariant interactions. In four (and any other even number of) dimensions, one makes the following intriguing observation: tensor-invariant interactions possess a $\mathbb{Z}_2$-symmetry, which is not immediately reflected in the theory before integrating out all but one color. It is conceivable, that this $\mathbb{Z}_2$-symmetry is a remnant of the more involved $\mathbb{Z}_2$-symmetries of the colored model discussed at the beginning of this section. A definite clarification of this point would require an extension of the analysis of~\cite{Bonzom:2012hw,Gurau:2011tj} to TGFT models defined on a non-compact domain subject to geometricity constraints, see also the remarks in Ref.~\cite{Carrozza:2016vsq}. We leave this intriguing task to future research. 

\paragraph{Causally complete colored model.} We have now studied separately the Landau-Ginzburg analysis of the causally complete BC model and the colored BC model restricted to spacelike tetrahedra. A full description would take into account a combination of these two extensions: (i) the causal properties of the quantum geometry and (ii) a colorization that yields Feynman diagrams which are bijective to $4d$ discrete manifolds. This model, which reflects both of these features, constitutes the to-date most physically compelling quantum geometric TGFT model with a Lorentzian signature. In Ref.~\cite{Jercher:2022mky} such a complete and colored model was set up, however, a further study of the interplay between the causal character and the colorization of the microscopic building blocks might be needed still. Leaving the latter point aside for now, the Landau-Ginzburg analysis can be performed along the lines presented in this work: Instead of having either a causal or a color index, the group field carries both indices $\Phi_{\alpha,i}$ such that the kinetic kernels, correlation functions, and the Ginzburg-$Q$ take the form of $15\times 15$ matrices. After a suitable identification of the parameter ranges in which the correlation function exhibits an exponential decay, we expect from previous results~\cite{Dekhil:2024ssa,Marchetti:2022igl,Marchetti:2022nrf} and the results of this work, that the qualitative scaling behavior of $Q$ is similar to what is shown here. In particular, the hyperbolic part of the Lorentz group will most likely lead to an exponential suppression in the limit $\mu\rightarrow 0$, and timelike faces will not affect the critical behavior. Thus, we are drawn to the reasonable conclusion that combining colorization and the causal completion within the same TGFT model is naturally expected to lead to the same qualitative results as above since the critical behavior is essentially determined by the hyperbolic structure of $\SL$. Hence, mean-field theory provides a trustworthy description of the phase transition towards a causally rich non-perturbative vacuum. Since such a state consists of many TGFT quanta, this result adds further support for the existence of a sensible continuum gravitational regime for realistic TGFT models for Lorentzian quantum gravity.

\section{Summary and Conclusion}\label{sec:Conclusion}

The main goal of this article was to further advance the application of Landau-Ginzburg mean-field theory to Lorentzian TGFT models for quantum geometry with minimally coupled massless and free scalar fields started in~\cite{Marchetti:2022igl,Marchetti:2022nrf,Dekhil:2024ssa}, to reveal their basic phase structure, and to study phase transitions therein. More specifically, in Ref.~\cite{Marchetti:2022igl,Marchetti:2022nrf,Dekhil:2024ssa} this method was applied to the context of the Lorentzian BC model and related models with tensor-invariant interactions. These were based on spacelike tetrahedra only~\cite{Jercher:2021bie} and coupled to such matter degrees of freedom. Note that the Barrett-Crane (BC) model aims to provide a TGFT quantization of $4d$ Lorentzian Palatini gravity in the first-order formulation. The structure of the ground states of such models and their relation to the continuum limit is still an open issue, making such an analysis highly relevant also beyond the TGFT approach to quantum gravity.

Here, we went beyond the restriction to spacelike building blocks and applied the mean-field method to the more realistic and complex causal completion of the Lorentzian BC model which does not only include spacelike tetrahedra as basic building blocks but also timelike and lightlike ones~\cite{Jercher:2022mky}. After analyzing the impact of these additional configurations on the basic phase structure in the presence of a single interaction, we extended the theory space and considered the case of a model with two simplicial interactions that are well-known from CDT. We then ventured to the more involved analysis of colored simplicial interactions. For these distinguished cases, we calculated the corresponding correlation functions as well as the correlation lengths and extracted the Ginzburg parameter which allowed us to check the self-consistency of the mean-field analysis.

The central result of our work is that also in the causally extended setting, one can find a transition towards a causally rich non-perturbative vacuum state and this phase transition can always be self-consistently accounted for by mean-field theory. Curiously, this is induced by the non-compact hyperbolic part of the Lorentz group which leads to an exponential suppression of the Ginzburg-$Q$. Consequently, the critical behavior is entirely driven by the representation labels of spacelike faces. In contrast, we demonstrate that timelike faces do not influence critical behavior as they are characterized by the rotational and thus compact part of the Lorentz group. Since such non-perturbative configurations are highly excited by TGFT quanta, our work lends further strong support for the existence of a sensible continuum gravitational approximation for such realistic TGFT models. This also concerns the closely related lattice quantum gravity and spin foam models. Moreover, our results indirectly strengthen the extraction of effective cosmological dynamics~\cite{Gielen:2013kla,Gielen:2013naa,Gielen:2016dss,Oriti:2016acw,Pithis:2019tvp,Jercher:2021bie,Oriti:2021oux,Marchetti:2021gcv} and the recently improved derivation of scalar cosmological perturbations~\cite{Jercher:2023nxa,Jercher:2023kfr} within a mean-field approximation.

The causally complete setting required us to master new challenges that were not present in previous works restricted to spacelike tetrahedra~\cite{Marchetti:2022igl,Marchetti:2022nrf,Dekhil:2024ssa}, to mention a few:
\begin{enumerate}
    \item[i.] The presence of multiple fields accounting for the spacelike, timeline, and lightlike tetrahedra with more involved types of interactions: this resulted in more involved mean-field equations and a matrix-valued kinetic kernel that complicated the computation of correlations, a derivation of which we provided in Appendix~\ref{sec:Correlation functions with geometricity constraints}. As a consequence, we found that also the Ginzburg-$Q_{\alpha\beta}$ is matrix-valued for more than one causal character.
    \item[ii.] A more elaborate representation theory due to the presence of spacelike and timelike faces: this obscured the computation of the spin representation of the correlation function. We solved this problem by noticing that timelike faces can only be shared by timelike tetrahedra such that only the $(--)$-component of the correlation function can contain contributions of timelike faces. As a result, we could explicitly compute the non-local correlation function for all labels being spacelike and at least one label being timelike separately.
    \item[iii.] An enlarged theory space with multiple mass parameters $\mu_\alpha$ and functions $Z^\phi_\alpha,Z^g_\alpha$: this necessitated a careful analysis of the parameter dependence of the correlation functions and their asymptotic behavior. More precisely, the qualitative behavior of the local correlation function depends on the values of $Z^\phi(\vbi)$. We have shown that there exists a regime where the components of the local correlation function exhibit an asymptotic exponential decay where the correlation length is scaling as $\xiloc^{-2}\sim\mu$ in a homogeneous limit of the $\mu_\alpha$.  Similarly, the non-local correlation function qualitatively depends on the values $Z^g(\vb*{0})$. Also here, we have demonstrated the existence of a regime of exponential suppression.\newline\
\end{enumerate}

In the following, we give a more detailed summary of our results, contextualize them, and then comment on the limitations as well as potential extensions of our work.

For a single arbitrary interaction, we have shown in Section~\ref{sec:arbitrary but fixed Q} that in a limit of $\mu_\alpha$ scaling homogeneously to zero, the Ginzburg-$Q_{\alpha\beta}$ exhibits an exponential suppression in all of its non-vanishing components. Therefore, we extend the results of~\cite{Marchetti:2022igl,Marchetti:2022nrf,Dekhil:2024ssa} to the causally complete setting. Importantly, the contributions to the non-local correlation function from timelike labels $(0,\nu)$ do not enter $Q_{\alpha\beta}$. This is a consequence of integrating out group variables which only maps onto the trivial representation $(i,0)$. Heuristically, this is in agreement with the fact that there are no phase transitions on compact domains since spacelike representation labels are conjugate to the non-compact hyperbolic part of holonomies while the timelike labels are conjugate to the compact rotational part. This intriguingly ties the generation of non-perturbative vacua in TGFT and their tentative continuum geometric interpretation to essential features of the Lorentz group.

In Section~\ref{sec:other interactions}, we extended the results of Section~\ref{sec:Fixed but arbitrary interaction} to other physically interesting types of interactions. First, in Section~\ref{sec:CDT-like model}, we considered the Landau-Ginzburg analysis of the CDT-like simplicial model that includes two interaction terms corresponding to the $(4,1)$ and $(3,2)$ simplices. In this case, the mean-field equations take a more complicated form than in Section~\ref{sec:Fixed but arbitrary interaction}, yielding an involved dependence on the couplings of the final result. With the methods developed in the previous section, we computed the Ginzburg-$Q_{\alpha\beta}$ which shows an exponential decay in the homogeneous limit $\mu\rightarrow 0$, thus validating the mean-field approximation in this regime. The CDT-like TGFT model and actual CDT show numerous differences, conceptually, operationally, and mathematically, see~\cite{Jercher:2022mky} for a discussion. Still, it is conceivable that the transition observed here and the transitions observed in CDT~\cite{Loll:2019rdj} are related. However, to illuminate this point further, a geometrical characterization of the TGFT mean-field vacuum studied here is required for a comparison.

The methods developed in this work allowed us to consider in Section~\ref{sec:colored model} a colored simplicial model with all tetrahedra being spacelike. Assuming the same weights $Z^\phi$ of the local Laplace operator for every color led to a local correlation function with an exponentially decaying and an oscillating term. This suggested to introduce a posteriori different weights $Z^\phi_i$ for the colors $i$, which implies a non-trivial interplay of color and the scalar matter degrees of freedom. For the non-local correlation function, we obtained complex effective masses, whose real part could be consistently chosen to be positive, leading to an asymptotic exponential suppression paired with oscillations. We have shown that the Ginzburg-$Q_{ij}$ for the colored simplicial BC model is exponentially suppressed close to the phase transition $\mu\rightarrow 0$, validating the mean-field approximation also in this case. Based on the results gathered on this work, we can stringently argue that they straightforwardly transfer to the colored version of the causally complete Lorentzian BC TGFT model. The reader may appreciate that this is by far the most involved and most realistic model considered so far in the literature.

For all the interactions we considered in this article, we observed that the qualitative behavior of local and non-local correlation functions depends on the choice of parameters. In particular the weights of the Laplace operators $Z^g$ and $Z^\phi$ determine the sign of the resulting effective masses. For the CDT-like model specifically, also the relative sign of couplings $\lambda_{(4,1)}$ and $\lambda_{(3,2)}$ played an important role. We have identified a regime of parameters where correlations are exponentially decaying and where mean-field theory is applicable. 

Our results are limited in that we restricted to a regime of parameters where local and non-local correlation functions exhibit an exponential decay. However, we emphasize that beyond that, correlations in TGFTs can exhibit exotic behavior such as oscillations or even exponential divergence which differs from local field theories. Potentially, a source for these differences lies in the fact that the non-local correlation function is defined on the space of geometries, i.e. on superspace, rather than on spacetime. On this level, intuition from local field theories might not be applicable. Examining these regimes further might require to consider different types of order parameters or to go beyond a perturbative treatment by employing methods from the FRG, see for instance~\cite{Geloun:2023ray}.

Landau-Ginzburg theory was originally developed to study second order phase transitions in phenomenologically motivated models that exhibit spontaneous symmetry breaking~\cite{Goldenfeld:1992qy}. In this work, we took a different starting point in that we studied models motivated from quantum gravity, including spin foam models, CDT, and tensor models. The proposed TGFT actions do not necessarily exhibit spontaneous symmetry breaking such as the uncolored simplicial or the CDT interaction. The absence of symmetry breaking does not imply that there is no phase transition at $\mu\rightarrow 0$. However, without symmetry breaking, determining the order of transition and the overall phase structure seems to require the evocation of arguments beyond the mean-field approach. Still, we emphasize that the exponential suppression of $Q_{\alpha\beta}$ proves the mean-field approximation valid in the regime $\mu\rightarrow 0$. 

Note that in~\cite{Freidel:2005qe} it was suggested that the physical inner product of spin network boundary states should be defined via the tree-level TGFT partition sum and that that would be sufficient to give an approximate description of continuum geometries. Following~\cite{Oriti:2006se}, this idea goes in fact back to the concept of \textit{third quantization}, where a quantum field theory on the space of geometries is proposed~\cite{Giddings:1988wv}. Therein, the classical field equations of motion correspond to a non-linear extension of the Wheeler-de Witt equation. Our results on the phase transitions and mean-field vacua as well as those on the extraction of effective cosmological dynamics within a mean-field approximation~\cite{Gielen:2013kla,Gielen:2013naa,Gielen:2016dss,Oriti:2016acw,Pithis:2019tvp,Jercher:2021bie,Oriti:2021oux,Marchetti:2021gcv} clearly support the conjecture of~\cite{Freidel:2005qe} as they demonstrate that the classical equations of motion of TGFT may indeed be sufficient to capture effective continuum geometric information. Moreover, it puts this idea on firmer grounds since the suppression of fluctuations validating the mean-field ansatz is induced by the boost part of the Lorentz group and is thus rooted in the relativistic nature of the underlying quantum geometry. To illuminate the latter point, it might be useful to clarify the physical meaning of the skirt radius $a$ which naively seems to be related to the speed of light $c$ via the rapidity parameter. We leave further investigations of these tentative yet intriguing points to a later stage.

Another highly important point is to extend the application of the Landau-Ginzburg method to the TGFT formulation of the Lorentzian Engle-Pereira-Rovelli-Livine (EPRL) spin foam model~\cite{Engle:2007wy,Rovelli:2011eq,Perez:2012wv}. This model is based on spacelike tetrahedra as elementary building blocks. At the current stage, only an EPRL-like TGFT model is available in the literature~\cite{Oriti:2016qtz} in which the detailed form of the interaction kernel (where the Lorentzian data are encoded) is not yet spelled out. This gap in the literature will have to be closed before mean-field theory can be applied. We note that since the EPRL model is based on $\text{SU}(2)$ data embedded into $\SL$ we expect that much of the application of Landau-Ginzburg theory to the BC model restricted to spacelike tetrahedra can be carried over and that essentially the same behavior for the $Q$-parameter can be found. In fact, results from area Regge calculus~\cite{Dittrich:2021kzs} and TGFT condensate cosmology~\cite{Jercher:2021bie} already suggest that the EPRL and the BC model lie in the same universality class from the point of view of continuum gravitational physics. This conjecture would be strongly supported if a Landau-Ginzburg analysis yields the same scaling exponents near criticality. In a second step, it would be important to generalize the model to also include timelike tetrahedra using the Conrady-Hnybida extension~\cite{Conrady:2010kc,Conrady:2010vx} of the EPRL spin foam model. Given the flexibility of the TGFT machinery, one may hope that based on this, the inclusion of lightlike configurations could also be facilitated. Once these points have been resolved, the phase structure of this causally completed EPRL TGFT could then be investigated with the same techniques as here to analyze if the two models lie in the same universality class.

Finally, beyond the identification of potential universality classes via critical exponents at mean-field level and the phenomenological exploration of the corresponding condensate states obtained here, a deeper understanding of the emergent geometries is highly desirable. The main challenge to make progress on this front lies in the identification of suitable operators and observables which would allow us to extract for instance the Hausdorff and spectral dimensions of such states. We leave the investigation of these important points to future investigations.

\subsection*{Acknowledgements}

The authors thank T. Angrick, S. Fl\"{o}rchinger, L. Marchetti, J. Sim\~{a}o, I. Soler, and in particular D. Oriti and R. Schmieden for helpful discussions. AGAP acknowledges funding from the Deutsche Forschungsgemeinschaft (DFG, German Research Foundation) research grants OR432/3-1 and OR432/4-1 and the John-Templeton Foundation via research grant 6242.  AFJ acknowledges support by the DFG under Grant No 406116891 within the Research Training Group RTG 2522/1 and under Grant No 422809950. RD, AFJ and AGAP are grateful for the generous financial support by the MCQST via the seed funding Aost 862933-9 granted to AGAP and the seed funding Aost 862981-8 granted to Jibril Ben Achour by the DFG under Germany’s Excellence Strategy – EXC-2111 – 390814868. AGAP in particular acknowledges funding by the DFG under the author’s project number 527121685 as a Principal Investigator. 

\appendix

\section{$\SL$, its representation theory and homogeneous spaces}\label{sec:Aspects of SL2C and its Representation Theory}

The unitary irreducible representation spaces of $\SL$ are realized as the space of homogeneous functions on $\C^2$ with degree $(\lambda,\mu)$, where $\lambda,\mu\in\C$ and $\lambda-\mu\in\mathbb{Z}$~\cite{Ruehl1970}. We parametrize the pair $(\lambda,\mu)$ as
\begin{equation}\label{eq:(lambda,mu)}
(\lambda,\mu) = (i\rho+\nu-1,i\rho-\nu-1)\,,
\end{equation}
with $\nu\in\mathbb{Z}/2$. For the remainder, we restrict to the principal series which is given by the restriction $\rho\in\R$. Denoting the representation space by $\mathcal{D}^{(\rho,\nu)}$, the group action of $\SL$ on this space is defined as~\cite{Ruehl1970}
\begin{equation}\label{eq:group action}
\left(\mathbf{D}^{(\rho,\nu)}(g)F\right)(\vb*{z}) = F(g^T\vb*{z}),
\end{equation}
where $g\in\SL$, $\vb*{z}\in\C^2$ and $\mathbf{D}^{(\rho,\nu)}$ is the representation matrix commonly referred to as Wigner matrix. In the canonical basis, the Wigner matrices have components $D^{(\rho,\nu)}_{jmln}$, which form an orthogonal basis of $L^2\left(\SL\right)$, 
\begin{equation}\label{eq:orthogonality relation of SL2C wigner matrices}
\int\limits_{\SL}\dd{h}\overline{D^{(\rho_1,\nu_1)}_{j_1 m_1 l_1 n_1}(h)}D^{(\rho_2,\nu_2)}_{j_2 m_2 l_2 n_2}(h)
=
\frac{\delta(\rho_1-\rho_2)\delta_{\nu_1, \nu_2}\delta_{j_1, j_2}\delta_{l_1, l_2}\delta_{m_1, m_2}\delta_{n_1, n_2}}{\rho_1^2+\nu_1^2},
\end{equation}
with \textit{magnetic indices} $(j,m,l,n)$ in the range 
\begin{equation}
j,l\in\{\abs{\nu},\abs{\nu}+1,...\}, \quad m\in\{-j,...,j\},\quad n\in\{-l,...,l\}.
\end{equation}
The Cartan decomposition of group elements $g\in\SL$ is used frequently in Section~\ref{sec:Fixed but arbitrary interaction} and given by~\cite{Ruehl1970} 
\begin{equation}\label{eq:Cartan decomp g}
g = u\e^{\frac{\eta}{2}\sigma_3}v,\qquad u,v\in\SUT,\quad \eta\in\R_+\,.
\end{equation}
Therein we have $\eta$ as the rapidity parameter of a boost along the $z$ axis while $u$ and $v$ are arbitrary $\SUT$-rotations. This induces a decomposition of the Haar measure on $\SL$
\begin{equation}\label{eq:Cartan decomp measure}
\dd{g} = \frac{1}{\pi}\dd{u}\dd{v}\dd{\eta}\sinh^2(\eta),
\end{equation}
as well as a decomposition of $\SL$-Wigner matrices
\begin{equation}\label{eq:Cartan decomp D}
D^{(\rho,\nu)}_{jmln}(g) = \sum_{q=-\mathrm{min}(j,l)}^{\mathrm{min}(j,l)}D^{j}_{mq}(u)d^{(\rho,\nu)}_{jlq}(\eta)D^{l}_{qn}(v),
\end{equation}
with $d^{(\rho,\nu)}$ being the reduced $\SL$-Wigner matrix~\cite{Ruehl1970}.\footnote{As explained below in Appendix~\ref{sec:Empty integrals}, the skirt radius $a$ of the hyperbolic part of $\SL$ is included by the substitution $\tilde{\eta} = a\eta$.}

Square-integrable functions can be expressed in the so-called \textit{spin representation} by exploiting that $\SL$-Wigner matrices form an orthogonal basis on $L^2(\SL)$. It is explicitly given by
\begin{equation}
\varphi(g) = \int\dd{\rho}\sum_\nu (\rho^2+\nu^2)\sum_{j,m,l,n}\varphi^{\rho,\nu}_{jmln}D^{(\rho,\nu)}_{jmln}(g).
\end{equation}
where $(\rho^2+\nu^2)$ is the Plancherel measure on $\SL$. Functions $\varphi\in L^2(\SL)$ which are class functions, i.e. satisfying $\varphi(g) = \varphi(hgh^{-1})$ for all $h\in\SL$, are expanded in terms of traces
\begin{equation}\label{eq:class function}
\varphi(g) = \int\dd{\rho}\sum_\nu(\rho^2+\nu^2)\varphi^{\rho,\nu}\Tr_{(\rho,\nu)}(g)\,,
\end{equation}
where here, $\Tr_{(\rho,\nu)}$ is the trace in the representation $(\rho,\nu)$, also referred to as \textit{character}.

The Casimir operators of $\SL$ act on states in the canonical basis $\ket{(\rho,\nu);jm}\in\mathcal{D}^{(\rho,\nu)}$ as\footnote{If the skirt radius $a$ is included, the two Casimirs come with a pre-factor of $1/a^2$.}
\begin{align}
\cas_1\ket{(\rho,\nu);jm} & =  -\rho^2+\nu^2-1\ket{(\rho,\nu);jm}\label{eq:cas1}\\[7pt]
\cas_2\ket{(\rho,\nu);jm} & = \rho\nu\ket{(\rho,\nu);jm}\label{eq:cas2}.
\end{align}
The second equation shows that the imposition of simplicity, which in the Barrett-Crane model corresponds to a vanishing second Casimir, leads to either $\rho$ or $\nu$ vanishing. Upon simplicity, the first Casimir, associated to the squared area operator of triangles, has either a positive or negative spectrum, depending on whether $\nu = 0$ or $\rho = 0$, respectively. Consequently, for the BC model, the labels $\rho$ are associated to spacelike faces while the labels $\nu$ are associated to timelike faces.

\paragraph{A note on conventions.} There is a variety of textbooks and articles on the unitary irreducible representation theory of $\SL$~\cite{Naimark1964,Ruehl1970,Martin-Dussaud:2019ypf} using different conventions.\footnote{The authors would like to thank Jos\'{e} Sim\~{a}o for clarifying discussions on this matter and refer to~\cite{Simao:2024chp} for further reading.} There are three choices of conventions one has to make:
\begin{enumerate}
    \item The parametrization of $(\lambda,\mu)$ in terms of $\rho$ and $\nu$: In our case this is given by Eq.~\eqref{eq:(lambda,mu)}. Different conventions are for example used in~\cite{Ruehl1970} with labels $(\rho_R,\nu_R)$ that are related to our choice by $(\rho_R,\nu_R) = (2\rho,2\nu)$.
    \item Haar measure: Since $\SL$ is non-compact, the Haar measure is determined up to a multiplicative factor. In Eq.~\eqref{eq:Cartan decomp measure}, this factor is given by $1/\pi$, which is the same choice as in~\cite{Dona:2021ldn}. In contrast, Refs.~\cite{Ruehl1970,Martin-Dussaud:2019ypf} use a pre-factor of $1/4\pi$.
    \item Group action: The group action defined in Eq.~\eqref{eq:group action} has been chosen as the left action of the transpose of $g$ on the argument of the function of $F\in\mathcal{D}^{(\rho,\nu)}$, corresponding to the conventions of~\cite{Ruehl1970}. Other choices, such as a right action or the action by the inverse group element are conceivable.
\end{enumerate}
Conventions 1. and 2. are important for the orthogonality relation of $\SL$-Wigner matrices and thus for the precise Plancherel measure and spin representation. Also, empty $\SL$ integrals and the projection onto the trivial representation, discussed in Sections~\ref{sec:Empty integrals} and~\ref{sec:trivial represenation}, respectively, depend on the choice of the first and second convention. The third point determines the precise form of the Wigner matrices. For instance, the formula of the reduced $\SL$-Wigner matrix coefficients in terms of hypergeometric functions, given in Eq.~\eqref{eq:reduced d with 2F1}, would change for a different group action.

\subsection{Homogeneous spaces}

The group $\SL$ acts on Minkowski space $\R^{1,3}$ via the isomorphism between the space of hermitian $2\times 2$ matrices and $\R^{1,3}$, see for instance~\cite{Martin-Dussaud:2019ypf}. To timelike, lightlike and spacelike vectors in $\R^{1,3}$, one can consequently associate stabilizer subgroups of $\SL$, being isomorphic to $\SUT, \ISO$ and $\SUO$, respectively, denoted in short with $\mathrm{U}^{(\alpha)}$ with $\alpha\in\{+,0,-\}$. The quotient spaces $\mathrm{H}_{\alpha} = \SL/\mathrm{U}^{(\alpha)}$ are the two-sheeted hyperboloid, the light cone and the one-sheeted hyperboloid, respectively, which can be described as submanifolds of $\R^{1,3}$
\begin{subequations}\label{eq:hom spaces in R13}
\begin{align}
\mathrm{H}_{\plus} &\defeq \left\{(t,x,y,z)\in\R^{1,3}\;\middle\vert\; t^2-x^2-y^2-z^2 = 1,\; t\gtrless 0\right\},\label{eq:defH3}\\[7pt]
\mathrm{H}_\zero &\defeq \left\{(t,x,y,z)\in\R^{1,3}\;\middle\vert\; t^2-x^2-y^2-z^2 = 0,\; t\gtrless 0\right\},\label{eq:defC}\\[7pt]
\mathrm{H}_{\minus} &\defeq \left\{(t,x,y,z)\in\R^{1,3}\;\middle\vert\; t^2-x^2-y^2-z^2 = -1\right\}.\label{eq:defH21}
\end{align}
\end{subequations}
Their line-element, obtained by inserting the defining relations of Eqs.~\eqref{eq:hom spaces in R13} into the Minkowski line-element, are given by
\begin{subequations}
\begin{align}
\dd{\mathrm{H}_{\plus}^2} &= \dd{\eta}^2+\sinh^2(\eta)\dd{\Omega_2^2},\label{eq:line element H+}\\[7pt]
\dd{\mathrm{H}_0^2} &= r^2\dd{\Omega_2^2},\label{eq:line element H0}\\[7pt]
\dd{\mathrm{H}_{\minus}^2} &= \dd{\eta}^2-\cosh^2(\eta)\dd{\Omega_2^2},\label{eq:line element H-}
\end{align}
\end{subequations}
with $\eta\in\R, r\in\R_+$ and $\dd{\Omega_2}$ being the normalized measure on the $2$-sphere $S^2$.\footnote{We omitted the skirt radius for $\mathrm{H}_{\plus}$ and $\mathrm{H}_{\minus}$ and discuss its inclusion at the end of the next section.} Taking the square root of the determinant of the resulting induced metric, we obtain the measure for the spaces $\mathrm{H}_{\plus}$ and $\mathrm{H}_{\minus}$,
\begin{subequations}
\begin{align}
\dd{X_{\plus}} &= \frac{1}{2\pi}\dd{\Omega_2}\dd{\eta}\sinh^2(\eta),\label{eq:measure on H+}\\[7pt]
\dd{X_{\minus}} &= \frac{1}{2\pi}\dd{\Omega_2}\dd{\eta}\cosh^2(\eta)\label{eq:measure on H-}.
\end{align}
\end{subequations}
Since $\mathrm{H}_{\plus}$ and $\mathrm{H}_{\minus}$ are obtained by forming the quotient space of $\SL$ with respect to the groups $\mathrm{U}^{(\alpha)}$, the choice of Haar measure on $\SL$, given in Eq.~\eqref{eq:Cartan decomp measure}, induces a measure on $\mathrm{H}_{\plus}$ and $\mathrm{H}_{\minus}$. Thus, we added a factor of $1/\pi$ to account for that and a factor of $1/2$ since here, $\eta\in\R$.

The light cone, equipped with topology $\R\times S^2$, exhibits a degenerate line element in Eq.~\eqref{eq:line element H0}, which is characteristic for a null surface. Correspondingly, the integration measure obtained by naively taking the determinant of the induced metric is independent of the degenerate non-compact direction. However, by parametrizing null vectors as 
\begin{equation}
X_\zero = (\lambda,\lambda\hat{\vb*{r}}),\qquad \lambda\in\R_+,\hat{\vb*{r}}\in S^2,
\end{equation}
which takes the degenerate direction into account, the measure on the light cone is in fact given by
\begin{equation}\label{eq:measure on H0}
\dd{X_\zero} = \frac{1}{2\pi}\dd{\Omega_2}\dd{\lambda}\lambda.
\end{equation}
For further details on null hypersurfaces and their geometric treatment, see~\cite{Ciambelli:2019lap}.

\subsection{Empty integrals}\label{sec:Empty integrals}

In Sections~\ref{sec:Fixed but arbitrary interaction} and~\ref{sec:other interactions}, the projection onto constant field configurations frequently yields empty integrals over $\SL$ and the homogeneous spaces $\mathrm{H}_{\alpha}$, yielding diverging volume factors $V_{\alpha}$. Consequently, a regularization of these terms is required. When removing the regulator from the final result, it is therefore important to understand to which degree these volume factors diverge.

From the Cartan decomposition of the Haar measure on $\SL$ in Eq.~\eqref{eq:Cartan decomp measure} one concludes that the volume factor of $\SL$ and $\mathrm{H}_\plus$ diverge in the same way. This can also be seen by observing that $\SL$ is of topology $\mathrm{H}_{\plus}\times S^3$, where the measure on $S^3$ is normalized, such that the non-compact direction is in fact the hyperboloid $\mathrm{H}_{\plus}$. Introducing a cutoff $L$, the regularized volume factor $V_{\plus}$ is therefore defined as
\begin{equation}
V_{\plus} = \frac{1}{\pi}\int\limits_{0}^{L}\dd{\eta}\sinh^2(\eta)\:\underset{L\gg 1}{\longrightarrow} \:\frac{1}{4\pi}\e^{2L}.
\end{equation}
Similarly, we find for the empty integral on the one-sheeted hyperboloid $\mathrm{H}_{\minus}$ the asymptotic behavior $V_{\minus}\:\longrightarrow \:\frac{1}{4\pi}\e^{2L}$. 

On the light-cone with measure given in Eq.~\eqref{eq:measure on H0}, we perform a coordinate transformation $\lambda\rightarrow\e^\eta$ and introduce an upper cutoff $L$, such that $V_\zero$ is given by
\begin{equation}
V_\zero = \frac{1}{2\pi}\int\limits_{\R_+}\dd{\lambda}\int\dd{\Omega_2}\lambda = \frac{1}{2\pi}\int\limits_{0}^L\dd{\eta}\int\dd{\Omega_2}\e^{2\eta} \:\underset{L\gg 1}{\longrightarrow} \:\frac{1}{4\pi}\e^{2L}.  
\end{equation}

From the scaling behavior of the volume factors in terms of the upper cutoff $L$, we find that the $V_\alpha$ all diverge to the same degree. In particular, expressions like $V_\alpha/V_\beta = 1$ after regularizing.  

Notice that in this Appendix, we parametrized the non-compact direction of the $\mathrm{H}_\alpha$ with the dimensionless variable $\eta$. When extracting the scaling behavior of the Ginzburg-$Q_{\alpha\beta}$ in the main body of this work, one would like to work with a dimensionful variable for the non-compact direction. This can be straightforwardly implemented by the substitution $\tilde{\eta} = a\eta$. Notice that assume the same scale $a$ for all three cases. For the hyperboloids $\mathrm{H}_\plus$ and $\mathrm{H}_\minus$, $a$ has the interpretation of a skirt radius. On the light cone $\mathrm{H}_\zero$, $a$ acts as a multiplicative factor of vectors along the degenerate direction without affecting its geometry. Then, the scaling of the volume factors for large cutoffs is given by
\begin{equation}
    V_\alpha \sim\frac{a^3}{4\pi}\e^{2L/a}.
\end{equation}
We explicitly utilize this scaling behavior in Secs.~\ref{sec:arbitrary but fixed Q},~\ref{sec:CDT-like model} and~\ref{sec:colored model}.

\subsection{Constant function and the trivial representation}\label{sec:trivial represenation}

Since $\SL$ is non-compact, the constant function is not part of the $L^2$-space. Thus, to be able to project onto constant field configurations, one has to extend the space of functions to that of distributions, as proposed by~\cite{Ruehl1970}. This allows us to define a pseudo-projector\footnote{\enquote{Pseudo} because applying the map twice yields again the pseudo-projector but multiplied with a volume factor of $\SL$.} onto the trivial representation, given by the integral expression
\begin{equation}\label{eq:zero mode projection pre integration}
\int\dd{g}D^{(\rho,\nu)}_{jmln}(g).
\end{equation}
Exploiting the Cartan-decomposition in Eqs.~\eqref{eq:Cartan decomp g}, \eqref{eq:Cartan decomp measure} and~\eqref{eq:Cartan decomp D}, we find
\begin{equation}
\int\dd{g}D^{(\rho,\nu)}_{jmln}(g) = \sum_q\int\dd{u}D^j_{mq}(u)\int\dd{v}D^l_{qn}(v^{-1})\frac{1}{\pi}\int\limits_{\R^+}\dd{\eta}\sinh^2(\eta)d^{(\rho,\nu)}_{jlq}(\eta).
\end{equation}
One can perform the $\SUT$-integrals explicitly, yielding Kronecker-$\delta$'s on the magnetic indices $(j,m,l,n)$. The condition that $j\geq\abs{\nu}$ (see above) together with $j=0$ forces the discrete representation label $\nu = 0$. As discussed in Sections~\ref{sec:Local correlation function} and~\ref{sec:arbitrary but fixed Q}, this bears important consequences for the timelike label contributions to the correlation function.  

Following~\cite{Ruehl1970} with the conventions as in Eqs.~\eqref{eq:(lambda,mu)} and~\eqref{eq:Cartan decomp measure}, the reduced Wigner matrix $d^{(\rho,0)}_{000}(\eta)$, also referred to as character, is given by
\begin{equation}\label{eq:d0}
d^{(\rho,0)}_{000}(\eta) = \frac{\sin(\rho\eta)}{\rho\sinh(\eta)}.
\end{equation}
Then, we can compute the above integral 
\begin{equation}
\begin{aligned}
 \frac{1}{\pi}\int\limits_{\R^+}\dd{\eta}\sinh^2(\eta)\frac{\sin(\rho\eta)}{\rho\sinh(\eta)} =& \frac{1}{\pi}\frac{1}{4i\rho}\int\limits_{\R}\dd{\eta}\left[\e^{i\eta(\rho-i)}-\e^{i\eta(\rho+i)}\right]\\[7pt]
=&
\frac{1}{2i\rho}\left[\delta(\rho-i)-\delta(\rho+i)\right]\longrightarrow -\delta(\rho-i).
\end{aligned}
\end{equation}
In the last step, we used the unitary equivalence of $(\rho,0)$ and the $(-\rho,0)$ representations, yielding a factor of $2$ and that $\frac{1}{i\rho}\delta(\rho-i)$ acts as a distribution as $-\delta(\rho-i)$, yielding a minus sign.\footnote{We emphasize that the explicit evaluation of Eq.~\eqref{eq:zero mode projection pre integration} depends on the conventions used in Eqs.~\eqref{eq:(lambda,mu)} and~\eqref{eq:Cartan decomp measure}. Both of these conventions differ from Ref.~\cite{Ruehl1970}, where the same computation would yield in total $-\frac{1}{4}\delta(\rho_R-2i)$.} 

In total, the integral of Eq.~\eqref{eq:zero mode projection pre integration} yields
\begin{equation}\label{eq:zero mode projection}
\int\dd{g}D^{(\rho,\nu)}_{jmln}(g) = -\delta(\rho-i)\delta_{\nu,0}\delta_{j,0}\delta_{m,0}\delta_{l,0}\delta_{n,0}.
\end{equation}
This makes manifest that the identity in the space of distributions on $L^2(\SL)$ is written as
\begin{equation}\label{eq:trivial rep}
\one = D^{(i,0)}_{0000}(g),
\end{equation}
such that any constant function $f$ can be written as $f\one$. Notice furthermore that Eq.~\eqref{eq:trivial rep} is consistent with the orthogonality relation in Eq.~\eqref{eq:orthogonality relation of SL2C wigner matrices}, i.e.
\begin{equation}
\int\dd{g}D^{(\rho,\nu)}_{jmln}(g)\one = \int\dd{g}D^{(\rho,\nu)}_{jmln}(g)D^{(i,0)}_{0000}(g) = \frac{\delta(\rho-i)\delta_{\nu,0}}{\rho^2+\nu^2}\delta_{j,0}\delta_{m,0}\delta_{l,0}\delta_{n,0}
\end{equation}

\begin{figure}
    \centering
    \includegraphics[width=0.6\linewidth]{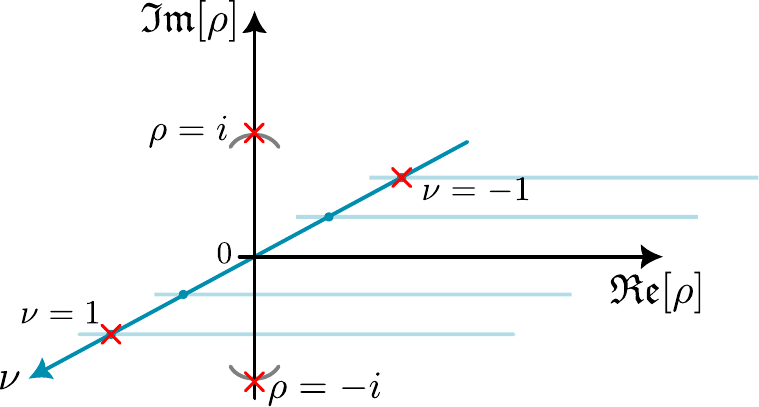}
    \caption{A visualization of the space of all irreducible unitary $\SL$-representations. The principal series is parametrized by $(\rho,\nu)\in\R\times\mathbb{Z}/2$ and the complementary series is given by $\nu = 0$ and $\rho\in (-i,i)$. Values of representations for which both Casimir operators vanish are drawn in red. These are also the points which are obtained from the complementary series in a limit $\rho\rightarrow\pm i$~\cite{Naimark1964}. Since the limits converge to two points, respectively, the space of irreducibles is in general equipped with a non-Hausdorff topology.}
    \label{fig:SL2C-irreps}
\end{figure}

In order to appreciate this result further, one needs to consider the full space of irreducible unitary representations of $\SL$, as presented in~\cite{Naimark1964} and depicted in Fig.~\ref{fig:SL2C-irreps}. Detailed therein, all of the irreducible unitary representations are captured by the principal series, the complementary series and the trivial representation. Clearly, the pseudo-projector defined above maps to the trivial representation which cannot be part of the principal series containing $(0,\pm 1)$. Consequently, constant field configurations will only involve $\rho = \pm i$ representations and vanishing magnetic indices $j=0,m=0$.

\subsection{Regularization of Dirac delta function}\label{sec:Regularization of Dirac delta function}

The matrix $\chi_{\alpha\beta}$ capturing the properties of the Hessian contains projections onto the trivial representation which lead to products of $\delta(\rho-i)$, normalized by volume factors of $\SL.$ We show in this subsection that these expressions are indeed regularized in the sense that they are objects similar to Kronecker-deltas.

Consider Eq.~\eqref{eq:zero mode projection} evaluated on vanishing magnetic indices,
\begin{equation}
-\delta(\rho-i) = \int\dd{g}D^{(\rho,0)}_{0000}(g),
\end{equation}
and evaluate both sides on $\rho = i$, yielding
\begin{equation}
\eval{-\delta(\rho-i)}_{\rho = i} = \int\dd{g}D^{(i,0)}_{0000}(g) = \int\dd{g} = V_{\plus}\, ,
\end{equation}
where we used that $D^{(i,0)}_{0000}(g)$ is the identity for all $g\in\SL$. Then,
\begin{equation}
\eval{-\frac{\delta(\rho-i)}{V_{\plus}}}_{\rho=i} = 1
\end{equation}
and we have in total
\begin{equation}\label{eq:regularized delta}
\delta_{\rho,i}\defeq -\frac{\delta(\rho-i)}{V_{\plus}} =
\begin{cases}
1,&\quad\text{for }\rho = i,\\[7pt]
0,&\quad\text{for }\rho\neq i.
\end{cases}
\end{equation}

Importantly, the symbol $\delta_{\rho,i}$ is obtained independent of conventions, i.e. choosing a different Haar measure or a different convention for Eq.~\eqref{eq:(lambda,mu)} changes the expressions for the volume factors and the pre-factors in front of $\delta(\rho-i)$ but not their ratio entering $\delta_{\rho,i}$.

\section{Asymptotics of reduced Wigner matrix with timelike labels}\label{sec:asymptotics of d}

In this section, we derive the asymptotics of the reduced Wigner matrix $d^{(0,\nu)}_{jlq}(\eta)$ in the limit of large $\eta$. To that end, the reduced Wigner matrix is expressed through the hypergeometric function ${}_2F_1$,
\begin{equation}\label{eq:reduced d with 2F1}
\begin{aligned}
& d^{(0,\nu)}_{jlq}(\eta) = c(j,l,q)\sqrt{\prod_{+,-}(j\pm\nu)!(j\pm q)!(l\pm \nu)!(l\pm q)!} \\[7pt]
&\times
\sum_{s,t}\frac{(-1)^{s+t}(\nu+q+s+t)!(j+l-\nu-q-s-t)!}{s!(j-\nu-s)!(j-q-s)!(\nu+q+s)!t!(l-\nu-t)!(l-q-t)!(\nu+q+t)!}\\[7pt]
&\times
\e^{-\eta(\nu+q+1+2t)} {}_2F_1(l+1, \nu+q+1+s+t, j+l+2,1-\e^{-2\eta}).
\end{aligned}
\end{equation}
The integers $s$ and $t$ are constrained such that the factorials in the above formula are defined, yielding the inequalities
\begin{equation}
s,t\geq 0,\quad j-s\geq\nu,\quad j-s\geq q,\quad \nu+q+s\geq 0,\quad \nu+q+t\geq 0.
\end{equation}

The hypergeometric function is diverging for $\eta\rightarrow\infty$ if $j-\nu-q-s-t \leq 0$, which is generically true in the magnetic index space spanned by $j,\nu,q,s$ and $t$. This can be seen more explicitly from the following expansion of the hypergeometric function in the case where $a,b,c\in\mathbb{Z}$,
\begin{equation}
{}_2F_1(a,b,c;x)
=
\frac{\log(1-x)}{x^{c-1}}\sum_{p}\alpha_p x^p + \frac{1}{x^{c-1}}\sum_{q}\beta_q x^q + (1-x)^{c-a-b}\sum_{n}\gamma_n x^n.
\end{equation}
Since we are interested in the behavior near $x=1$, we consider only the logarithmic and the $(1-x)^{c-a-b}$ term as relevant. Importantly, ${}_2F_1$ enters the reduced Wigner matrix in Eq.~\eqref{eq:reduced d with 2F1} not isolated but as a product with $\e^{-\eta(\nu+q+1+2t)}$, which is written as $(1-x)^{\frac{1}{2}(\nu+q+1+2t)}$ for $x = 1-\e^{-2\eta}$. Combining these two factors, we observe that the logarithmic term is suppressed, while only 
\begin{equation}
\sim (1-x)^{\frac{1}{2}(2j+1-\nu-q-2s)}= \e^{-\eta(2j+1-\nu-q-2s)}
\end{equation}
remains. Clearly, as $x\rightarrow 1$, the term for which the exponent is minimal will dominate. As a result, we find numerically as well as analytically that the minima are given by the configuration
\begin{equation}
q = \nu,\quad  s = j-\nu,\quad t = 0. 
\end{equation}
This result is supported by the numerical findings where the dominating terms of the rescaled hypergeometric function have been obtained. Inserting these labels into Eq.~\eqref{eq:reduced d with 2F1}, we find 
\begin{equation}
\begin{aligned}
& \sum_q\sum_{s,t}f(q,s,t)(1-x)^{\frac{1}{2}(2j+1-\nu-q-2s)}{}_2F_1(l+1, \nu+q+1+s+t, j+l+2, x)\\[7pt]
&\overset{x\rightarrow 1}{\longrightarrow}f(\nu,j-\nu,0)(1-x)^{\frac{1}{2}}\sim \e^{-\eta}.
\end{aligned}
\end{equation}
Remarkably, the dominant term of $d^{(0,\nu)}(\eta)$ in the limit $\eta\rightarrow \infty$ is independent of the representation label $\nu$. This behavior has crucial consequences for the Landau-Ginzburg analysis as it renders timelike faces irrelevant for the critical behavior. For a detailed explanation we refer the reader to Section~\ref{sec:Contributions with timelike faces}.

\section{Correlation functions with geometricity constraints}\label{sec:Correlation functions with geometricity constraints}

In this appendix, we describe in detail how a correlation function is obtained in the presence of non-local variables which are subject to geometricity constraints. This forms the basis of the analysis in Sections~\ref{sec:Fixed but arbitrary interaction} and~\ref{sec:other interactions}.

\subsection{Derivation of the correlation function}\label{sec:A derivation of the correlation function}

The perturbations $\delta\Phi_\alpha$ with the normal integrated out satisfy the equation
\begin{equation}
\sum_\beta\int\dd{\vbg'}\dd{\vbf'}G_{\alpha\beta}(\vb*{g},\vb*{\phi};\vb*{g}',\vb*{\phi}')\delta\Phi_\beta(\vb*{g}',\vb*{\phi}')=0,
\end{equation}
with $G_{\alpha\beta}$ being the modified propagator composed of the kinetic kernel and the Hessian of the linearized interaction. 

In the following, we consider those components of $G_{\alpha\beta}$, which only contain spacelike labels $\rho$ in their spin representation. The $(--)$-component, which contains also timelike labels $\nu$, is summarized thereafter. $G_{\alpha\beta}$ satisfies the symmetries of a two-point function on a domain with local and non-local variables which are summarized in Eq.~\eqref{eq:2point G}. Consequently, its spin representation is given in terms of traces,
\begin{equation}
\eval{G_{\alpha\beta}(\vbg,\vbf;\vbg',\vbf')}_{\mathrm{sl}} = \int\dd{\mu(\vbr,\vbk)}\e^{i\vbk (\vbf-\vbf')}\prod_c D^{(\rho_c,0)}_{j_c m_c j_c m_c}(g_c^{-1}g_c')G_{\alpha\beta}^{\vbr}(\vbk),
\end{equation}
where $\dd{\mu}$ is short-hand notation that contains the Plancherel measure and $2\pi$-factors of the $\vbk$-integrations. Summation over repeated magnetic indices is understood. Using the expansion of $G_{\alpha\beta}$, the equations of motion in spin representation are given by
\begin{equation}
\sum_\beta G_{\alpha\beta}^{\vbr}(\vbk)\delta\Phi^{\vbr,\beta}_{\vb*{j}\vb*{m}}(\vbk)B^{\vbr,\beta}_{\vb*{l}\vb*{n}} = 0,
\end{equation}
where the magnetic indices in this equation are uncontracted. An important detail here is that the Barrett-Crane intertwiner cannot be erased from the equation as it enters the sum over the signature label $\beta$. 

Starting from this equation, our goal is to obtain the correlation function first in spin representation and ultimately in group representation. To that end, we set up an \textit{effective action} for the field $\delta\Phi$,
\begin{equation}
S_{\mathrm{eff}}=\frac{1}{2}\sum_{\alpha,\beta}\int\dd{\mu(\vbr,\vbk)}\delta\Phi^{\vbr,\alpha}_{\vb*{j}\vb*{m}}(\vbk)B^{\vbr,\alpha}_{\vb*{l}\vb*{n}}G_{\alpha\beta}^{\vbr}(\vbk)\Phi^{\vbr,\beta}_{\vb*{j}\vb*{m}}(\vbk)B^{\vbr,\beta}_{\vb*{l}\vb*{n}},
\end{equation}
which, upon variation, yields the equations of motion above. The generating functional for $n$-point functions is given by
\begin{equation}\label{eq:Z[J] before int}
Z[J] = \int\mathcal{D}[\delta\Phi_\alpha]\e^{-S_{\mathrm{eff}}[\delta\Phi_\alpha]}\exp\left(\sum_\alpha\int\delta\Phi^{\vbr,\alpha}_{\vb*{j}\vb*{m}}(\vbk)B^{\vbr,\alpha}_{\vb*{l}\vb*{n}}J^{\vbr,\alpha}_{\vb*{j}\vb*{m}\vb*{l}\vb*{n}}(\vbk)\right).
\end{equation}
The fact that the source $J$ couples to both the field $\delta\Phi$ and the Barrett-Crane intertwiner is necessary to ensure the correct equations of motion that explicitly incorporate the Barrett-Crane intertwiner. Due to the Gaussian form of $Z[J]$, one can perform a completion of the square by introducing new variables
\begin{equation}
V^{\vbr,\alpha}_{\vb*{j}\vb*{m}\vb*{l}\vb*{n}}(\vbk)\defeq \delta\Phi^{\vbr,\alpha}_{\vb*{j}\vb*{m}}(\vbk)B^{\vbr,\alpha}_{\vb*{l}\vb*{n}}-\sum_\beta \left(G^{\vbr}(\vbk)^{-1}\right)_{\alpha\beta}J^{\vbr,\beta}_{\vb*{j}\vb*{m}\vb*{l}\vb*{n}}(\vbk),
\end{equation}
and explicitly perform the integration. As a result, $Z[J]$ takes the form
\begin{equation}\label{eq:Z[J] after int}
Z[J] = \frac{1}{\mathrm{D}}\exp\left(-\frac{1}{2}\sum_{\alpha,\beta}\int J^{\vbr,\alpha}_{\vb*{j}\vb*{m}\vb*{l}\vb*{n}}(\vbk)\left(G^{\vbr}(\vbk)^{-1}\right)_{\alpha\beta}J^{\vbr,\alpha}_{\vb*{j}\vb*{m}\vb*{l}\vb*{n}}(\vbk)\right),
\end{equation}
where $\mathrm{D}$ is the determinant factor resulting from the Gaussian integration. This factor drops out when computing expectation values and is therefore irrelevant for the correlation function.

Taking the second functional derivative of $\ln Z[J]$ as defined by Eq.~\eqref{eq:Z[J] before int}, one finds
\begin{equation}\label{eq:2nd der after int}
\eval{\frac{\delta^2\ln Z[J]}{\delta J^{\vbr,\alpha}_{\vb*{j}\vb*{m}\vb*{l}\vb*{n}}(\vbk)\delta J^{\vbr',\beta}_{\vb*{j}'\vb*{m}'\vb*{l}'\vb*{n}'}(\vbk')}}_{J = 0} = \left(G^{\vbr}(\vbk)^{-1}\right)_{\alpha\beta}\delta(\vbk+\vbk')\prod_{c}\delta(\rho_c-\rho_c')\delta_{j_c,j_c'}\delta_{m_c,m_c'}\delta_{l_c,l_c'}\delta_{n_c,n_c'}.
\end{equation}
One the other hand, if we use the expression of Eq.~\eqref{eq:Z[J] after int} for $Z[J]$, then same derivative is instead given by
\begin{equation}\label{eq:2nd der before int}
\eval{\frac{\delta^2\ln Z[J]}{\delta J^{\vbr,\alpha}_{\vb*{j}\vb*{m}\vb*{l}\vb*{n}}(\vbk)\delta J^{\vbr',\beta}_{\vb*{j}'\vb*{m}'\vb*{l}'\vb*{n}'}(\vbk')}}_{J = 0} = \expval{\delta\Phi^{\vbr,\alpha}_{\vb*{j}\vb*{m}}(\vbk)B^{\vbr,\alpha}_{\vb*{l}\vb*{n}}\delta\Phi^{\vbr',\beta}_{\vb*{j}'\vb*{m}'}(\vbk')B^{\vbr',\beta}_{\vb*{l}'\vb*{n}'}}.
\end{equation}

In the last equation, one can identify the correlation function in spin representation, which in this setting enters with two additional Barrett-Crane intertwiners. To finally obtain the correlation function $C_{\alpha\beta}(\vbg,\vbf;\vbg',\vbf')$, we start with the defining expression
\begin{equation}
C_{\alpha\beta}(\vbg,\vbf;\vbg',\vbf') = \expval{\delta\Phi_\alpha(\vbg,\vbf)\delta\Phi_\beta(\vbg',\vbf')}.
\end{equation}
Going to spin representation of the right-hand side of the equation, we find as the integrand Eq.~\eqref{eq:2nd der before int}. Using Eq.~\eqref{eq:2nd der after int}, the correlation function is therefore finally given by
\begin{equation}
C_{\alpha\beta}(\vbg,\vbf;\vbg',\vbf') = \int\dd{\mu(\vbr,\vbk)}\e^{i\vbk (\vbf-\vbf')}\left(G^{\vbr}(\vbk)^{-1}\right)_{\alpha\beta}\prod_c D^{(\rho_c,0)}_{j_c m_c j_c m_c}(g_c^{-1}g_c').
\end{equation}

Now, written in this form, it is apparent that the correlation function satisfies the same symmetries as $G_{\alpha\beta}$. Consequently, the form of $C_{\alpha\beta}$ is given by
\begin{equation}
C_{\alpha\beta}(\vbg,\vbf;\vbg',\vbf') = C_{\alpha\beta}(\vbg^{-1}\vbg',\abs{\vbf-\vbf'}).
\end{equation}
This allows to replace $(\vb*{g}',\vb*{\phi}')$ with $(\vb*{e},\vb*{0})$, finally yielding  the function $C_{\alpha\beta}(\vb*{g},\vb*{\phi})$.

Summarizing, the components $(\alpha\beta)\neq(--)$ of the correlation function in group representation are given by
\begin{equation}\label{eq:spinrepCab}
C_{\alpha\beta}(\vbg,\vbf) = \int\dd{\mu}(\vbr,\vbk)\e^{i\vbk\vbf}\prod_{c}D^{(\rho_c,0)}_{j_cm_cj_cm_c}(g_c)\left(G^{\vbr}(\vbk)^{-1}\right)_{\alpha\beta}
\end{equation}
whereas the $(--)$-component is given by
\begin{equation}\label{eq:spinrepC--}
\begin{aligned}
&C_{\minus\minus}(\vb*{g},\vb*{\phi})
=
\int\dd{\mu}(\vbr,\vbk)\e^{i\vb*{k}\vbf}\prod_c D^{(\rho_c,0)}_{j_cm_cj_cm_c}(g_c)\left(G^{\vbr}(\vbk)^{-1}\right)_{\minus\minus}\\[7pt]
+&
\sum_{t=1}^4\sum_{(c_1,...,c_t)}\sumint\dd{\mu((\vbr\vbn)_t,\vbk)}\e^{i\vb*{k}\vbf}\prod_{c=c_1}^{c_t} D^{(0,\nu_c)}_{j_c m_c j_c m_c}(g_c)\prod_{c'=c_{t+1}}^{c_4}D^{(\rho_{c'},0)}_{j_{c'}m_{c'}j_{c'}m_{c'}}(g_{c'})\frac{1}{G_{\minus\minus}^{(\vbr\vbn)_t}(\vbk)},
\end{aligned}
\end{equation}
with $(\vbr\vbn)_t \equiv (\nu_{c_1},...,\nu_{c_t},\rho_{c_{t+1}},...\rho_{c_4})$ and where the sum-integral symbol denotes integration over the $\rho$'s and summation over the $\nu$'s.

\section{Explicit expressions for the matrix $\chi$}\label{sec:Explicit expressions for chi}

For the type of interactions considered in this work, the form of the matrix $\chi_{\alpha\beta}$ crucially depends on the interplay of combinatorics and the details of the causal structure. This is summarized with the notion of a causal vertex graph. This appendix intends to give an exhaustive list of all the possible causal vertex graphs if the underlying vertex graph is given by double-trace melonic, quartic melonic, necklace and standard (i.e. not colored) simplicial combinatorics. To every configuration, we provide the associated matrix $\chi_{\alpha\beta}$.

In order to keep the notation clean, we present the fields in the interactions as 
\begin{equation}
\Phi_{1234}^\alpha\equiv \int\dd{X_\alpha}\Phi(g_1,g_2,g_3,g_4,\vbf,X_\alpha),
\end{equation}
where repeated numbers imply a contraction via group integration. Also, we suppress the dependence on the scalar fields $\vbf$, as this is not of relevance for what is about to follow. The resulting $\chi_{\alpha\beta}$ matrices contain regularized symbols $\delta_{\rho,i}$ and we write for convenience
\begin{equation}
\delta_{\rho_1,i}\equiv\delta_1,\qquad \delta_{\rho_1,i}\delta_{\rho_2,i}\equiv \delta^2_{12},\qquad \prod_{c=1}^3\delta_{\rho_c,i}\equiv \delta^3_{123},\qquad \prod_{c=1}^4\delta_{\rho_c,i} \equiv\delta^4.
\end{equation}

Notice also that it suffices to perform the computation of a certain combination of signatures and combinatorics for one exemplary set of signatures, e.g. the simplicial case with $(n_\plus,n_0,n_\minus) = (2,2,1)$ will be similar to $(1,2,2)$, but with rows exchanged.\footnote{We remind the reader that $n_\alpha$ is the number of tetrahedra with signature $\alpha$, so $\np,\nz$ and $\nm$ are the numbers of spacelike, lightlike and timelike tetrahedra, respectively.} The form of the determinant of the effective kernel $G_{\alpha\beta}$ does not change and thus, the pole structure of the correlator remains unaffected. 

\newpage
\begin{table}[]
    \vspace{-5mm}
    \centering
    \begin{tabular}{c|c|c}
    \hline
    \multicolumn{3}{c}{$(\np,\nz,\nm) = (4,0,0)$}\\
    \hline
     $\Phi^{\plus}_{1234}\Phi^{\plus}_{1234}\Phi^{\plus}_{5678}\Phi^{\plus}_{5678}$  & {\scriptsize \cvftpppp} & $\chi = 4\left(2\delta^4+1\right)$  \\
    \hline
    $\Phi_{1234}^\plus\Phi_{4567}^\plus\Phi_{5678}^\plus\Phi_{8123}^\plus$   & {\scriptsize \cvfpppp} & $\chi = 4\delta^4+2\delta^3_{123}+2\delta^3_{234}+2\delta_1+2\delta_4$ \\
    \hline
    $\Phi^\plus_{1234}\Phi^\plus_{3456}\Phi^\plus_{5678}\Phi^\plus_{7812}$ & {\scriptsize \cvfnpppp} & $\chi = 4(\delta^4+\delta^2_{12}+\delta^2_{34})$\\
    \hline
    \multicolumn{3}{c}{$(\np,\nz\nm) = (5,0,0)$}\\
    \hline
    $\Phi^\plus_{1234}\Phi^\plus_{4567}\Phi^\plus_{7389}\Phi^\plus_{9620}\Phi^\plus_{0851}$ & {\scriptsize \cvfsppppp} & $\chi = 5(\delta^3_{123}+\delta^3_{234}+\delta^3_{341}+\delta^3_{412})$\\
    \hline
    \end{tabular}
    \caption{Interactions with a single type of signature (taken here as an example to be spacelike), their pictorial representation and the resulting $\chi$, which is in this case a scalar. Notice that the $\chi$'s agree with the functions $\mathcal{X}$ of~\cite{Marchetti:2020xvf,Marchetti:2022igl}.}
    \label{tab:400}
\end{table}

\begin{table}[]
    \hspace{-7mm}
    \begin{tabular}{c|c|c}
    \hline
    \multicolumn{3}{c}{$(\np,\nz,\nm) = (3,0,1)$}\\
    \hline
    $\Phi^{\plus}_{1234}\Phi^{\plus}_{1234}\Phi^{\plus}_{5678}\Phi^{\minus}_{5678}$     &  {\scriptsize \cvftpppm} & $    \chi = 
    \begin{pmatrix}
     4\delta^4+2\ & 2\delta^4+1\\[7pt]
     2\delta^4+1\ & 0
    \end{pmatrix}$\\
    \hline
    $\Phi_{1234}^\plus\Phi_{4567}^\plus\Phi_{5678}^\plus\Phi_{8123}^\minus$ & {\scriptsize \cvfpppm} & $ \chi = 
    \begin{pmatrix}
    2\delta^4+\delta^3_{123}+\delta_4+\delta^3_{234}+\delta_1\ & \delta^4+\delta^3_{123}+\delta_4\\[7pt]
    \delta^4+\delta^3_{234}+\delta_1\ & 0
    \end{pmatrix}$\\
    \hline
    $\Phi^\plus_{1234}\Phi^\plus_{3456}\Phi^\minus_{5678}\Phi^\plus_{7812}$    & {\scriptsize \cvfnpppm} & $\chi = 
    \begin{pmatrix}
    2(\delta^4+\delta^2_{12}+\delta^2_{34})\ & \delta^4+\delta^2_{12} +\delta^2_{34}\\[7pt]
    \delta^4+\delta^2_{12}+\delta^2_{34}\ & 0
    \end{pmatrix}$\\
    \hline
    \multicolumn{3}{c}{$(\np,\nz,\nm) = (4,0,1)$}\\
    \hline
    $\Phi^\plus_{1234}\Phi^\plus_{4567}\Phi^\plus_{7389}\Phi^\plus_{9620}\Phi^\minus_{0851}$ & {\scriptsize \cvfsppppm} & $\chi = \begin{pmatrix}
    3\sum_c\prod_{c'\neq c}\delta_{\rho_{c'},i}\ & \sum_c\prod_{c'\neq c}\delta_{\rho_{c'},i}\\[7pt]
    \sum_c\prod_{c'\neq c}\delta_{\rho_{c'},i}\ & 0
    \end{pmatrix}$\\
    \hline
    \end{tabular}
    \caption{Interactions with all but one tetrahedron of the same signature. As an example, all tetrahedra are spacelike except one timelike tetrahedron.}
    \label{tab:301}
\end{table}

\newpage
\begin{table}[]
\hspace{-15mm}
    \begin{tabular}{c|c|c}
    \hline
    \multicolumn{3}{c}{$(\np,\nz,\nm) = (2,0,2)$}\\
    \hline
    $\Phi^{\plus}_{1234}\Phi^{\plus}_{1234}\Phi^{\minus}_{5678}\Phi^{\minus}_{5678}$     & {\scriptsize \cvftppmma} & $\chi = 
   \begin{pmatrix}
    2\ & 4\delta^4\\[7pt]
    4\delta^4\ & 2    
   \end{pmatrix}$ \\
   \hline
   $\Phi^{\plus}_{1234}\Phi^{\minus}_{1234}\Phi^{\plus}_{5678}\Phi^{\minus}_{5678}$   & {\scriptsize \cvftppmmb} & $\chi = 
   \begin{pmatrix}
    2\delta^4\ & 2\delta^4+2\\[7pt]
    2\delta^4+2\ & 2\delta^4    
   \end{pmatrix}$\\
   \hline
    $\Phi_{1234}^\plus\Phi_{4567}^\minus\Phi_{5678}^\minus\Phi_{8123}^\plus$ & {\scriptsize \cvfppmma} & $ \chi = 
        \begin{pmatrix}
        \delta_1+\delta_4\ & 2\delta^4+\delta^3_{123}+\delta^3_{234}\\[7pt]
        2\delta^4+\delta^3_{123}+\delta^3_{234}\ & \delta_1+\delta_4
        \end{pmatrix}$\\
    \hline
     $\Phi_{1234}^\plus\Phi_{4567}^\plus\Phi_{5678}^\minus\Phi_{8123}^\minus$ & {\scriptsize \cvfppmmb}  & $\chi = 
        \begin{pmatrix}
        \delta^3_{123}+\delta^3_{234}\ & 2\delta^4+\delta_1+\delta_4\\[7pt]
        2\delta^4+\delta_1+\delta_4\ & \delta^3_{123}+\delta^3_{234},
        \end{pmatrix}$\\
    \hline
    $\Phi_{1234}^\plus\Phi_{4567}^\minus\Phi_{5678}^\plus\Phi_{8123}^\minus$ & {\scriptsize \cvfppmmc} & $\chi = 
        \begin{pmatrix}
        2\delta^4\ & 2\delta^3_{123}+2\delta_4\\[7pt]
        2\delta^3_{234}+2\delta_1\ & 2\delta^4
        \end{pmatrix}$\\
        \hline
    $\Phi^\plus_{1234}\Phi^\minus_{3456}\Phi^\minus_{5678}\Phi^\plus_{7812}$ & {\scriptsize \cvfnppmma} & $\chi = 
    \begin{pmatrix}
    \delta^2_{12}+\delta^2_{34}\ & 2\delta^4+\delta^2_{12} +\delta^2_{34}\\[7pt]
    2\delta^4+\delta^2_{12} +\delta^2_{34}\ & \delta^2_{12}+\delta^2_{34}
    \end{pmatrix}$\\
    \hline
    $\Phi^\plus_{1234}\Phi^\minus_{3456}\Phi^\plus_{5678}\Phi^\minus_{7812}$ & {\scriptsize \cvfnppmmb} & $    \chi = 
    \begin{pmatrix}
    2\delta^4\ & 2(\delta^2_{12} +\delta^2_{34})\\[7pt]
    2(\delta^2_{12} +\delta^2_{34})\ & 2\delta^4
    \end{pmatrix}$\\
    \hline
    \multicolumn{3}{c}{$(\np,\nz,\nm) = (3,0,2)$}\\
    \hline
    $\Phi^\plus_{1234}\Phi^\plus_{4567}\Phi^\plus_{7389}\Phi^\minus_{9620}\Phi^\minus_{0851}$ & {\scriptsize \cvfspppmm} & $ \begin{pmatrix}
    2(\delta^3_{123}+\delta^3_{234})+\delta^3_{341}+\delta^3_{412}\ & \delta^3_{123}+\delta^3_{234}+2(\delta^3_{341}+\delta^3_{412})\\[7pt]
    \delta^3_{123}+\delta^3_{234}+2(\delta^3_{341}+\delta^3_{412})\ & \delta^3_{123}+\delta^3_{234}
    \end{pmatrix}$\\
    \hline
    \end{tabular}
    \caption{Interactions with two types of signature, here $\np,\nm$, with $\np,\nm\geq 2$.\vspace{40mm}}
    \label{tab:202}
\end{table}

\newpage
\begin{table}[]
    \centering
    \begin{tabular}{c|c|c}
    \hline
    \multicolumn{3}{c}{$(\np,\nz,\nm) = (2,1,1)$}\\
    \hline
    &&\\[-7pt]
    $\Phi^{\plus}_{1234}\Phi^{\plus}_{1234}\Phi^{0}_{5678}\Phi^{\minus}_{5678}$     & {\scriptsize \cvftppmza} & $\chi = 
   \begin{pmatrix}
    2\ & 2\delta^4\ & 2\delta^4 \\[7pt]
    2\delta^4\ & 0\ & 1 \\[7pt]
    2\delta^4\ & 1\ & 0
   \end{pmatrix}$\\
   &&\\[-7pt]
   \hline
   &&\\[-7pt]
    $\Phi^{\plus}_{1234}\Phi^{0}_{1234}\Phi^{\plus}_{5678}\Phi^{\minus}_{5678}$     &  {\scriptsize \cvftppmzb} & $ \chi = 
   \begin{pmatrix}
    2\delta^4\ & \delta^4+1\ & \delta^4+1 \\[7pt]
    \delta^4+1\ & 0\ & \delta^4 \\[7pt]
    \delta^4+1\ & \delta^4\ & 0
   \end{pmatrix}$\\
   &&\\[-7pt]
   \hline
   &&\\[-7pt]
   $\Phi^\plus_{1234}\Phi_{4567}^0\Phi_{5678}^\minus\Phi_{8123}^\plus$ & {\scriptsize \cvfppmza} & $\chi = 
        \begin{pmatrix}
        \delta_1 + \delta_4\ & \delta^4+\delta^3_{123}\ & \delta^4+\delta^3_{234}\\[7pt]
        \delta^4+\delta^3_{234}\ & 0\ & \delta_1\\[7pt]
        \delta^4+\delta^3_{123}\ & \delta_4\ & 0
        \end{pmatrix}$\\
    &&\\[-7pt]
    \hline
    &&\\[-7pt]
    $\Phi_{1234}^\plus\Phi_{4567}^\plus\Phi_{5678}^0\Phi_{8123}^\minus$ & {\scriptsize \cvfppmzb} & $        \chi = 
        \begin{pmatrix}
        \delta^3_{123}+\delta^3_{234}\ & \delta^4+\delta_1\ & \delta^4+\delta_4\\[7pt]
        \delta^4+\delta_4\ & 0\ & \delta^3_{123}\\[7pt]
        \delta^4+\delta_1\ & \delta^3_{234}\ & 0
        \end{pmatrix}$\\
    &&\\[-7pt]
    \hline
    &&\\[-7pt]
    $\Phi_{1234}^\plus\Phi_{4567}^0\Phi_{5678}^\plus\Phi_{8123}^\minus$ & {\scriptsize \cvfppmzc} & $    \chi = 
    \begin{pmatrix}
    2\delta^4\ & \delta^3_{123}+\delta_4\ & \delta^3_{123}+\delta_4\\[7pt]
    \delta_1+\delta^3_{234}\ & 0\ & \delta^4\\[7pt]
    \delta_1+\delta^3_{234}\ & \delta^4\ & 0
    \end{pmatrix}$\\
    &&\\[-7pt]
    \hline
    &&\\[-7pt]
    $ \Phi^\plus_{1234}\Phi^0_{3456}\Phi^\minus_{5678}\Phi^\plus_{7812}$ & {\scriptsize \cvfnppmza} & $    \chi = 
    \begin{pmatrix}
    \delta^2_{12}+\delta^2_{34}\ & \delta^4+\delta^2_{12}\ & \delta^4+\delta^2_{34}\\[7pt]
    \delta^4+\delta^2_{34}\ & 0\ & \delta^2_{12}\\[7pt]
    \delta^4 + \delta^2_{12}\ & \delta^2_{34}\ & 0
    \end{pmatrix}$\\
    &&\\[-7pt]
    \hline
    &&\\[-7pt]
    $\Phi^\plus_{1234}\Phi^0_{3456}\Phi^\plus_{5678}\Phi^\minus_{7812}$ & {\scriptsize \cvfnppmzb} & $    \chi = 
    \begin{pmatrix}
    2\delta^4\ & \delta^2_{12}+\delta^2_{34}\ & \delta^2_{12}+\delta^2_{34}\\[7pt]
    \delta^2_{12}+\delta^2_{34}\ & 0\ &\delta^4\\[7pt]
    \delta^2_{12}+\delta^2_{34}\ & \delta^4\ & 0
    \end{pmatrix}$\\
    \hline
    \end{tabular}
    \caption{Quartic interactions with three different signatures.\vspace{20mm}}
    \label{tab:211}
\end{table}

\newpage
\begin{table}[]
    \centering
    \begin{tabular}{c c}
    \hline
    \multicolumn{2}{c}{$(\np,\nz,\nm) = (3,1,1)$}\\
    \hline
    &\\[-7pt]
    $\Phi^\plus_{1234}\Phi^\plus_{4567}\Phi^\plus_{7389}\Phi^0_{9620}\Phi^\minus_{0851}$ & {\scriptsize \cvfspppmz}\\
    \multicolumn{2}{c}{ $\chi = \begin{pmatrix}
    2(\delta^3_{123}+\delta^3_{234})+\delta^3_{341}+\delta^3_{412}\ & \delta^3_{123}+\delta^3_{341}+\delta^3_{412}\ &\delta^3_{234}+\delta^3_{341}+\delta^3_{412}\\[7pt]
    \delta^3_{123}+\delta^3_{341}+\delta^3_{412}\ & 0\ & \delta^3_{123}\\[7pt]
    \delta^3_{234}+\delta^3_{341}+\delta^3_{412}\ & \delta^3_{234}\ & 0
    \end{pmatrix}$}\\
    &\\[-7pt]
    \hline
    \multicolumn{2}{c}{$(\np,\nz,\nm) = (2,2,1)$}\\
    \hline
    &\\[-7pt]
    $\Phi^\plus_{1234}\Phi^\plus_{4567}\Phi^0_{7389}\Phi^0_{9620}\Phi^\minus_{0851}$ & {\scriptsize \cvfsppmzz}\\
    \multicolumn{2}{c}{$\chi = \begin{pmatrix}
    \delta^3_{123}+\delta^3_{234}\ & 2\delta^3_{412}+\delta^3_{123}+\delta^3_{341}\ &\delta^3_{234}+\delta^3_{341}\\[7pt]
    2\delta^3_{341}+\delta^3_{234}+\delta^3_{412}\ & \delta^3_{123}+\delta^3_{234}\ & \delta^3_{123}+\delta^3_{412}\\[7pt]
    \delta^3_{123}+\delta^3_{412}\ & \delta^3_{234}+\delta^3_{341}\ & 0
    \end{pmatrix}$}\\
    \hline
    \end{tabular}
    \caption{Simplicial interactions with all three signatures.}
    \label{tab:221}
\end{table}

\newpage
\bibliographystyle{JHEP}
\bibliography{references.bib}

\end{document}